\documentclass[aps,prx,notitlepage,superscriptaddress,longbibliography,nofootinbib,twocolumn]{revtex4-2}
\usepackage[utf8]{inputenc}
\usepackage{braket}
\usepackage{amsmath,amsthm,amssymb,amsfonts}
\usepackage{mathtools}
\usepackage{graphicx}
\usepackage{hyperref}
\usepackage{comment}
\usepackage{color}
\usepackage{makecell}
\usepackage{enumitem}
\usepackage{mathdots}
\usepackage{tikz}
\usepackage{epsfig}
\usepackage{blkarray}
\usepackage{colortbl}
\usetikzlibrary{patterns,patterns.meta}
\usepackage{xcolor}
\usepackage{bbm}
\usepackage{bm}
\usetikzlibrary{arrows.meta}
\usepackage[bb=boondox]{mathalfa}
\usepackage{adjustbox}
\usepackage{float}
\usepackage{tikz,tikz-3dplot,pgfplots}
\usetikzlibrary{shapes,arrows,patterns}
\usetikzlibrary{3d}
\usetikzlibrary{positioning}
\tikzset{
  diagrambox/.style={
    draw,
    minimum width=10mm,
    minimum height=10mm,
    inner sep=1pt,
    align=center,
    font=\footnotesize
  }
}
\usepackage{tikz-cd}
\numberwithin{equation}{section}
\usetikzlibrary{positioning,calc}

\newlength{\OVboxW}   
\newlength{\OlogboxW} 
\newlength{\OoneAboxW}
\newlength{\OoneBboxW}
\newlength{\OoneCboxW}
\newcommand{\boxedwithlabelW}[3]{%
  \tikz[baseline=(main.base)]{%
    \node[
      draw, rectangle,
      inner xsep=2pt, inner ysep=7pt,
      text height=2.5ex, text depth=1ex,
      text width=#1, align=center,
      outer sep=0pt
    ] (main) {$#3$};

    \path[use as bounding box] (main.north west) rectangle (main.south east);

    \begin{scope}[overlay]
      \node[
        draw, circle, minimum size=4.5mm, inner sep=0.2pt, fill=white,
        anchor=east
      ] at ([xshift=8pt,yshift=5pt] main.south east) {$\scriptstyle #2$};
    \end{scope}
  }%
}

\newlength{\GapColW}
\newcommand{\OoneSumThree}[3]{%
  \begingroup
  \setlength{\tabcolsep}{0pt}%
  \begin{tabular}[c]{@{}l@{\kern\GapColW$\mathord{+}$\kern\GapColW}l@{\kern\GapColW$\mathord{+}$\kern\GapColW}l@{}}
    \makebox[\OoneAboxW][c]{#1} &
    \makebox[\OoneBboxW][c]{#2} &
    \makebox[\OoneCboxW][c]{#3}
  \end{tabular}%
  \endgroup
}
\newcommand{\OoneSumTwo}[2]{%
  \begingroup
  \setlength{\tabcolsep}{0pt}%
  \begin{tabular}[c]{@{}l@{\kern\GapColW$\mathord{+}$\kern\GapColW}l@{}}
    \makebox[\OoneAboxW][c]{#1} &
    \makebox[\OoneBboxW][c]{#2}
  \end{tabular}%
  \endgroup
}
\newcommand{\symrow}{\rule[1ex]{0pt}{4.0ex}}

\newcommand{\Tr}{\operatorname{Tr}}

\newcommand{\ie}{{\it i.e.},\ }

\newcommand{\e}{\operatorname{e}}

\newcommand{\ro}{\rho}
\newcommand{\V}{V}

\begin{document}
\title{Random matrix prediction of average entanglement entropy\\
in non-Abelian symmetry sectors}
	
\author{Anwesha Chakraborty}
\email{anwesha.chakraborty@unimelb.edu.au}
\affiliation{School of Mathematics and Statistics, The University of Melbourne, Parkville, VIC 3010, Australia}
	
\author{Lucas Hackl}
\email{lucas.hackl@unimelb.edu.au}
\affiliation{School of Mathematics and Statistics, The University of Melbourne, Parkville, VIC 3010, Australia}
\affiliation{School of Physics, The University of Melbourne, Parkville, VIC 3010, Australia}
	
\author{Mario Kieburg}
\email{mkieburg@unimelb.edu.au}
\affiliation{School of Mathematics and Statistics, The University of Melbourne, Parkville, VIC 3010, Australia}
	
\begin{abstract}
We study the average bipartite entanglement entropy of Haar-random pure states in quantum many-body systems with global $\mathrm{SU}(2)$ symmetry, constrained to fixed total spin $J$ and magnetization $J_z = 0$. Focusing on spin-$\tfrac12$ lattices and subsystem fractions $f < \tfrac12$, we derive a asymptotic expression for the average entanglement entropy up to constant order in the system volume $\V$. In addition to the expected leading volume law term, we prove the existence  of a $\tfrac12\log \V$ finite-size correction resulting from the scaling of the Clebsch-Gordon coefficients and compute explicitly the $O(1)$ contribution reflecting angular-momentum coupling within magnetization blocks. Our analysis uses features of random matrix ensembles and provides a fully analytical treatment for arbitrary spin densities, thereby extending Page type results to non-Abelian sectors and clarifying how $\mathrm{SU}(2)$ symmetry shapes average entanglement.
\end{abstract}	
	
\maketitle

\section{Introduction}\label{sec:intro}

Entanglement captures non-classical correlations which is central to quantum theory. Among many measures, \textit{bipartite entanglement entropy} serves as a versatile diagnostic of this correlation across quantum information~\cite{PhysRevA.54.3824,PhysRevLett.78.2275}, phases of matter~\cite{RevModPhys.82.277}, black-hole information~\cite{page,Ryu_2006,PhysRevD.110.L061901}, and holography~\cite{PhysRevLett.130.053602}. Beyond its evaluation for individual states, a key question concerns the \textit{typical} entanglement across ensembles -- a research direction initialized by Page’s seminal result~\cite{PhysRevLett.44.301}, which established the celebrated ``Page curve’’ for uniformly distributed random pure states (also called Haar-random pure states) and showed that typical entanglement in large Hilbert spaces is close to maximal. The Page curve serves nowadays as a benchmark for chaotic quantum systems.

A major research direction now concerns the behavior of bipartite entanglement entropy in highly excited eigenstates of many-body quantum systems. In both, integrable and quantum-chaotic models, the average eigenstate entanglement obeys a leading volume-law scaling, in stark contrast to the area law~\cite{RevModPhys.82.277} familiar from ground states and low-energy excitations~\cite{PRXQuantum.3.030201}. Crucially, the \textit{coefficient} of the volume law acts as a sharp diagnostic. It is maximal in fully chaotic systems and systematically lower in integrable ones~\cite{PhysRevE.105.014109,PhysRevE.107.064119,Langlett:2025fam,Ruiz:2024mei}. Growing evidence~\cite{PhysRevE.100.062134,PhysRevB.108.245101,PRXQuantum.3.030201,Swietek:2023fka,Russotto:2025cpn,Yauk:2023wbu,PhysRevB.107.045102,PhysRevX.14.031014,PhysRevE.105.014109,mishra2025quantum} indicates that typical eigenstate entanglement reliably distinguishes chaos from integrability. From a broader perspective, such  entanglement measures may also provide statistical insight into quantum geometric states relevant to loop quantum gravity~\cite{Bianchi:2023avf,Ashtekar:2004eh,Ashtekar:2021kfp}.

While most physical Hamiltonian studies remain numerical, significant analytical progress has been made using \textit{random matrix theory} (RMT). Random matrices accurately describe universal eigenstate statistics of many-body quantum systems~\cite{Buijsman:2025gle,Fitter:2022vpr,Kim:2023ykr}, with quantum chaos famously associated to Wigner--Dyson spectral statistics~\cite{PhysRevE.82.031130}. Likewise, entanglement entropy~\cite{PhysRevLett.119.220603,PhysRevE.100.022131,Huang:2017std,PhysRevD.100.105010,PhysRevLett.124.050602,Russotto:2025cpn,Bianchi:2024aim,Patil:2025ump,Yauk:2023wbu} and the spectral form factor~\cite{PhysRevX.8.041019,PhysRevLett.123.210603,PhysRevX.8.021062} match RMT predictions to leading order. Haar-random pure states and uniformly distributed Gaussian fermionic states (with or without particle-number conservation) have enabled sharp analytical derivations of volume-law scaling~\cite{PhysRevLett.119.020601,lydzba2021single,lydzba2021entanglement,PhysRevB.103.L241118}, with deviations from maximality encoded only in $O(1)$ subleading corrections.

An especially active line of research concerns global \textit{symmetry constraints}. For systems with conserved particle number or magnetization, corresponding to a global ${\rm U}(1)$ symmetry, average entanglement entropy acquires a symmetry-induced correction scaling as $\sqrt{\V_A}$~\cite{PhysRevLett.119.220603}, with $\V_A$ the volume of the smaller of the two subsystems. The same effect appears in Haar-random states restricted to fixed charge sectors~\cite{PRXQuantum.3.030201,Yauk:2023wbu}, and similar constraints arise under energy conservation~\cite{PhysRevD.100.105010}. 

These developments naturally raise the question of how \textit{non-Abelian} symmetries modify entanglement entropy.
Recently, focus has shifted to the simplest non-Abelian subgroup ${\rm SU}(2)$, see~\cite{PhysRevB.108.245101,Russotto:2024pqg}, which are prevalent in systems with spin and angular momentum conservation, and play a significant role in quantum thermodynamics~\cite{Guryanova_2016,Yunger_Halpern_2016,PRXQuantum.3.010304,PRXQuantum.4.020318,Ma:2022vnd} and the eigenstate thermalization hypothesis~\cite{PhysRevLett.130.140402,PhysRevE.107.014130,PhysRevE.102.062113}. Understanding the average entanglement entropy in this constrained setting is not only of theoretical interest but it is also crucial for benchmarking entanglement behavior in spin systems with conserved angular momentum, as encountered in atomic, nuclear, and condensed matter physics. Moreover, unlike systems with only ${\rm U}(1)$ symmetry,  ${\rm SU}(2)$ symmetry introduces an additional layer of non-Abelian structure, entangling states across symmetry sectors in a more intricate manner.
	 
In this work, we study quantum many-body systems with \textit{global ${\rm SU}(2)$ symmetry}, where the total spin $J$ and its magnetic quantum number $J_z$ are conserved with the following central question: what is the average entanglement of pure states that are simultaneous eigenstates of the operators $\widehat{J}^2$ and $\widehat{J}_z$? Previously, some of us~\cite{PhysRevB.108.245101} analyzed random pure states with spin-$\tfrac12$ in sectors of $J$ and $J_z=0$. For $J=0$, it was shown that the average entanglement equals to that of a maximally mixed state within the ${\rm SU}(2)$ symmetric subspace—mirroring the leading volume law seen in highly excited eigenstates of quantum-chaotic Hamiltonians. For $J>0$, the problem was studied in the large volume $\V$ limit at fixed spin density $j=2J/\V$. Using two approximations justified by numerics, it predicted the leading order volume law of the entanglement entropy and indicated that there may be a subleading logarithmic correction. These entropies were compared to the entanglement entropy of eigenstates of integrable and quantum chaotic Hamiltonians with $\mathrm{SU}(2)$ symmetry. Consistent with earlier studies of systems with Abelian symmetries~\cite{PhysRevLett.119.220603,PhysRevE.100.062134,PRXQuantum.3.030201,Yauk:2023wbu}, eigenstates of quantum chaotic systems behave as Haar random states, while in integrable systems the entropy is consistently smaller for large subsystems. Summarizing, average entanglement entropy emerges as a robust diagnostic distinguishing quantum-chaotic systems from integrable systems in the presence, even in systems with non-Abelian ${\rm SU}(2)$ symmetry. 

Our aim here is to obtain a fully analytic derivation for the average entanglement entropy for Haar–random pure states restricted to the non-Abelian $\rm{SU}(2)$ sector of total spin $J$ and $J_z = 0$ and and give comparative discussion with previously obtained results in~\cite{PhysRevB.108.245101} for the total angular momentum of order $J=O(1), O(\V)$. We observe that unlike the singlet case $(J=0)$, where subsystem spins must match ($J_A = J_B$) and the Hilbert space admits a direct sum of tensor products structure, the general case $J>0$ introduces a highly nontrivial coupling between many admissible pairs ($J_A,J_B$). This destroys any simple tensor product decomposition and produces a reduced density matrix built from many correlated rectangular blocks, each weighted by Clebsch–Gordan (CG) coefficients and multiplicities that grow exponentially with system size. The central challenge is thus to understand which contributions dominate among this blocks.

To overcome this, we reorganize the reduced density matrix into fixed-magnetization sectors, as those are still good quantum numbers reflected in a direct sum of the density matrix. Then, we lift the fixed-trace constraint to a coupled and correlated Wishart ensemble. The spectrum of the resulting density matrix can then be accessed using Wigner’s moment method~\cite{Wigner}, \ie computing the asymptotic behavior of the averaged moments $\langle \Tr\rho^L\rangle$ as $\V\to\infty$. In this approach, each moment naturally expands into a weighted sum over rooted planar trees, where edges encode the admissible spin transitions and vertex weights are given by the subsystem's Hilbert space multiplicities. A crucial ingredient of our analysis is a dimensional selection principle -- because the multiplicities and, thence, the dimensions scale exponentially in the system size $\V$, almost all trees except two extremal ones give exponentially suppressed contributions.  The resulting eigenvalue density yields a clean volume-law term, a universal $\frac12 \log V$ correction, and an explicit $O(1)$ contribution arising from the Clebsch–Gordan structure. 
This provides an analytical derivation of average entanglement in ${\rm SU}(2)$-symmetric sectors with arbitrary total spin $J>0$, extending and justifying earlier numerical and other analyses. \\

The paper is organized as follows. In section~\ref{sec:review}, we review the model by recalling the ${\rm SU}(2)$ multiplicity structure and the bipartite decomposition of spin 1/2 system of $V$ total sites and their corresponding Hilbert space structure, followed by a derivation of the reduced density matrix for $J=0$ and for generic $J>0$ for random pure states. In subsection~\ref{sec:dim.asymp}, we present the asymptotic evaluation of the relevant Hilbert-space dimensions and ${\rm SU}(2)$ multiplicities required for the subsequent analysis. The concepts of dimensional selection and the scaling behavior of Clebsch--Gordan coefficients are outlined in subsections~\ref{sec:dim.select} and~\ref{sec:scaling}, respectively. Section~\ref{sec:new calc} contains the core analytical computations and results. In particular, subsection~\ref{sec:lift fix trace} introduces a mapping from random matrices drawn from a fixed-trace ensemble to an equivalent coupled Wishart ensemble, rendering the calculations analytically tractable. In subsections~\ref{sec:planar} and~\ref{sec:selection}, we employ Wigner's moment method to compute the leading and subleading asymptotic contributions to the average entanglement entropy and represent the resulting terms as rooted tree diagrams with vertex weights determined by ${\rm SU}(2)$ multiplicities. Finally, in section~\ref{sec:final result}, we summarize our main results and provide a comparative discussion with existing literature. We conclude in section~\ref{sec:conclusion} with an overview of the implications of our findings and an outlook for future directions.

\section{Review of the model}\label{sec:review}

Understanding the entanglement properties of quantum many-body systems with global symmetries requires a careful examination of the structure and scaling of their Hilbert spaces. In the case of spin lattices divided into subsystems, the multiplicities associated with different spin sectors play a central role in shaping the entanglement features of random pure states. In this section, we review the key aspects of subsystem Hilbert space dimensions incorporating the additional constraints imposed by the ${\rm SU}(2)$ symmetry, such as the restriction to the singlet sector with total angular momentum $J=0$ and then generalizing to arbitrary $J$ value. By characterizing the asymptotic behavior of multiplicities and their dependence on spin density, we establish the groundwork for the subsequent analysis of average entanglement entropy in symmetry-resolved ensembles.
    
\subsection{Hilbert space decomposition}\label{sec:Hilbert.space}
To quantify the entanglement in our spin lattice of $V$ sites, each with spin $\frac{1}{2}$, we perform a bipartite partition of the total Hilbert space $\mathcal{H}$. We consider distinguishable spins so that no (anti-) symmetrization is performed. The dimensionality of the spin lattice does not play any role in our discussion as we do not study any couplings between spins. Hence, the notion of locality is not of any importance here.
    
The system is divided into an observed subsystem $A$, containing $\V_A$ sites, and its complement $B$, containing the remaining $V - \V_A$ sites, such that
\begin{equation}
	\mathcal{H} ={\bigotimes}^V_{a=1}\mathcal{H}^a_{\frac{1}{2}}= \mathcal{H}_A \otimes \mathcal{H}_B.\label{tensor structure}
\end{equation}
where $\mathcal{H}_{\frac{1}{2}}=\text{span}\{\ket{\frac{1}{2}},\ket{-\frac{1}{2}}\}$ is a two dimensional Hilbert space of an individual site. For a pure state $\ket{\psi} \in \mathcal{H}$ of the total system, all information about subsystem $A$ is encoded in its reduced density matrix, $\rho_A$, obtained by tracing out the degrees of freedom of $B$ as $\rho_A=\text{Tr}_B(\ket{\psi}\bra{\psi})$. The entanglement between these two subsystems is then measured by the von Neumann entropy of subsystem $A$, defined by
\begin{equation}
	S_A(|\psi\rangle) = -\text{Tr}(\rho_A \log \rho_A).\label{entropy}
\end{equation}
This quantity serves as a direct measure of the quantum correlations between the two parts of the spin lattice. 
	
To investigate the effect of non-Abelian symmetry, namely ${\rm SU}(2)$ symmetry on average entanglement entropy, we restrict the pure state in a particular symmetry sector. For that we shift from the microscopic spin basis to the total angular momentum basis associated with the vector operator $\vec{J}=(\widehat J_x,\widehat J_y,\widehat J_z)$ defined by
\begin{equation}
	\Vec{J}\equiv {\sum}_{a=1}^V \vec{J}^{(a)},\qquad 
	\widehat J_z \equiv {\sum}_{a=1}^V \widehat J^{(a)}_z ,
\end{equation}
where $\vec{J}^{(a)}$ and $\widehat J_z^{(a)}$ are the angular momentum operator and its $z$-component, respectively, which act on the Hilbert space of the $a^{\mathrm{th}}$ spin site. The simultaneous eigenstates of $\vec{J}^2$ and $\widehat J_z$, denoted as $\ket{J,J_z}$ satisfy 
\begin{equation}
	\vec{J}^2 \ket{J,J_z}= J(J+1)\ket{J,J_z},\qquad
	\widehat J_z \ket{J,J_z} = J_z\ket{J,J_z} ,
\end{equation}
where $J$ is the total spin and $J_z$ is the corresponding total magnetic quantum number of the $V$ sites. This structure induces a decomposition of the full Hilbert space $\mathcal{H}$ into irreducible ${\rm SU}(2)$ sectors of fixed total spin,
	\begin{equation}
		\mathcal{H} \;=\; \bigoplus_{J}\, \mathcal{H}_J^{\oplus n_J^V}=\; \bigoplus_{J} \,\underbrace{\mathcal{H}_J \oplus \cdots \oplus \mathcal{H}_J}_{n^V_J\ \text{copies}}
		\;\,.
		\label{eq:2.4}
	\end{equation}
The allowed values of integer (half-integer) $J$ are from $0~(1/2)$ to $V/2$ for an even (odd) number of total sites $V$. Each Hilbert space $\mathcal{H}_J$ is the $(2J+1)$-dimensional irreducible subspace corresponding to a total spin $J$ which appears with a multiplicity $n_J$. For a chain of $V$ spin-$1/2$ degrees of freedom, the multiplicities admit a closed form~\cite{PhysRevB.108.245101}
	\begin{align}
		n_J^V \;=\; \frac{2(1+2J)}{2+V+2J}\binom{V}{V/2+J}.
		\label{eq:nJL}
	\end{align}
One can take the Haar-measure average of the entanglement entropy~\eqref{entropy} over random pure states restricted to simultaneous eigenstates of $\vec{J}^2$ and $\widehat{J}_z$. In particular, we concentrate on the total magnetic quantum number $J_z=0$ sector in the present work, meaning we only consider paramagnetic states. Surely, a non-zero $J_z$ will have an impact on the entanglement entropy as a maximal magnetic quantum number increases the information about the subsystems. For example, choosing $J=J_z=V/2$ implies that the ensemble of states consists only of a single state with all spins being up, which clearly is a product with respect to any system decomposition yielding zero entanglement entropy.
	
	When we split the system into two parts, see~\eqref{tensor structure}, each subsystem also decomposes into its own total spin sectors $ J_A $ and $ J_B $ such that the total spin $J$ is restricted by the triangle inequalities $ |J_A - J_B| \leq J \leq J_A + J_B $, according to principle of angular momentum addition.
	So the total Hilbert space $\mathcal{H}$ can be rewritten as
	\begin{equation}
		\bigoplus_{J=0}^{V/2} \mathcal{H}_J = \bigoplus_{J_A = 0}^{\V_A/2} \mathcal{H}_{J_A}^{\oplus n_{J_A}^A} \otimes \bigoplus_{J_B = 0}^{\V_B/2} \mathcal{H}_{J_B}^{\oplus n_{J_B}^B},\label{eq:2.1.2}
	\end{equation}
	The multiplicity of states in each of the subspaces $\mathcal{H}_{J_A}$ and $\mathcal{H}_{J_B}$ are respectively equal to $n^A_{J_A}=n_{J_A}^{\V_A} $ and $ n^B_{J_B}=n_{J_B}^{\V_B}$ (calculated as $n_J^V$ from~\eqref{eq:nJL})\footnote{Like~\eqref{eq:2.4}, the lower bound of total angular momenta in each subspace $\mathcal{H}_{J_A(J_B)}$ starts from either $0$ or $1/2$ when $\V_A$ is respectively even or odd. Since we shall be dealing with asymptotic limits for large $V$ the difference due to the lower bound is expected to be insignificant, which is why we shall be using $\min(J)=0$.}.
    
    We will denote the set of eligible pairs of angular momentum quantum numbers $(J_A,J_B)$ by
    \begin{equation}\label{J.set}
    \begin{split}
     \mathfrak{J}=&\ \{(J_A,J_B)\in(\mathbb{N}_0/2)^2:J_A\leq \tfrac{V_A}{2},\ J_B\leq \tfrac{V_B}{2},\\
     &|J_A-J_B|\leq J\leq J_A+J_B,\\
     &{\rm and}\quad {\rm mod}_1(J_A+J_B)={\rm mod}_1(J)\}.
     \end{split}
    \end{equation}
The modulus guarantees that $J_A+J_B$ is a (half-)integer if $J$ is one. In the present work, we will actually consider the case $J_z=0$ which is only possible when $J$ is an integer. Under this restriction it is ${\rm mod}_1(J_A+J_B)={\rm mod}_1(J)=0$, meaning $J_A$ and $J_B$ must both either integers or half-integers.

We also introduce two other index sets which come in handy later on, namely the first is all $J_B$ under the condition of a fixed $J_A$,
    \begin{equation}\label{JB.cond.JA}
    \begin{split}
     \mathfrak{J}_B^{(J_A)}=&\ \biggl\{J_B\in\mathbb{N}_0/2:|J-J_A|\leq J_B\leq \min\left[\frac{\V_B}{2},J+J_A\right]\\
     &{\rm and}\quad {\rm mod}_1(J_A+J_B)={\rm mod}_1(J)\}
     \end{split}
    \end{equation}
    with $J_A\in[\max\{0,J-V_B/2\},\min\{V_A/2,J+V_B/2\}]$,
    and the set of all possible $J_A$ under the condition of a given $J_B$,
    \begin{equation}\label{JA.cond.JB}
    \begin{split}
     \mathfrak{J}_A^{(J_B)}=&\ \biggl\{J_A\in\mathbb{N}_0/2:|J-J_B|\leq J_A\leq \min\left[\frac{\V_A}{2},J+J_B\right]\\
     &{\rm and}\quad {\rm mod}_1(J_A+J_B)={\rm mod}_1(J)\}\\
     \end{split}
    \end{equation}
    with $J_B\in[\max\{0,J-V_A/2\},\min\{V_B/2,J+V_A/2\}]$,
    see~\cite{PhysRevB.108.245101} for the derivation of the restrictions.
	
\subsection{Case $J=0$}\label{sec:J0}

To understand the difference to the more challenging case $J>0$, we will briefly review the construction of the reduced density matrix for subsystem $A$ for the case $J=0$, based on~\cite{PhysRevB.108.245101}.
    
For total spin $J=0$, the only possible condition on subsystem's total spins is $J_A=J_B$ and the corresponding Hilbert space sector $\mathcal{H}_{J=0, J_z=0} $ can be written as~\eqref{eq:2.1.2} 
\begin{equation}
		\mathcal{H}_{J=0,J_z=0} = {\bigoplus}_{J_A = 0}^{\V_A/2} \mathcal{H}_{J_A}^{J=0,J_z=0},\label{eq:2.1.4}
\end{equation}
where $ \mathcal{H}^{J=0,J_z=0}_{J_A} \subset \mathcal{H}^{\oplus n^{V_A}_{J_A}}_{J_A} \otimes \mathcal{H}^{\oplus n^{V_B}_{J_A}}_{J_A}$ contains $ n^{V_A}_{J_A} \times n^{V_B}_{J_A} $ states that have zero total spin\footnote{We are considering $\V_A< \V_B$, so that the upper bound of $J_A$ in table~\ref{tab:bounds} reduces to $\V_A/2$.}. It shows how the total spin $J$ sector is built entirely from matched local spin sectors in subsystems $A$ and $B$.
    
We see in~\eqref{eq:2.1.4} that the corresponding Hilbert space sector $\mathcal{H}_{J=0,J_z=0}$ can be expressed as a direct sum of subspaces of tensor products. The ability to write the Hilbert space in this form greatly simplifies subsequent calculations and is unique to $J=0$. It means that the entanglement entropy can be computed in the same way, as if the Hilbert space were a direct sum of tensor products, which was discussed in~\cite{PhysRevD.100.105010}. In ensuing sections, we will observe that for $J>0$, such a clean decomposition in a direct sum of tensor products is no longer possible.
	
Recalling that the total magnetic quantum number $J_z=0$ vanishes, it must be that the two subsystems have also opposite magnetic quantum number $m$ and $-m$. Hence, the basis for Hilbert space~\eqref{eq:2.1.4} is
\begin{equation}
		\ket{\psi_{ab}^{J_A}} = {\sum}_{m=-J_A}^{J_A} c_m(J_A) \ket{J_A, m}_a \otimes \ket{J_A, -m}_b, \label{eq:2.1.5}
\end{equation}
where $ m $ is the  eigenvalue of $ \widehat{J}_z $ within subsystem A and $ a, b $  labels the degeneracy of the state $\ket{J_A,m}$ and $\ket{J_A,-m}$ in their respective Hilbert spaces given by the numbers $n^A_{J_A}, n^B_{J_B} $ and $ c_m(J_A) $ is the Clebsch-Gordan (CG) coefficient given by $\braket{J_A, m; J_A, -m | J = 0, J_z = 0}$.
A random pure state in the subspace $ \mathcal{H}^0_{J_A} $ is the superposition of the base states $ |\psi_{ab}\rangle $ in~\eqref{eq:2.1.5}, \ie
\begin{equation}
	\ket{\psi^0_{J_A}} = {\sum}_{a=1}^{n^A_{J_A}} {\sum}_{b=1}^{n^B_{J_A}} W_{ab}^{J_A} \ket{\psi_{ab}^{J_A} }, \label{eq:2.1.7}
\end{equation}
where $ W=\{W_{ab}^{J_A}\} $ is a complex $n^A_{J_A}\times n^B_{J_A}$ random matrix drawn from the fixed trace ensemble, in particular it is $ \Tr(W W^\dagger) = 1 $ and $W$ is uniformly distributed on the sphere defined by this condition.
    
The reduced density matrix of subsystem $A$ can be written in terms of this random matrix as follows
\begin{equation}
	\rho_A =\Tr_B | \psi^0_{J_A} \rangle \langle \psi^0_{J_A} |= \sum_{a,a',m} R^{(m)}_{aa'} \ket{J_A, m}_a \bra{J_A, m}_{ a'} . \label{eq:2.1.8}
\end{equation}
The matrix is block-diagonal in the subspaces of fixed $m$, with each block given by $R^{(m)}_{aa'} = |c_m|^2 (W W^\dagger)_{aa'}$, which eventually gives two contribution in the entropy formula -- one due to the CG coefficients and the other due to the fixed-trace ensemble~\cite{PRXQuantum.3.030201,PhysRevB.108.245101}. For large $\V$, this yields~\cite{PhysRevB.108.245101}
\begin{align}\label{case J=0}
    \hspace{-4mm}\braket{S_A}_{J=0,J_z=0}=\log(2)\V+\tfrac{3(f+\log(1-f))}{2}-\tfrac{1}{2}\delta_{f,\frac{1}{2}}\,.
\end{align}
In the following, it will be our goal to rigorously derive the equivalent formula for $J=O(\V)$ and $0<f<\frac{1}{2}$.

\subsection{Case $J>0$}\label{sec:J}
        
In contrast to the $J=0$ case~\eqref{eq:2.1.4}, the random states can no longer be expressed as direct sums in $J_A$ due to a multitude of pairings between $J_A$ and $J_B$ to produce a total spin $J\ne 0$. This creates interactions between different $J_A$ sectors. Yet, the magnetic quantum number of the two subsystems must still add to zero due to our choice $J_z=0$ implying an integer $J$. Therefore, a basis $\{ |\phi_{ab}\rangle \}$ of the constrained Hilbert space $\mathcal{H}_{J, J_z=0}$, expressed in terms of $|J_A, m \rangle \otimes |J_B, -m \rangle$, can be written as
		\begin{equation}
			|\phi_{ab}^{J_AJ_B}\rangle =\sum_{m\in \mathfrak{M}^{(J_A,J_B)}} c_m( J_A, J_B) |J_A, m\rangle_a \otimes |J_B, -m\rangle_b,\label{eq:2.1.26}
		\end{equation}
		where $ c_m( J_A, J_B) = \langle J_A, m; J_B, -m | J, 0 \rangle $ is the corresponding CG coefficient
        and we sum over the set
        \begin{equation}\label{M.JAJB.def}
        \begin{split}
            \mathfrak{M}^{(J_A,J_B)}=&\ \{m\in\mathbb{Z}/2: |m|\leq\min\{J_A,J_B\},\\
            & {\rm mod}_1(m)={\rm mod}_1(J_A)={\rm mod}_1(J_B)\\
            &{\rm and}\ m\neq0\ {\rm if}\ {\rm mod}_2(J-J_A-J_B)=1\}.
        \end{split}
        \end{equation}
        Once again, the first modulus guarantees that $m$ is a (half-)integer if $J_A$ and $J_B$ are. We note that $J_A$ and $J_B$ must be both integers or half-integers as their sum must be an integer because $J$ is. The second modulus reflects that the Clebsch-Gordan (CG) coefficients are antisymmetric in the magnetic quantum number $m=-\min\{J_A,J_B\},\ldots,\min\{J_A,J_B\}$ of the subsystems when $J-J_A-J_B$ is an odd integer. Hence, the corresponding wavefunctions for $m=0$ are not part of the Hilbert space with fixed total quantum numbers $J$ and $J_z=0$. This is the reason why they are excluded.  
        
        A random pure state can be written as a linear superposition of the basis vectors $\ket{\phi_{ab}}$,
		\begin{equation}
			\ket{\phi^J}=\sum_{(J_A,J_B)\in\mathfrak{J}}\sum_{a=1}^{n_{J_A}^A}\sum_{b=1}^{n_{J_B}^B} W_{ab}^{(J_A,J_B)} \ket{\phi_{ab}^{J_AJ_B}}\label{phi_J},
		\end{equation}
		where the bounds on $J_A$ and $J_B$ are given in~\eqref{J.set}. The corresponding reduced density matrix can, essentially, be decomposed into irreducible spin sectors labeled by $J_A$ as $\rho_{J_A}$. The corresponding Hilbert space decomposition according to~\eqref{eq:2.1.2} is given by $\mathcal{H}_A={\bigoplus}_{J_A}\mathcal{H}_{J_A}$. Each of these $\mathcal{H}_{J_A}$ is a well-defined representation space of dimension $2J_A+1$, spanned by the magnetic quantum numbers $m=-J_A,\dots,J_A$.
		This is the natural decomposition dictated by angular momentum representation theory. 
        
        In contrast, we can also reorganize the reduced density matrix $\rho_A$ to reflect that it is a direct sum over fixed values of $m$ ranging from $-\V_A/2$ to $\V_A/2$,
		\begin{equation}
			\rho_A={\bigoplus}_{m=-\V_A/2}^{\V_A/2} ~~\rho_A^{(m)},\label{block diag}
		\end{equation}
		considering $\V_A< \V_B$. The direct sum over $m$ comprises integers and half-integers. Since $|m|\leq \text{min}[J_A, J_B]$, together with $|J_A-J_B|\leq J\leq J_A+J_B$, and $J_A \leq \frac{\V_A}{2} ,J_B\leq \frac{\V_B}{2}$, the resulting admissible range of $J_A,J_B$ for a given $m$ is 
		\begin{align}
			&\max(|m|,\phantom{|}J-\V_B/2\phantom{|}\!)\leq J_A\leq \V_A/2\,,\hfill\label{JA_JB_new bound}\\
			&\max(|m|,|J-J_A|)\leq J_B\leq \min(J+J_A,\V_B/2),\nonumber
		\end{align}
see table~\ref{tab:bounds}.

		\begin{table}[t!]
			\centering
			\renewcommand{\arraystretch}{1.6}
			\begin{tabular}{cll}
				\hline
				\hline
				\textbf{Label}\hspace{5mm} & \textbf{Lower bound}\hspace{10mm}  & \textbf{Upper bound} \\
				\hline
				$m$ & $-\V_A/2$ & $\V_A/2$ \\
				$J_A$   & $\max(|m|,J-\tfrac{\V_B}{2})$ & $\V_A/2$\\
				$J_B$   &     $\max(|m|,|J-J_A|)$ & $\min(J+J_A,\tfrac{\V_B}{2})$\\
                $\mu$   & $-f\sqrt{V}/2$    &   $f\sqrt{V}/2$\\
				$j_A$    & $\max(2|\mu|/\sqrt{V},|j+f-1|)$  & $f$\\
				$j_B$   & $\max(2|\mu|/\sqrt{V},|j-j_A|)$ & $\min(j+j_A,1-f)$\\
				$a$    & 1 & $n_{J_A}^A$\\
				$b$   & 1 & $n_{J_B}^B$\\
				\hline
				\hline
			\end{tabular}
			\caption{List of upper and lower bounds of various labels used to define the reduced density matrix. We rescale the quantities as follows: $V_A=f V$, $V_B=(1-f)V$, $J=j V/2$, $J_A=j_AV/2$, $J_B=j_BV/2$, and $m=\mu\sqrt{V}$.}
		\label{tab:bounds}
		\end{table}
        
        The corresponding set of eligible pairings $(J_A,J_B)$ is denoted by
        \begin{equation}\label{Jm.def}
        \begin{split}
            \mathfrak{J}_m=&\ \{(J_A,J_B)\in\mathfrak{J}:\ \min\{J_A,J_B\}\geq|m|\\
        &{\rm and}\ {\rm mod}_2(J-J_A-J_B)=0\ {\rm if}\ m=0\},
        \end{split}
        \end{equation}
        where we have, anew, taken into account that some states do not exist in the considered Hilbert space for $m=0$.
        When we also condition a single $J_B$ or $J_A$ to a given $J_A$ or $J_B$, respectively, we will use the notation
        \begin{equation}\label{JBm.def}
        \begin{split}
            \mathfrak{J}_{B,m}^{(J_A)}=&\ \{J_B\in\mathfrak{J}_B^{(J_A)}:\ \min\{J_A,J_B\}\geq|m|,\\
        &{\rm mod}_1(J_A)={\rm mod}_1(m)\\
        &{\rm and}\ {\rm mod}_2(J-J_A-J_B)=0\ {\rm if}\ m=0\}
        \end{split}
        \end{equation}
        and
        \begin{equation}\label{JAm.def}
        \begin{split}
            \mathfrak{J}_{A,m}^{(J_B)}=&\ \{J_A\in\mathfrak{J}_A^{(J_B)}:\ \min\{J_A,J_B\}\geq|m|,\\
        &{\rm mod}_1(J_B)={\rm mod}_1(m)\\
        &{\rm and}\ {\rm mod}_2(J-J_A-J_B)=0\ {\rm if}\ m=0\}.
        \end{split}
        \end{equation}
        
        So writing $\rho_A$ as a direct sum over $m$ does not change the physics, it simply reorders the state into fixed magnetization blocks and correspondingly  reshuffles ($J_A ,J_B$) pairs populating each block, subject to constant total $J$. It is advantageous in the computation of the average entanglement entropy because the trace of the whole density matrix reduces to a sum of traces of density matrices with fixed a magnetic quantum number $m$. It also solves various other technical analytical issues later on. 
        
        We use this representation to write a random wave vector in $\mathcal{H}_{J,J_z=0}$ and its corresponding reduced density matrix can be as $\ket{\phi^J}=\oplus^{V_A/2}_{m=-V_A/2}\ket{\phi}^{(m)}$ and $\rho_A=\oplus^{V_A/2}_{m=-V_A/2}\rho_A^{(m)}$ with~\cite{PhysRevB.108.245101}
		\begin{small}
			\begin{align}
				\ket{\phi}^{(m)}&=\sum_{\substack{J_A,J_B\\a,b}}\!\!\! W_{ab}^{(J_A, J_B)} c_m( J_A, J_B) |J_A, m\rangle_a \otimes |J_B, -m\rangle_b,\label{eq:2.1.27} \\
				\rho^{(m)}_A&=\sum_{\substack{J_A, J_A'\\a,a'}}\sum_{J_B\in\mathfrak{J}_{B,m}^{(J_A)}\cap\mathfrak{J}_{B,m}^{(J'_A)}}\{\widehat{W}_m^{(J_A,J_B)}(\widehat{W}_m^{(J_A',J_B)})^\dagger\}_{aa'}\nonumber\\[-15pt]
				&\hspace{110pt}\times|J_A, m\rangle_a \langle J_A', m |_{a'}\,.\label{eq:2.1.28}
			\end{align}
		\end{small}
        where we abbreviated with the help of the matrix blocks
        \begin{equation}\label{W.hat.def}
            \widehat{W}_m^{(J_A,J_B)}=\{c_m( J_A, J_B)W_{ab}^{(J_A,J_B)}\}_{\substack{a=1,\ldots,n_{J_A}^{A}\\ b=1,\ldots,n_{J_B}^{B}}}.
        \end{equation}
        The coefficients $W_{ab}^{(J_A,J_B)} \in \mathbb{C}$ are complex random variables with zero mean. They are uniformly drawn from the sphere that is defined by the normalization of the state $|\phi^J\rangle$ in~\eqref{phi_J} given explicitly as follows
		\begin{equation}
			\begin{split}
				1=\ &\Tr\rho_A={\sum}_{m}\Tr \rho_A^{(m)}\\
				=\ &{\sum}_m{\sum}_{J_A, J_B} {\sum}_{a,b} \lambda_m^{(J_A,J_B)} |W_{ab}^{(J_A, J_B)}|^2\\
				=\ &{\sum}_{J_A, J_B} {\sum}_{a,b} |W_{ab}^{(J_A, J_B)}|^2.
			\end{split}
			\label{eq:2.1.29}
		\end{equation}
        with the abbreviation
        \begin{equation}\label{lambda.def}
            \lambda_m^{(J_A,J_B)}=|c_m(J_A,J_B)|^2.
        \end{equation}
In the last line of~\eqref{eq:2.1.29}, we interchanged the sums so that the one of $m$ is carried out first. Then, we employed the normalization of the CG coefficients for fixed $J,J_A,J_B$, \ie
\begin{align}
    \sum_{m\in \mathfrak{M}^{(J_A,J_B)}}\lambda_m^{(J_A,J_B)}=1\,.\label{eq:CG-normalization}
\end{align}

Next, we discuss the asymptotic behavior of various Hilbert space dimensions that will be relevant in our calculation.

\subsection{Dimensions and their asymptotic}\label{sec:dim.asymp}
    
One paramount ingredient in analyzing the average entanglement entropy are the asymptotic behavior of the dimensions
\begin{align}
    n_{J_A}^A={}&\frac{\alpha(j_A/f)}{\sqrt{f\V}}\e^{\beta(j_A/f)f\V}(1+O(\tfrac{1}{\V}))\,,\label{eq:n1-asymptotics}\\
	n_{J_B}^B={}&\frac{\alpha(j_B/(1-f))}{\sqrt{(1-f)\V}}\e^{\beta(j_B/(1-f))(1-f)\V}(1+O(\tfrac{1}{\V})),\label{eq:n2-asymptotics}
\end{align}
where we used the double-scaling $j=2J/V$, $j_A=2J_A/V$,  $j_B=2J_B/V$ and $V_A=f V$ when $V\to\infty$,
see table~\ref{tab:bounds}. The functions introduced above are
\begin{align}
	\alpha(\tilde j)&=\sqrt{\frac{2}{\pi}}\frac{2\tilde j}{(1+\tilde j)\sqrt{1-\tilde j^2}}\,,\label{alpha.gamma.def}\\
	\beta(\tilde j)&=-\frac{1+\tilde j}{2}\log\frac{1+\tilde j}{2}-\frac{1-\tilde j}{2}\log\frac{1-\tilde j}{2}\,.\label{eq:beta}
\end{align}
Especially, the function $\beta(\tilde j/\tilde f)\tilde f$ plays an important role in selecting the leading contributions. It is symmetric about the origin and concave in $\tilde j$. Additionally, it is strictly decreasing in $\tilde j\in [0,\tilde f]$.
        
This we see already for the total dimension of the Hilbert space $\mathcal{H}_{J,J_z=0}$ which is
\begin{equation}
    d={}\sum_{(J_A,J_B)\in\mathfrak{J}}n_{J_A}^An_{J_B}^B=\frac{\alpha(j)}{\sqrt{\V}}\e^{\beta(j)\V}(1+O(\tfrac{1}{\V}))\,.\label{eq:d-asymptotics}
\end{equation}
It can be calculated by introducing the variable
\begin{align}
    \tilde{n}^B_{J_A}=\sum_{J_B\in\mathfrak{J}_B^{(J_A)}}n_{J_B}^B\,,\label{eq:ntilde}
\end{align}
which allows us to write the sum as $d={}\sum_{J_A}n_{J_A}^A\tilde{n}_{J_A}^B$. 

We can evaluate~\eqref{eq:ntilde} and approximate the sum at its lower terminal $J_B^{\min}=|J-J_A|$ as $\beta(j_B/(1-f))$ is strictly decreasing. In particular, we substitute $j_B=|j-j_A|+2k/V$ with $k\in\mathbb{N}_0$ an integer. Then, we Taylor-expand in $k$ yielding a geometric series,
\begin{align}
	\tilde{n}^B_{J_A}&=\sum^{J_B^{\max}-J_B^{\min}}_{k=0}\frac{\alpha(|j-j_A|/[1-f]+2k/\V_B)}{\sqrt{\V_B}}\nonumber\\
    &\quad\times\e^{\V_B\beta(|j-j_A|/[1-f]+2k/\V_B)}(1+O(\tfrac{1}{\V}))\nonumber\\
	&=n_{|J-J_A|}^B\sum^{\infty}_{k=0}\e^{2\beta'(|j-j_A|/[1-f])k}(1+O(\tfrac{1}{\V}))\nonumber\\
	&=\frac{n_{|J-J_A|}^B}{1-\gamma(|j-j_A|/[1-f])}(1+O(\tfrac{1}{\V}))\,.\label{eq:pre-final-n2}
\end{align}
with $J_B^{\max}=\min[J+J_A,V_B/2]$ and
\begin{equation}
    \gamma(\tilde j)=\e^{2\beta'(\tilde j)}=\frac{1-\tilde j}{1+\tilde j}\,.\label{eq:gamma}
\end{equation}
The prefactor $1/[1-\gamma(\tilde j)]=[1+\tilde j]/[2\tilde j]$ is evidently only algebraic so that it cannot compete with the exponential decay in the term $n_{J_A}^An_{|J-J_A|}^B$ given by the function in the exponent
\begin{equation}\label{B0.5}
    B_{1/2}(j_A,f)=\beta\left(\frac{j_A}{f}\right)f+\beta\left(\frac{j-j_A}{1-f}\right)(1-f).
\end{equation}
This function has a maximum at $j_A=fj$ as can be readily checked from its saddle point equation
\begin{equation}
    \begin{split}
    0&=\beta'\left(\frac{j_A}{f}\right)-\beta'\left(\frac{j-j_A}{1-f}\right)\\
    &=\frac{1}{2}\log\left(\frac{(f-j_A)(1-f+j-j_A)}{(f+j_A)(1-f-j+j_A)}\right).
    \end{split}
\end{equation}
Expanding in $2J_A/V=j_A=fj+\delta j_A/\sqrt{V}$ yields a quadratic approximation
\begin{equation}
    B_{1/2}(j_A,f)=\beta\left(j\right)-\frac{\delta j_A^2}{2f(1-f)(1-j^2)V}+O\left(\frac{\delta j^3}{V^{3/2}}\right).
\end{equation}
The prefactor gives
\begin{equation}
\begin{split}
&\,\frac{\alpha(j_A/f)\alpha(|j-j_A|/[1-f])}{\sqrt{f(1-f)}V(1-\gamma(|j-j_A|/[1-f])}\\
=&\sqrt{\frac{2}{\pi}}\frac{\alpha(j)}{\sqrt{f(1-f)(1-j^2)}V}\left(1+\Big(\frac{\delta j_A}{\sqrt{V}}\Big)\right)\,
\end{split}
\end{equation}
for which we could use
\begin{align}
	\alpha(\tilde j)&=\sqrt{\frac{2}{\pi}}\frac{1-\gamma(\tilde j)}{\sqrt{1-\tilde j^2}}.
\end{align}
After taking the limit of the Riemannian sum in terms of a Gaussian integral we arrive at~\eqref{eq:d-asymptotics}. A useful byproduct of this computation is the Gaussian approximation
\begin{align}\label{gaussian-peak}
 &\frac{n_{J_A}^{A} \tilde{n}^B_{J_A}}{d}=\frac{n_{J_A}^{A} n_{|J-J_A|}^B}{d\big(1-\gamma(|j-j_A|/[1-f])\big)}\\
 ={}&\frac{2}{\V}\frac{1}{\sqrt{2\pi\sigma^2}}\exp\left[-\frac{(j_A-f j)^2}{2\sigma^2}\right]\left(1+O\Big(\frac{|j_A-fj|^3}{\sqrt{V}}\Big)\right),\nonumber
\end{align}
with the variance $\sigma^2=\frac{f(1-f)(1-j^2)}{V}$. In sections~\ref{sec:T1} and~\ref{sec:T2}, we will rewrite this Gaussian in the rescaled variables $\delta j_A=\sqrt{\V}(j_A-fj)$, for which the variance becomes independent of $\V$.

The formulas presented in this subsection will be important for the asymptotic analysis of the entanglement entropy at large $\V$.

\subsection{Idea of the dimensional selection }\label{sec:dim.select}

We will compute the entanglement entropy via a replica approach based on Wigner's moment method~\cite{Wigner}. In this approach, we will evaluate the moments in a diagrammatic way where the diagrams are weighted with the dimensions of combinations  $(n_{J_A}^A)^{2\kappa}(n_{J_B}^B)^{2(1-\kappa)}$ at $J_B=|J-J_A|$ or $J_A=|J-J_B|$ with specific $\kappa\in(0,1)$.

The exponential behavior of these powers is reflected in the function
\begin{equation}\label{Bepsilon}
    B_{\kappa}(\tilde j,\tilde f)=2\kappa\beta\left(\frac{\tilde j}{\tilde f}\right)\tilde f+2(1-\kappa)\beta\left(\frac{j-\tilde j}{1-\tilde f}\right)(1-\tilde f)
\end{equation}
with any $\tilde f\in(0,1)$. The function $B_{\kappa}(\tilde j,\tilde f)$ is concave on $\tilde j\in[\max\{0,j-1+\tilde f\},\min\{\tilde f,j+1-\tilde f\}]$ because of the second derivative
\begin{equation}\label{2nd.der.B}
    \partial_{\tilde j}^2B_{\kappa}(\tilde j,\tilde f)=-\frac{\kappa}{\tilde f^2-\tilde j^2}-\frac{1-\kappa}{(1-\tilde f)^2-(j-\tilde j)^2}\leq0.
\end{equation}
Additionally, its first derivative
\begin{equation}\label{saddle.B.a}
    \partial_{\tilde j}B_{\kappa}(\tilde j,\tilde f)=2\kappa\beta'\left(\frac{\tilde j}{\tilde f}\right)-2(1-\kappa)\beta'\left(\frac{j-\tilde j}{1-\tilde f}\right)
\end{equation}
is positive at the lower terminal $0$ when $ j-1+\tilde f<0$ or diverges to positive infinity when $ j-1+\tilde f\geq0$ and goes to to negative infinity at the upper terminal $\tilde f$ when $j+1-\tilde f>\tilde f$ as well  when $\tilde j\to j+1-\tilde f$ for $j+1-\tilde f<\tilde f$. To see this one needs to use that the symmetry $\beta(\tilde j)=\beta(-\tilde j)$ implies the antisymmetry in its first derivative, $\beta'(\tilde j)=-\beta'(-\tilde j)$. Summarizing, there is a unique maximum given by the saddle point equation
\begin{equation}\label{saddle.B}
    \partial_{\tilde j}B_{\kappa}(\tilde j,\tilde f)|_{\tilde j=j_{\kappa,\tilde f}}=0,
\end{equation}
which we denote by $j_{\kappa,\tilde f}\in[\max\{0,j-1+\tilde f\},\min\{\tilde f,j+1-\tilde f\}]$. This is true regardless whether $\tilde f$ is larger or smaller than $1/2$.

Moreover, let us mention the useful reflection symmetry
\begin{equation}\label{B.symmetry}
   B_{\kappa}(\tilde j,\tilde f)=B_{1-\kappa}(j-\tilde j,1-\tilde f)
\end{equation}
which originates in the equivalence of considering subsystem $A$ or $B$.
From this we find that the unique maxima satisfy 
\begin{equation}\label{saddle-relation}
    j_{\kappa,\tilde f}=j-j_{1-\kappa,1-\tilde f}
\end{equation}
relating those for system $A$ with $\tilde j=j_A$ and $\tilde f=f$ with those for system $B$ with the choice $\tilde{j}=j_B$ and $\tilde f=1-f$.

To understand how the maximum value 
$B_{\kappa}(j_{\kappa,\tilde f},\tilde f)$ behaves when varying $\kappa$, we have a closer look at the saddle point equation~\eqref{saddle.B},
which can be rewritten as follows
\begin{equation}\label{saddle.B.b}
    \kappa\log\left(\frac{\tilde f-j_{\kappa,\tilde f}}{\tilde f+j_{\kappa,\tilde f}}\right)=(1-\kappa)\log\left(\frac{1-\tilde f-j+j_{\kappa,\tilde f}}{1-\tilde f+j-j_{\kappa,\tilde f}}\right).
\end{equation}
When differentiating~\eqref{saddle.B} in $\kappa$ we find the following equation for $\partial_\kappa j_{\kappa,\tilde f}$,
\begin{align}
    0&=\partial_\kappa [\partial_{\tilde j}B_{\kappa}(\tilde j,\tilde f)|_{\tilde j=j_{\kappa,\tilde f}}]\nonumber\\
    &=\log\left(\frac{(\tilde f-j_{\kappa,\tilde f})(1-\tilde f-j+j_{\kappa,\tilde f})}{(\tilde f+j_{\kappa,\tilde f})(1-\tilde f+j-j_{\kappa,\tilde f})}\right)\nonumber\\
    &\quad+\partial_{\tilde j}^2B_{\kappa}(\tilde j,\tilde f)|_{\tilde j=j_{\kappa,\tilde f}}\partial_\kappa j_{\kappa,\tilde f}.
\end{align}
Exploiting~\eqref{saddle.B.b} to obtain
\begin{align}\label{der.jkf}
   \partial_\kappa j_{\kappa,\tilde f}=-\frac{1}{(1-\kappa)\partial_{\tilde j}^2B_{\kappa}(\tilde j,\tilde f)|_{\tilde j=j_{\kappa,\tilde f}}}\log\left(\frac{\tilde f-j_{\kappa,\tilde f}}{\tilde f+j_{\kappa,\tilde f}}\right)\leq0
\end{align}
and recalling that $0\leq j_{\kappa,\tilde f}\leq\tilde f$ as well as~\eqref{2nd.der.B}, it becomes immediate that the position $j_{\kappa,\tilde f}$ of the maximum  is monotonously decreasing in $\kappa$. This implies a restricted domain for $j_{\kappa,\tilde f}$ since we know that $j_{1/2,\tilde f}=\tilde f j$ and $j_{0,\tilde f}=\min\{j,\tilde{f}\}$, where the latter means that the maximum can also lie at the boundary which is not necessarily a critical point.  We obtain
\begin{equation}\label{red.int.saddle}
    j_{\kappa,\tilde f}\in[\tilde f j,\min\{j,\tilde{f}\}]
\end{equation}
for $0\leq\kappa\leq 1/2$ and find the respective domain for $\kappa\geq1/2$ when combining this with the relation~\eqref{saddle-relation}.

 Furthermore, when differentiating the maximum $B_{\kappa}(j_{\kappa,\tilde f},\tilde f)$ twice in $\kappa$, we find
\begin{align}
    \partial_\kappa^2B_{\kappa}(j_{\kappa,\tilde f},\tilde f)&=2\partial_\kappa\left[\beta\left(\frac{j_{\kappa,\tilde f}}{\tilde f}\right)\tilde f-\beta\left(\frac{j-j_{\kappa,\tilde f}}{1-\tilde f}\right)(1-\tilde f)\right]\nonumber\\
    &\hspace*{-2cm}=\log\left(\frac{(\tilde f-j_{\kappa,\tilde f})(1-\tilde f-j+j_{\kappa,\tilde f})}{(\tilde f+j_{\kappa,\tilde f})(1-\tilde f+j-j_{\kappa,\tilde f})}\right)\partial_\kappa j_{\kappa,\tilde f}\\
    &\hspace*{-2cm}=-\frac{1}{(1-\kappa)^2\partial_{\tilde j}^2B_{\kappa}(\tilde j,\tilde f)|_{\tilde j=j_{\kappa,\tilde f}}}\left[\log\left(\frac{\tilde f-j_{\kappa,\tilde f}}{\tilde f+j_{\kappa,\tilde f}}\right)\right]^2\geq0
\end{align}
where we employed the saddle point equation~\eqref{saddle.B} in the first line and  its explicit representation~\eqref{saddle.B.b} as well as~\eqref{der.jkf} in the last one. This means that $B_{\kappa}(j_{\kappa,\tilde f},\tilde f)$ is a convex function in $\kappa$ so that its most extreme values can be found always for the most extreme choices of $\kappa\in[0,1]$. 

The question that remains is when we choose two $\kappa_1<\kappa_2$ which maximum $B_{\kappa}(j_{\kappa,\tilde f},\tilde f)$ is larger. Actually, we are only interested in cases where either $\kappa_1<\kappa_2\leq1/2$ or $1/2\leq\kappa_1<\kappa_2$. Yet, we already know that $j_{1/2,\tilde f}=\tilde f j$, see the above discussion for deriving the asymptotic of $d$. Plugging this into the first derivative 
\begin{equation}
 \partial_\kappa B_{\kappa}(j_{\kappa,\tilde f},\tilde f)=\beta\left(\frac{j_{\kappa,\tilde f}}{\tilde f}\right)\tilde f-\beta\left(\frac{j-j_{\kappa,\tilde f}}{1-\tilde f}\right)(1-\tilde f)
\end{equation}
we find
\begin{equation}
 \partial_\kappa B_{\kappa}(j_{\kappa,\tilde f},\tilde f)|_{\kappa=1/2}=\beta\left(j\right)(2\tilde f-1),
\end{equation}
which is negative for $\tilde f<1/2$ and positive for $\tilde f>1/2$. In combination with the convexity of $B_{\kappa}(j_{\kappa,\tilde f},\tilde f)$ in $\kappa$, we conclude
\begin{align}
\kappa_1<\kappa_2\leq\frac{1}{2} &\Rightarrow& B_{\kappa_1}(j_{\kappa_1,\tilde f},\tilde f)>B_{\kappa_2}(j_{\kappa_2,\tilde f},\tilde f),\\
\frac{1}{2}\leq\kappa_1<\kappa_2 &\Rightarrow& B_{\kappa_1}(j_{\kappa_1,\tilde f},\tilde f)<B_{\kappa_2}(j_{\kappa_2,\tilde f},\tilde f).
\end{align}
In other words, when $\tilde f\leq 1/2$, meaning for the case of system $A$ with $\tilde f=f$, the smaller $\kappa$ the larger the maximal value. Yet, it is the opposite for system $B$ where $\tilde f=1-f\geq 1/2$ where the larger $\kappa$ the larger the maximal value.

We have gone through this exercise, as we will come across sums of combinations of the form $n_{J_A}^{A} (n_{|J-J_A|}^B)^{l}$ and $(n_{|J-J_B|}^{A})^{l} n_{J_B}^B$ with $l\in\mathbb{N}$. What we have shown is that the ratios
\begin{equation}\label{vanishing.ratios}
\begin{split}
    \frac{n_{J_A}^{A} (n_{|J-J_A|}^B)^{l}}{(n_{J_A}^{A} (n_{|J-J_A|}^B)^{L})^{(l+1)/(L+1)}}&\overset{V\to\infty}{\longrightarrow}0,\\
    \frac{ (n_{|J-J_B|}^A)^{l}n_{J_B}^{B}}{((n_{|J-J_B|}^A)^{L}n_{J_B}^{B})^{(l+1)/(L+1)}}&\overset{V\to\infty}{\longrightarrow}0
\end{split}
\end{equation}
vanish exponentially in $V$ whenever $L>l$ since the corresponding $\kappa$ in the numerator is $1/(l+1)$ while the one in the denominator is $1/(L+1)$ once we take the $(l+1)$st root of the ratio. This vanishing gives us the selection rules when studying the terms in the higher purities of $\rho_A$.

Actually, we also need to understand the difference
\begin{equation}
    \Delta B(\kappa,\tilde f)=B_\kappa(j_{\kappa,\tilde f},\tilde f)-B_\kappa(j_{\kappa,1-\tilde f},1-\tilde f)
\end{equation}
This quantity carries the information of the behavior of the ratio $[(n_{|J-j_{\kappa,f}V/2|}^B)^{L}n_{j_{\kappa,f}V/2}^{A}]/[(n_{|J-j_{\kappa,1-f}V/2|}^A)^{L}n_{j_{\kappa,1-f}V/2}^{B}]$ with $\kappa=1/(L+1)$ and identifying $\tilde f=f$ or $\tilde f=1-f$.

The relation~\eqref{saddle-relation} tells us that $j_{\kappa,1-\tilde f}=j-j_{1-\kappa,\tilde f}$ which in combination with~\eqref{B.symmetry} immediately implies $\Delta B(1/2,\tilde f)=0$. To measure how much this changes when varying to $\kappa<1/2$, we consider its derivative in $\kappa$ which is equal to
\begin{equation}\label{der.kappa.Delta-B}
\begin{split}
    \partial_\kappa \Delta B(\kappa,\tilde f)&=\partial_\kappa B_\kappa(\tilde j,\tilde f)|_{\tilde j=j_{\kappa,\tilde f}}+\partial_\kappa B_\kappa(\tilde j,\tilde f)|_{\tilde j=j_{1-\kappa,\tilde f}}
\end{split}
\end{equation}
because of the saddle point equation~\eqref{saddle.B}, the relation~\eqref{saddle-relation} and the symmetry $\partial_\kappa B_\kappa(\tilde j,\tilde f)=-\partial_\kappa B_\kappa(j-\tilde j,1-\tilde f)$.

The derivative~\eqref{der.kappa.Delta-B} vanishes for $\tilde f=1/2$ which becomes clear from $\partial_\kappa B_\kappa(\tilde j,1/2)=\beta(2\tilde j)-\beta(2j-2\tilde j)$ and $j_{1-\kappa,1/2}=j-j_{\kappa,1/2}$. Furthermore, $\partial_\kappa B_\kappa(\tilde j,\tilde f)$ is an increasing function in $\tilde f$ when $\tilde j\leq \min\{\tilde f,j\}$ because its derivative in $\tilde f$ is
\begin{equation}
\begin{split}
    \partial_{\tilde f}\partial_\kappa B_\kappa(\tilde j,\tilde f)&=\beta\left(\frac{\tilde{j}}{\tilde f}\right)+\beta\left(\frac{j-\tilde{j}}{1-\tilde f}\right)\\
    &\hspace*{-1cm} -\log\left(\frac{\tilde f-\tilde j}{\tilde f+\tilde j}\right)\frac{\tilde j}{\tilde f}-\log\left(\frac{1-\tilde f-j+\tilde j}{1-\tilde f+j-\tilde j}\right)\frac{j-\tilde j}{1-\tilde f}.
\end{split}
\end{equation}
All four summands are positive when $\max\{0,j+\tilde f-1\}<\tilde j< \min\{\tilde f,j\}$. For $0<\kappa<1/2$, this is the case for $j_{\kappa,\tilde f}$, see~\eqref{red.int.saddle} as well as $j_{1-\kappa,\tilde f}=j-j_{\kappa,1-\tilde f}\in(\max\{0,j+\tilde f-1\},\tilde f j)$.

We conclude that $\partial_\kappa \Delta B(\kappa,\tilde f)$ is negative for $\tilde f<1/2$ and positive when $\tilde f>1/2$. Thence, the difference  $\Delta B(\kappa,\tilde f)$ is positive for $\kappa,\tilde f<1/2$ and negative for $1-\kappa,\tilde f>1/2$. In terms of the dimensions we have shown that
\begin{equation}\label{vanishing.ratios.b}
    \frac{(n_{|J-j_{\kappa,f}V/2|}^B)^{L}n_{j_{\kappa,f}V/2}^{A}}{(n_{|J-j_{\kappa,1-f}V/2|}^A)^{L}n_{j_{\kappa,1-f}V/2}^{B}}\overset{V\to\infty}{\longrightarrow}
    \begin{cases}
    \infty, & f<1/2,\\ 0, & f>1/2,
    \end{cases}
\end{equation}
where the divergence to infinity and convergence to zero is exponential in $V$. This dimensional selection also becomes important. It is exactly here where it becomes clear that something drastically changes at half-system size $f=1/2$. We will not go into details in this case. We only intend to highlight that the following method fails because the limit of the ratio~\eqref{vanishing.ratios.b} neither vanishes nor diverges exponentially then. We believe this lies at the heart of the non-analytic behavior of the results at $\kappa=1/2$ that arise for $f=1/2$.

\subsection{Scaling behavior of the Clebsch-Gordon coefficients}\label{sec:scaling}

In our analysis below, we will come across sums of the form
\begin{equation}\label{Lambda.def}
    \Lambda_{{\bf J}_A,{\bf J}_B}=\sum_{m=-{\rm Min}_{{\bf J}_A,{\bf J}_B}}^{{\rm Min}_{\widehat{J}_A,\widehat{J}_B}}\prod_{k=1}^K\lambda_m^{(J_{A,k},J_{B,k})}
\end{equation}
with $K\geq 1$ and
\begin{equation}\label{min.m}
{\rm Min}_{{\bf J}_A,{\bf J}_B}=\min\{J_{A,1},\ldots,J_{A,K},J_{B,1},\ldots,J_{B,K}\}
\end{equation}
with the possibility of excluding $m=0$ in the sum if there is some pair $(J_{A,k},J_{B,k})$ so that $J-J_{A,k}-J_{B,k}$ is odd. To guarantee that this term is not competing with the analysis of the dimensions in the previous subsection, we need to bound these sums away from infinity as well as from zero.

A bound from above is simply obtained as the moduli of the CG-coefficients are bound by $1$. Thus, $\Lambda_{\widehat{J}_A,\widehat{J}_B}$ can never grow faster than $2{\rm Min}_{{\bf J}_A,{\bf J}_B}+1=O(V)$.

When ${\rm Min}_{{\bf J}_A,{\bf J}_B}>0$, we can also bound from below
\begin{equation}
    \Lambda_{{\bf J}_A,{\bf J}_B}\geq \prod_{k=1}^K\lambda_{{\rm Min}_{{\bf J}_A,{\bf J}_B}}^{(J_{A,k},J_{B,k})}.
\end{equation}
We note that for the particular case ${\rm Min}_{{\bf J}_A,{\bf J}_B}=0$, we have excluded those sums where the triplet $(J_{A,l},J_{B,l},m)$ corresponds to a zero CG coefficient, \ie it is $\lambda_0^{(J_{A,k},J_{B,k})}>0$ for those case. Hence, it must be $\prod_{k=1}^K\lambda_0^{(J_{A,k},J_{B,k})}>0$.
For $\lambda_m^{(J_{A,k},J_{B,k})}>0$ with either $m=0,1/2,1$ it is known~\cite{1999JMP....40.4782R,rowe2010shifted} that they scale like $1/\sqrt{V}$. Therefore, the lower bound is also only algebraic in $V$, namely of order $1/V^{K/2}$, regardless which behavior of ${\rm Min}_{{\bf J}_A,{\bf J}_B}$ we consider. 

As both bounds are algebraic we can indeed rely on the dimensional selection of the terms in the following discussion which has been studied in the previous subsection.

\section{Analytical results with arbitrary total spin J}\label{sec:new calc}

A previous study~\cite{PhysRevB.108.245101} has already explored average entanglement entropy of subsystem for the total system having ${\rm SU}(2)$ symmetry. However, a full analytical proof was only restricted to system with $J=0$ total spin as shown in the previous section~\ref{sec:review}. In particular, it is not clearly known how the joint constraints of fixed arbitrary $J$ and $J_z$ shape the distribution of entanglement entropy among random pure states in the symmetric subspace. By investigating the ensemble of Haar-random pure states restricted to these irreducible ${\rm SU}(2)$ representations, we aim to uncover whether universal features of entanglement persist under such strong symmetry constraints, and how they differ from those in unconstrained or Abelian symmetric cases.
		
\subsection{Lifting the fixed trace condition}\label{sec:lift fix trace}

As aforementioned, the reduced density matrix constructed from random pure states must satisfy the normalization condition $\Tr \rho_A = 1$. Enforcing this constraint directly complicates the analysis. A powerful way to bypass this difficulty is to trace back the problem to a random matrix ensemble of positive-definite Hermitian matrices without normalization, namely coupled Wishart ensembles, where matrices are distributed according to a Gaussian weight in $W$ given by $\exp[-\Tr \ro]=\exp[-\Tr WW^\dagger]$. This ensemble is mathematically much more tractable and their eigenvalue distributions are well studied.

The physical reduced density matrix $\rho_A$ is then obtained by normalizing such a matrix,
		\begin{align}
			\rho_A = \frac{\ro}{\Tr \ro}\,.\label{eq:density-ro}
		\end{align}
This mapping effectively translates the fixed-trace ensemble into the Wishart ensemble and enables the use of well-developed tools from random matrix theory~\cite{PRXQuantum.3.030201}. The average of entanglement entropy over the uniform measure, as defined in~\eqref{entropy}, is then
		\begin{align}
			\braket{S_A}={}&-\langle\Tr (\rho_A\log\rho_A)\rangle\label{eq:3.3}\\
			={}& \frac{-\int_{\mathbb{C}^{d}}d[\ro]\Tr(\frac{\ro}{\Tr \ro}\log \frac{\ro}{\Tr \ro}) \e^{-\Tr \ro}}{\int_{\mathbb{C}^{d}} d[\rho_A]\quad \e^{-\Tr \ro}}\nonumber\\
			={}&-\int_0^{\infty}dt\bigl\langle\bigl(\Tr\ro\log[(1+t) \ro]\nonumber\\
            &\qquad- \Tr\ro \log[(1+t)\Tr \ro]\bigl) \e^{-{t\Tr \ro }}\bigl\rangle,\nonumber
		\end{align}
where the ensemble average in the last line is given by the average over $\rho_A$ with respect to the weight $\e^{-\Tr \ro}$. We recall that $d = \sum_{J_A} d_{J_A}$ is the total Hilbert space dimension determining the number of linearly independent matrix entries in $\rho_A$. We introduced an auxiliary integral over $t \in [0, \infty)$ to take care of the normalization terms given by $\Tr \rho$ via Laplace transform techniques. Actually, the second logarithmic term can be also taken care of  in the very same way when introducing the auxiliary parameters $\epsilon>0$ as follows
		\begin{align}
			&-\Tr\ro \log[(1+t)\Tr \ro] \e^{-{t\Tr \ro }}\\
            =&[\log(1+t)\partial_t-\partial_\epsilon(-\partial_t)^\epsilon \e^{-{t\Tr \ro}}|_{\epsilon=1}\,,\nonumber
		\end{align}
which is justified as the expression in an integer $\epsilon$ is actually analytic about $\epsilon=1$.

To proceed with the simplification, we rescale $\ro \to \ro/(1+t)$ in~\eqref{eq:3.3} where the shift by $1$ originates from the weight $\exp[-\Tr\ro]$. The Jacobian gives a factor of $1/(1+t)^d$.
		This leads to
		\begin{align}
			\braket{S_A}&=-\int_0^{\infty}dt\biggl[\frac{\left\langle\Tr \ro\log \ro\right\rangle}{1+t}\\
			&\qquad+\log(1+t)\partial_t-\partial_\epsilon(-\partial_t)^\epsilon\biggl]\frac{1}{(1+t)^d}\biggl|_{\epsilon=1}\nonumber
		\end{align}
		Exploiting the relation
		\begin{equation}
			(-\partial_t)^\epsilon\frac{1}{(1+t)^d}=\frac{\Gamma[d+\epsilon]}{\Gamma[d]}\frac{1}{(1+t)^{d+\epsilon}}
		\end{equation}
in terms of the Gamma function, where the analytic nature in $\epsilon$, the order of the derivative, is immediate, we see that the second term cancels with a term from the third one and we arrive at
		\begin{align}
			\braket{S_A}=\Psi(d+1)-\frac{\braket{\Tr (\ro\log \ro)}}{d}\,,\label{Sa.1}
		\end{align}
where $\Psi(x)=\Gamma'(x)/\Gamma(x)$ is the Digamma function. As aforementioned, everything has been exact so far without any approximation. Interestingly, the Digamma function $\Psi(d+1)$ is very universal and has appeared also in the computation of the Page curve, see~\cite{page}. It reflects the normalization of the pure state and, thus, only depends on the Hilbert space dimension of the total quantum system.

So far everything is exact without any approximation. However, when $V\to\infty$ the dimension $d\to\infty$ grows exponentially in $V$, see~\eqref{eq:d-asymptotics}. The Digamma function is well-approximated in this regime  by $\psi(d+1)=\log d+O(1/d)$ where the error is exponentially small. Using the fact that $\braket{\Tr\rho}=d$ we can write
\begin{equation}\label{SA.Digamma.approx}
\begin{split}
    \braket{S_A}&=-\Bigl\langle\Tr\frac{\rho}{d}\log\left(\frac{\rho}{d}\right)\Bigl\rangle+O\left(d^{-1}\right).
\end{split} 
\end{equation}
We will derive an expression for the leading contribution of the level density of the $d\times d$ matrix $\rho/d$ with the help of Wigner's moment method~\cite{Wigner} where we analyze the moments $\braket{\Tr\rho^L}$ with an integer power $L\in\mathbb{N}_0$.

\subsection{Planar diagram approximation}\label{sec:planar}

We note that the random matrix $\rho$ inherits the direct sum structure of $\rho_A$. Recalling the definitions of the sets~\eqref{JB.cond.JA} and~\eqref{JA.cond.JB} and the terminal~\eqref{min.m} for the magnetic quantum number $m$ under the condition of various $J_{A,l}$ and $J_{B,l}$, the expectation value $\braket{\Tr\rho^L}$ for integer $L$ can be written explicitly as follows
\begin{equation}
\begin{split}\label{higher.purity.a}
    &\braket{\Tr\rho^L}=\sum_{m}\braket{\Tr(\rho^{(m)})^L}\\    &=\sum_{(J_{A,1},J_{B,1})\in\mathfrak{J}}\sum_{J_{A,2}\in\mathfrak{J}_A^{(J_{B,1})}}\sum_{J_{B,2}\in\mathfrak{J}_B^{(J_{A,2})}}\cdots\\
    &\quad\sum_{J_{A,L}\in\mathfrak{J}_A^{(J_{B,L-1})}}\sum_{J_{B,L}\in\mathfrak{J}_B^{(J_{A,L})}\cap\mathfrak{J}_B^{(J_{A,1})}}\\
    &\quad\times\left(\sum_{m=-{\rm Min}_{{\bf J}_A,{\bf J}_B}}^{{\rm Min}_{{\bf J}_A,{\bf J}_B}}\prod_{l=1}^L c_m(J_{A,l},J_{B,l})c_m^*(J_{A,l+1},J_{B,l})\right)\\
    &\quad\times\left\langle\Tr \left[\prod_{l=1}^L W^{(J_{A,l},J_{B,l})}(W^{(J_{A,l+1},J_{B,l})})^\dagger\right]\right\rangle,
\end{split}
\end{equation}
where $J_{A,L+1}=J_{A,1}$ and the product is ordered starting with the term $l=1$ from the left and ending with the term $l=L$ at the right as we deal with matrices which are non-commutative. The nested sums are the tricky part we have to deal with as they are conditioned on the preceding sums unlike what Wigner~\cite{Wigner} has encountered when computing the level density of Gaussian random matrices.

Before we come to this we inductively integrate out the matrices with the help of the Isserlis-Wick theorem. The main idea is to pair $W^{(J_{A,l},J_{B,l})}$ with $(W^{(J_{A,l'+1},J_{B,l'})})^\dagger$ which traces higher moments of Gaussians to second moments, only. Doing so, we come across the two identities of second moments
\begin{equation}\label{cutting}
   \braket{\Tr [W^{(J_{A,l},J_{B,l})}\mathfrak{A}(W^{(J_{A,l},J_{B,l})})^\dagger\mathfrak B]}=\Tr [\mathfrak A] \Tr[\mathfrak B ]
\end{equation}
and
\begin{equation}\label{glueing}
   \braket{\Tr [W^{(J_{A,l},J_{B,l})}\mathfrak{C}]\Tr[(W^{(J_{A,l},J_{B,l})})^\dagger\mathfrak D]}=\Tr [\mathfrak C\mathfrak D] 
\end{equation}
 when averaging over a single block. Those hold true
for any matrices $\mathfrak A$, $\mathfrak B$, $\mathfrak C$ and $\mathfrak D$ of fitting dimensions. The first identity is the cutting of a loop into two while the second one is the glueing or merging of two loops to one. Considering the expectation values in~\eqref{higher.purity.a}, we could successively carry out all integrations of each $W^{(J_{A,l},J_{B,l})}$ until only identity matrices are left over in the traces. This shows that the glueing of loops reduces the contribution by a factor of two dimensions $n_{J_{A,l}}^A$ and $n_{J_{B,l}}^B$ and, hence, only contribute exponentially suppressed subleading terms. Therefore, we can neglect those. In a diagrammatic picture with loops for traces, this is called planar diagram approximation.
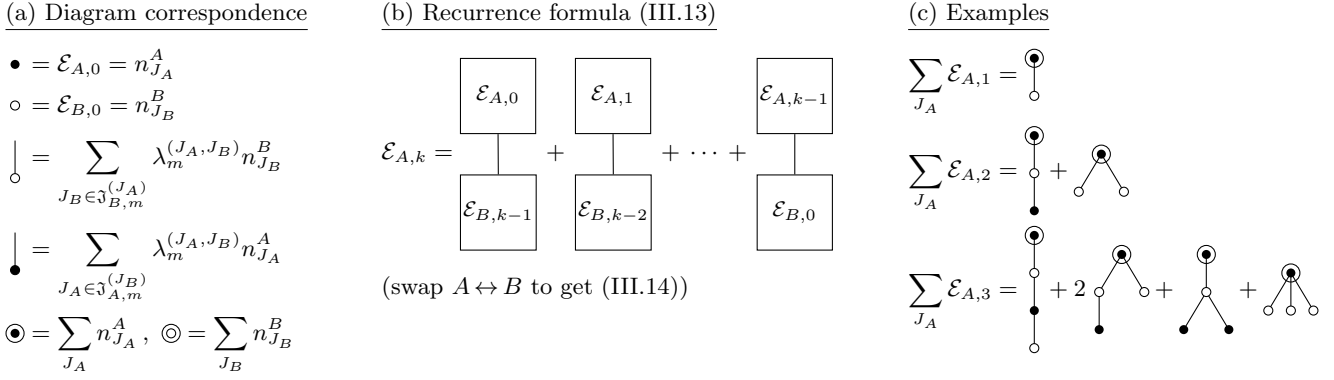
\begin{figure*}[t!]
\centering
\begin{tikzpicture}
	\draw (0,0) node[anchor=north west,text width=3cm]{\underline{(a) Diagram correspondence}\\[2mm]
		$\begin{aligned}[t]
			\begin{tikzpicture}[baseline={(shifted)}]
				\coordinate (P1) at (0,0);
				\fill (P1) circle (.6mm);
				\phantom{\draw (P1) circle (1.2mm);}
				\coordinate (shifted) at ($(current bounding box.center)+(0,-2.5pt)$);
			\end{tikzpicture}
			&=\mathcal{E}_{A,0}=n^A_{J_A}\\
			\begin{tikzpicture}[baseline={(shifted)}]
				\coordinate (P1) at (0,0);
				\draw (P1) circle (.6mm);
				\phantom{\draw (P1) circle (1.2mm);}
				\coordinate (shifted) at ($(current bounding box.center)+(0,-2.5pt)$);
			\end{tikzpicture}&=\mathcal{E}_{B,0}=n^B_{J_B}\\
			\begin{tikzpicture}[baseline={(shifted)}]
				\coordinate (P1) at (0,0);
				\coordinate (P2) at (0,-.5);
				\draw (P1) -- (P2);
				\phantom{\draw (P1) circle (1.2mm);}
				\filldraw[fill=white] (P2) circle (.6mm);
				\coordinate (shifted) at ($(current bounding box.center)+(0,-2.5pt)$);
			\end{tikzpicture}&=\sum_{J_B\in\mathfrak{J}_{B,m}^{(J_A)}}\lambda^{(J_A,J_B)}_m n^B_{J_B}\\
			\begin{tikzpicture}[baseline={(shifted)}]
				\coordinate (P1) at (0,0);
				\coordinate (P2) at (0,-.5);
				\draw (P1) -- (P2);
				\filldraw (P2) circle (.6mm);
				\phantom{\draw (P1) circle (1.2mm);}
				\coordinate (shifted) at ($(current bounding box.center)+(0,-2.5pt)$);
			\end{tikzpicture}&=\sum_{J_A\in\mathfrak{J}_{A,m}^{(J_B)}}\lambda^{(J_A,J_B)}_m n^A_{J_A}\\
			\begin{tikzpicture}[baseline={(shifted)}]
				\coordinate (P1) at (0,0);
				\fill (P1) circle (.6mm);
				\draw (P1) circle (1.2mm);
				\coordinate (shifted) at ($(current bounding box.center)+(0,-2.5pt)$);
			\end{tikzpicture}&=\sum_{J_A}n^A_{J_A}\,,\,\,
			\begin{tikzpicture}[baseline={(shifted)}]
				\coordinate (P1) at (0,0);
				\filldraw[fill=white] (P1) circle (.6mm);
				\draw (P1) circle (1.2mm);
				\coordinate (shifted) at ($(current bounding box.center)+(0,-2.5pt)$);
			\end{tikzpicture}=\sum_{J_B}n^B_{J_B}
		\end{aligned}$
	};
	\draw (5,0) node[anchor=north west,text width=6cm]{\underline{(b) Recurrence formula~\eqref{recc.JA}}\\[-2mm]
		\[
		\mathcal{E}_{A,k}
		=
		\begin{tikzpicture}[baseline=(mid.center)]
			\coordinate (mid) at (0,-.5);
			\node[diagrambox] (top) at (0,8.5mm) {$\mathcal{E}_{A,0}$};
			\node[diagrambox] (bot) at (0,-7mm) {$\mathcal{E}_{B,k-1}$};
			\draw ($(top.south)$) -- ($(bot.north)$);
		\end{tikzpicture}
		\;+\;
		\begin{tikzpicture}[baseline=(mid.center)]
			\coordinate (mid) at (0,-.5);
			\node[diagrambox] (top) at (0,8.5mm) {$\mathcal{E}_{A,1}$};
			\node[diagrambox] (bot) at (0,-7mm) {$\mathcal{E}_{B,k-2}$};
			\draw ($(top.south)$) -- ($(bot.north)$);
		\end{tikzpicture}
		\;+\;\cdots\;+\;
		\begin{tikzpicture}[baseline=(mid.center)]
			\coordinate (mid) at (0,-.5);
			\node[diagrambox] (top) at (0,8.5mm) {$\mathcal{E}_{A,k-1}$};
			\node[diagrambox] (bot) at (0,-7mm) {$\mathcal{E}_{B,0}$};
			\draw ($(top.south)$) -- ($(bot.north)$);
		\end{tikzpicture}
		\]
		(swap $A\hspace{-3mm}\mathrel{\leftrightarrow}\hspace{-3mm}B$ to get~\eqref{recc.JB})%
	};
	\draw (12,0) node[anchor=north west,text width=6cm]{\underline{(c) Examples}\\[2mm]
		$\begin{aligned}[t]
			\sum_{J_A} \mathcal{E}_{A,1} &=
			\begin{tikzpicture}[baseline={(shifted)}]
				\coordinate (P1) at (0,0);
				\coordinate (P2) at (0,-.5);
				\draw (P1) -- (P2);
				\fill (P1) circle (.6mm);
				\draw (P1) circle (1.2mm);
				\filldraw[black,fill=white] (P2) circle (.6mm);
				\coordinate (shifted) at ($(current bounding box.center)+(0,-2.5pt)$);
			\end{tikzpicture}\\
			\sum_{J_A} \mathcal{E}_{A,2} &=
			\begin{tikzpicture}[baseline={(shifted)}]
				\coordinate (P1) at (0,0);
				\coordinate (P2) at (0,-.5);
				\coordinate (P3) at (0,-1);
				\draw (P1) -- (P3);
				\fill (P1) circle (.6mm) (P3) circle (.6mm);
				\draw (P1) circle (1.2mm);
				\filldraw[black,fill=white] (P2) circle (.6mm);
				\coordinate (shifted) at ($(current bounding box.center)+(0,-2.5pt)$);
			\end{tikzpicture}
			+
			\begin{tikzpicture}[baseline={(shifted)}]
				\coordinate (P1) at (0,0);
				\coordinate (P2) at (-.3,-.5);
				\coordinate (P3) at (.3,-.5);
				\draw (P2) -- (P1) -- (P3);
				\fill (P1) circle (.6mm) ;
				\draw (P1) circle (1.2mm);
				\filldraw[black,fill=white] (P2) circle (.6mm) (P3) circle (.6mm);
				\coordinate (shifted) at ($(current bounding box.center)+(0,-2.5pt)$);
			\end{tikzpicture}\\
			\sum_{J_A} \mathcal{E}_{A,3} &=
			\begin{tikzpicture}[baseline={(shifted)}]
				\coordinate (P1) at (0,0);
				\coordinate (P2) at (0,-.5);
				\coordinate (P3) at (0,-1);
				\coordinate (P4) at (0,-1.5);
				\draw (P1) -- (P4);
				\fill (P1) circle (.6mm) (P3) circle (.6mm);
				\draw (P1) circle (1.2mm);
				\filldraw[black,fill=white] (P2) circle (.6mm) (P4) circle (.6mm);
				\coordinate (shifted) at ($(current bounding box.center)+(0,-2.5pt)$);
			\end{tikzpicture}
			+2\,\,
			\begin{tikzpicture}[baseline={(shifted)}]
				\coordinate (P1) at (0,0);
				\coordinate (P2) at (-.3,-.5);
				\coordinate (P3) at (.3,-.5);
				\coordinate (P4) at (-.3,-1);
				\draw (P3) -- (P1) -- (P2) -- (P4);
				\fill (P1) circle (.6mm) (P4) circle (.6mm);
				\draw (P1) circle (1.2mm);
				\filldraw[black,fill=white] (P2) circle (.6mm) (P3) circle (.6mm);
				\coordinate (shifted) at ($(current bounding box.center)+(0,-2.5pt)$);
			\end{tikzpicture}
			+
			\begin{tikzpicture}[baseline={(shifted)}]
				\coordinate (P1) at (0,0);
				\coordinate (P2) at (0,-.5);
				\coordinate (P3) at (-.3,-1);
				\coordinate (P4) at (.3,-1);
				\draw (P4) -- (P2) -- (P3) (P1) -- (P2);
				\fill (P1) circle (.6mm) (P3) circle (.6mm) (P4) circle (.6mm);
				\draw (P1) circle (1.2mm);
				\filldraw[black,fill=white] (P2) circle (.6mm) ;
				\coordinate (shifted) at ($(current bounding box.center)+(0,-2.5pt)$);
			\end{tikzpicture}
			+
			\begin{tikzpicture}[baseline={(shifted)}]
				\coordinate (P1) at (0,0);
				\coordinate (P2) at (0,-.5);
				\coordinate (P3) at (-.3,-.5);
				\coordinate (P4) at (.3,-.5);
				\draw (P2) -- (P1) -- (P3) (P1) -- (P4);
				\fill (P1) circle (.6mm);
				\draw (P1) circle (1.2mm);
				\filldraw[black,fill=white] (P2) circle (.6mm)  (P3) circle (.6mm) (P4) circle (.6mm);
				\coordinate (shifted) at ($(current bounding box.center)+(0,-2.5pt)$);
			\end{tikzpicture}
		\end{aligned}$
	};
\end{tikzpicture}

\vspace{-8pt}
\caption{We illustrate the diagrammatic representation of the sums as \emph{rooted tree diagrams} of this section, which can be constructed from the recurrence formulas~\eqref{recc.JA} and~\eqref{recc.JB}. (a) Vertices represent dimensions $n^A_{J_A}=\bullet$ or $n^B_{J_B}=\circ$, which arise as elementary building block from~\eqref{expect.start}, where we associate to each $\bullet$-vertex a different summation variable $J_A$ and to each $\circ$-vertex a variable $J_B$. Edges represent conditioned sums over $\lambda^{(J_A,J_B)}_m$, where the variable $J_A$ or $J_B$ of the lower vertex is summed over, which is conditioned by the variable of the upper vertex, \ie $J_A\in\mathfrak{J}_{A,m}^{(J_B)}$ or $J_B\in\mathfrak{J}_{B,m}^{(J_A)}$. (b) The aforementioned recurrence formulas then take the diagrammatic forms illustrated above, where we dropped the dependence of $\mathcal{E}$ on $J_A$ or $J_B$, as these are summed over. (c) When applying the recurrence formula using the newly introduced diagrammatic notation, we find that $\sum_{J_A}\mathcal{E}_{A,k}$ corresponds to the sum of \emph{all inequivalent rooted trees} with $(k+1)$ vertices, where each diagram is weighted by how many equivalent diagrams with re-ordered sub-trees one can draw. We illustrate the cases up to $k=3$, where $\sum_{J_A}\mathcal{E}_{A,2}$ is written in~\eqref{2nd.purity.a} and $\sum_{J_A}\mathcal{E}_{A,3}$ is written in~\eqref{3rd.purity.a}.}
	\label{fig:tree-diagrams} 
\end{figure*}

In the next step, we want to systematically find the leading contributions. For this purpose, we make use of the abbreviation~\eqref{W.hat.def} of matrix blocks $\widehat{W}_m^{(J_A,J_B)}$ and extend those to a large matrix $\widehat{W}_m=\{\widehat{W}_m^{(J_A,J_B)}\}$ with $J_A$ and $J_B$ ranging over all their possible values. We set $\widehat{W}_m^{(J_A,J_B)}=0$ whenever $(J_A,J_B)\notin\mathfrak{J}_m$, see~\eqref{Jm.def}, as those correspond to unphysical transitions. This notation helps us to define the expectation values
\begin{equation}
    \mathcal{E}_{A,k}^{(J_A)}=\left\langle\Tr \{(\widehat{W}_m\widehat{W}_m^\dagger)^k\}^{(J_A)}\right\rangle
\end{equation}
and 
\begin{equation}
    \mathcal{E}_{B,k}^{(J_B)}=\left\langle\Tr \{(\widehat{W}_m^\dagger \widehat{W}_m)^k\}^{(J_B)}\right\rangle,
\end{equation}
where $\mathfrak{X}^{(\tilde{J})}$ gives the $(\tilde{J},\tilde{J})$ diagonal block of the matrix $\mathfrak{X}$. With the help of this notation and the index sets~\eqref{JBm.def} and~\eqref{JAm.def}, the Isserlis-Wick theorem and the planar diagram approximation gives us the following recurrence relations
\begin{equation}\label{recc.JA}
    \begin{split}
    \mathcal{E}_{A,k}^{(J_A)}&=\sum_{l=1}^k \sum_{J_B\in\mathfrak{J}_{B,m}^{(J_A)}}\lambda_m^{(J_A,J_B)}\mathcal{E}_{A,k-l}^{(J_A)}\mathcal{E}_{B,l-1}^{(J_B)}+\text{NPT}
    \end{split}
\end{equation}
and
\begin{equation}\label{recc.JB}
    \begin{split}
    \mathcal{E}_{B,k}^{(J_B)}&=\sum_{l=1}^k \sum_{J_A\in\mathfrak{J}_{A,m}^{(J_B)}}\lambda_m^{(J_A,J_B)}\mathcal{E}_{A,k-l}^{(J_A)}\mathcal{E}_{B,l-1}^{(J_B)}+\text{NPT},
    \end{split}
\end{equation}
where $\text{NPT}$ stands for non-planar terms and $\lambda_m^{(J_A,J_B)}$ are the squared moduli of the CG coefficients, see~\eqref{lambda.def}. This leading of this recurrence relation is a weighted version of Segner's recurrence relation~\cite{Segner}.
The first formula can be derived by taking the first matrix $\widehat{W}_m$ and selecting the block $\widehat{W}_m^{(J_A,J_B)}$ in the product given by $\{(\widehat{W}_m\widehat{W}_m^\dagger)^k\}^{(J_A)}$ and then selecting any of the $k$ block matrices $(\widehat{W}_m^{(J_A,J_B)})^\dagger$ also comprised in the product which can be paired with it. This gives the sum over $l$. Since we can take any $J_B$  as long as $(J_A,J_B)\in\mathfrak{J}_m$ we also sum over $J_B$ under the condition of a given $J_A$ and $m$ yielding the sum over $J_B\in\mathfrak{J}_{B,m}^{(J_A)}$. The splitting into separate averages is only allowed when ignoring the glueing identity~\eqref{glueing} and only taking the cutting~\eqref{cutting} for the remaining integrations, meaning applying the planar diagram approximation.

The recurrence is terminated for $k=0$ where we take the trace of the corresponding identity matrix, especially they are equal to the dimensions
\begin{equation}\label{expect.start}
   \mathcal{E}_{A,0}^{(J_A)}= n_{J_A}^A\quad{\rm and}\quad \mathcal{E}_{B,0}^{(J_B)}= n_{J_B}^B.
\end{equation}

\subsection{Dimensional selection}\label{sec:selection}

With the help of the ingredients derived in the previous subsection, we can write the expected higher purity as follows
\begin{align}
    \braket{\Tr\rho^L}&=\sum_m\sum_{J_{A,1}}\mathcal{E}_{A,L}^{(J_{A,1})}.
\end{align}
The first sum in $J_{A,1}$ is unrestricted apart from a given $m$ and runs over all possible values which we highlight by not writing the set.
What now follows are nested sums which have a diagrammatic representation in terms of a tree with $L+1$ vertices and, hence, $L$ edges. The diagrammatic picture shall help us to overlook and structure the computation. The numbers of all these trees with a fixed $L$ are given by the Catalan numbers $(2L)!/[L!(L+1)!]$, see~\cite{Wigner,tHooft1974,Eynard:2016yaa}. Yet, in contrast to the derivation of the Mar\v{c}enko-Pastur distribution~\cite{Pastur:1967zca}, the edges and vertices of the graphs here carry weights. 

There are two types of vertices in the tree which represent the summing indices $J_{A,r}$ (black/full vertex) and $J_{B,r}$ (white/empty vertex). Those sums are conditioned to the previous summing index, only, due to the recurrence rules~\eqref{recc.JA} and~\eqref{recc.JB}. This is the reason why it yields a connected graph without loops, meaning a tree. We draw an edge between two vertices if the sum over an index is conditioned to its preceding summing index. For instance when considering the sum $\sum_{J_{B,r}\in\mathfrak{J}_B^{(J_{A,r'})}}$ we draw an edge between the black vertex representing $J_{A,r'}$ to the white vertex $J_{B,r}$.

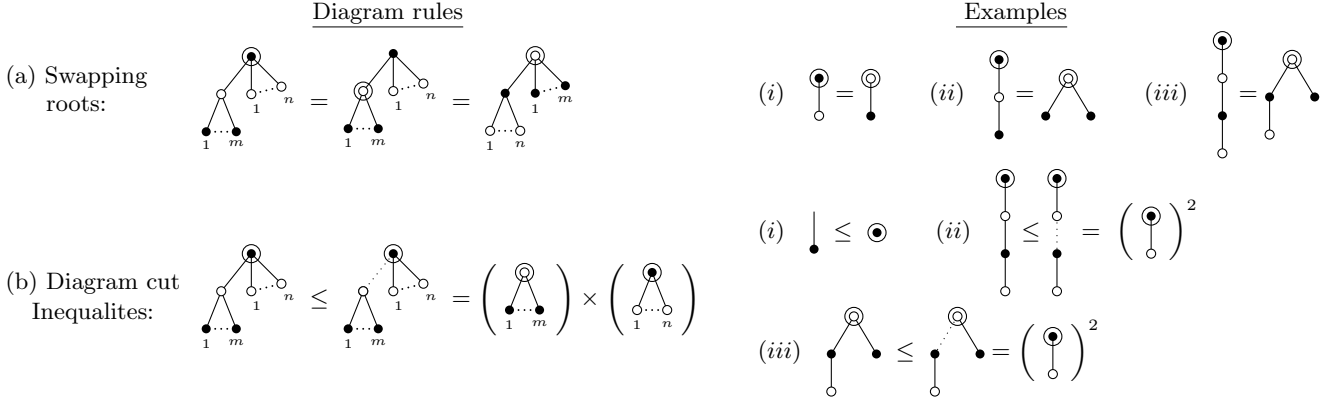
\begin{figure*}[t!]
	\centering
\begin{tikzpicture}
	\draw (-4,0) node[anchor=north west,text width=2cm]{
		~\\[8mm]
		$\begin{aligned}[t]
			&\text{\shortstack{(a)~Swapping\\roots:}}\\&~\\[14mm]
			&\text{\shortstack{(b)~Diagram cut\\Inequalites:}}    
		\end{aligned}$};
	\draw (-1.5,0) node[anchor=north west,text width=9cm]{\hspace{45pt}\underline{Diagram rules}\\[2mm]
		$\begin{aligned}[t]
			\begin{tikzpicture}[baseline={(shifted)}]
				\coordinate (R) at (0,0);
				\draw (R) circle (1.2mm);
				\fill (R) circle (0.6mm);
				\coordinate (A) at (-0.4,-0.5);
				\coordinate (B) at (0,-0.5);
				\coordinate (C) at (0.4,-0.4);
				\draw (R) -- (A);
				\draw (R) -- (B);
				\draw (R) -- (C);
				\filldraw[black,fill=white] (B) circle (0.6mm);
				\filldraw[black,fill=white] (C) circle (0.6mm);
				\foreach \t in {0.3,0.5,0.7} {%
					\fill ($(B)!\t!(C)$) circle (0.35pt); 
				};
				\node[below=0.2mm of B,xshift=0.7 mm]{\tiny{$1$}};
				\node[below=0.2mm of C,xshift=1 mm]{\tiny{$n$}};
				\coordinate (D) at(-0.6,-1);
				\coordinate (E) at(-0.2,-1);
				\fill (D) circle (0.6mm);
				\fill (E) circle (0.6mm);
				\draw (A) -- (D);
				\draw (A) -- (E);
				\filldraw[black,fill=white] (A) circle (0.6mm);
				\foreach \t in {0.3,0.5,0.7} {%
					\fill ($(D)!\t!(E)$) circle (0.35pt); 
				};
				\node[below=0.05mm of D]{\tiny{$1$}};
				\node[below=0.05mm of E]{\tiny{$m$}};
				\coordinate (shifted) at ($(current bounding box.center)+(0,-2pt)$);
			\end{tikzpicture}
			&= 
			\begin{tikzpicture}[baseline={(shifted)}]
				\coordinate (R) at (0,0);
				\fill (R) circle (0.6mm);
				\coordinate (A) at (-0.4,-0.5);
				\coordinate (B) at (0,-0.5);
				\coordinate (C) at (0.4,-0.4);
				\draw (R) -- (A);
				\draw (R) -- (B);
				\draw (R) -- (C);
				\filldraw[black,fill=white] (B) circle (0.6mm);
				\filldraw[black,fill=white] (C) circle (0.6mm);
				\foreach \t in {0.3,0.5,0.7} {%
					\fill ($(B)!\t!(C)$) circle (0.35pt); 
				};
				\node[below=0.2mm of B,xshift=0.7 mm]{\tiny{$1$}};
				\node[below=0.2mm of C,xshift=1 mm]{\tiny{$n$}};
				\draw (A) circle (1.2mm);
				\coordinate (D) at(-0.6,-1);
				\coordinate (E) at(-0.2,-1);
				\fill (D) circle (0.6mm);
				\fill (E) circle (0.6mm);
				\draw (A) -- (D);
				\draw (A) -- (E);
				\filldraw[black,fill=white](A) circle (0.6mm);
				\foreach \t in {0.3,0.5,0.7}{\fill ($(D)!\t!(E)$) circle (0.35pt);};
				\node[below=0.2mm of D]{\tiny{$1$}};
				\node[below=0.2mm of E]{\tiny{$m$}};
				\coordinate (shifted) at ($(current bounding box.center)+(0,-2pt)$);
			\end{tikzpicture}
			=
			\begin{tikzpicture}[baseline={(shifted)}]
				\coordinate (R) at (0,0);
				\coordinate (A) at (-0.4,-0.5);
				\coordinate (B) at (0,-0.5);
				\coordinate (C) at (0.4,-0.4);
				\foreach \t in {0.3,0.5,0.7} {%
					\fill ($(B)!\t!(C)$) circle (0.35pt); 
				};
				\node[below=0.3mm of B]{\tiny{$1$}};
				\node[below=0.3mm of C]{\tiny{$m$}};
				\draw (R) -- (A);
				\draw (R) -- (B);
				\draw (R) -- (C);
				\filldraw[black,fill=white] (R) circle (0.6mm);
				\draw (R) circle (1.2mm);
				\fill (B) circle (0.6mm);
				\fill (C) circle (0.6mm);
				\fill (A) circle (0.6mm);
				\coordinate (D) at(-0.6,-1);
				\coordinate (E) at(-0.2,-1);
				\draw (A) -- (D);
				\draw (A) -- (E);
				\foreach \t in {0.3,0.5,0.7} {%
					\fill ($(D)!\t!(E)$) circle (0.35pt); 
				};
				\filldraw[black,fill=white] (D) circle (0.6mm);
				\filldraw[black,fill=white](E) circle (0.6mm);
				\node[below=0.3mm of D]{\tiny{$1$}};
				\node[below=0.3mm of E]{\tiny{$n$}};
				\coordinate (shifted) at ($(current bounding box.center)+(0,-2pt)$);
			\end{tikzpicture}\\[10mm]
			\begin{tikzpicture}[baseline=(shifted)]
				\coordinate (R) at (0,0);
				\draw (R) circle (1.2mm);
				\fill (R) circle (0.6mm);
				\coordinate (A) at (-0.4,-0.5);
				\coordinate (B) at (0,-0.5);
				\coordinate (C) at (0.4,-0.4); 
				\draw (R) -- (A);
				\draw (R) -- (B);
				\draw (R) -- (C);
				\foreach \t in {0.3,0.5,0.7} {%
					\fill ($(B)!\t!(C)$) circle (0.35pt);
				};
				\node[below=0.2mm of B,xshift=0.7 mm]{\tiny{$1$}};
				\node[below=0.2mm of C,xshift=1 mm]{\tiny{$n$}};
				\filldraw[black,fill=white] (B) circle (0.6mm);
				\filldraw[black,fill=white] (C) circle (0.6mm);
				\coordinate (D) at(-0.6,-1);
				\coordinate (E) at(-0.2,-1);
				\draw (A) -- (D);
				\draw (A) -- (E); 
				\filldraw[black,fill=white] (A) circle (0.6mm);
				\foreach \t in {0.3,0.5,0.7} {%
					\fill ($(D)!\t!(E)$) circle (0.35pt); 
				};
				\node[below=0.2mm of D]{\tiny{$1$}};
				\node[below=0.2mm of E]{\tiny{$m$}};
				\fill(D) circle (0.6mm);
				\fill (E) circle (0.6mm);
				\coordinate (shifted) at ($(current bounding box.center)+(0,-2pt)$);
			\end{tikzpicture}
			&\leq
			\begin{tikzpicture}[baseline={(shifted)}]
				\coordinate (R) at (0,0);
				\draw (R) circle (1.2mm);
				\fill (R) circle (0.6mm);
				\coordinate (A) at (-0.4,-0.5);
				\coordinate (B) at (0,-0.5);
				\coordinate (C) at (0.4,-0.4);
				\draw (A) circle (0.6mm);
				\draw[dotted] (R) -- (A);
				\draw (R) -- (B);
				\draw (R) -- (C);
				\filldraw[black,fill=white] (B) circle (0.6mm);
				\filldraw[black,fill=white] (C) circle (0.6mm);
				\foreach \t in {0.3,0.5,0.7} {%
					\fill ($(B)!\t!(C)$) circle (0.35pt); 
				};
				\node[below=0.2mm of B,xshift=0.7 mm]{\tiny{$1$}};
				\node[below=0.2mm of C,xshift=1 mm]{\tiny{$n$}};
				\coordinate (D) at(-0.6,-1);
				\coordinate (E) at(-0.2,-1);
				\draw (A) -- (D);
				\draw (A) -- (E);
				\fill (D) circle (0.6mm);
				\fill (E) circle (0.6mm);
				\foreach \t in {0.3,0.5,0.7} {
					\fill ($(D)!\t!(E)$) circle (0.35pt); 
				};
				\filldraw[black,fill=white] (A) circle (0.6mm);
				\node[below=0.2mm of D]{\tiny{$1$}};
				\node[below=0.2mm of E]{\tiny{$m$}};
				\coordinate (shifted) at ($(current bounding box.center)+(0,-2pt)$);   
			\end{tikzpicture}=
			\left(\begin{tikzpicture}[baseline={(shifted)}]
				\coordinate (R) at (0,0);
				\draw (R) circle (1.2mm);
				\coordinate (D) at(-0.2,-0.5);
				\coordinate (E) at(0.2,-0.5);
				\fill (D) circle (0.6mm);
				\fill (E) circle (0.6mm);
				\draw (R) -- (D);
				\draw (R) -- (E);
				\filldraw[black,fill=white]  (R) circle (0.6mm);
				\foreach \t in {0.3,0.5,0.7} {%
					\fill ($(D)!\t!(E)$) circle (0.35pt); 
				};
				\node[below=0.2mm of D]{\tiny{$1$}};
				\node[below=0.2mm of E]{\tiny{$m$}};
				\coordinate (shifted) at ($(current bounding box.center)+(0,-2pt)$);
			\end{tikzpicture}\right)
			\times 
			\left(\begin{tikzpicture}[baseline={(shifted)}]
				\coordinate (R) at (0,0);
				\draw (R) circle (1.2mm);
				\fill (R) circle (0.6mm);
				\coordinate (D) at(-0.2,-0.5);
				\coordinate (E) at(0.2,-0.5);
				\draw (R) -- (D);
				\draw (R) -- (E);
				\filldraw[black,fill=white]  (D) circle (0.6mm);
				\filldraw[black,fill=white]  (E) circle (0.6mm);
				\foreach \t in {0.3,0.5,0.7} {%
					\fill ($(D)!\t!(E)$) circle (0.35pt); 
				};
				\node[below=0.2mm of D]{\tiny{$1$}};
				\node[below=0.3mm of E]{\tiny{$n$}};
				\coordinate (shifted) at ($(current bounding box.center)+(0,-2pt)$);
			\end{tikzpicture}\right)
		\end{aligned}$};
	\draw (6,0) node[anchor=north west,text width=8cm]{\hspace{75pt}\underline{ Examples} \\
		$\begin{aligned}[t]
			&(i)\quad\begin{tikzpicture}[baseline={(shifted)}]
				\coordinate (R) at (0,0);
				\fill (R) circle (0.6mm);
				\draw (R) circle (1.2mm);
				\coordinate (A) at (0,-0.5);
				\draw (R) -- (A);
				\filldraw[black,fill=white] (A) circle (0.6 mm);
				\coordinate (shifted) at ($(current bounding box.center)+(0,-2pt)$);
			\end{tikzpicture}
			=
			\begin{tikzpicture}[baseline={(shifted)}]
				\coordinate (R) at (0,0);
				\draw (R) circle (1.2mm);
				\coordinate (A) at (0,-0.5);
				\draw (R) -- (A);
				\filldraw[black,fill=white]  (R) circle (0.6mm);
				\fill (A) circle (0.6 mm);
				\coordinate (shifted) at ($(current bounding box.center)+(0,-2pt)$);
			\end{tikzpicture}\qquad
			(ii)\quad\begin{tikzpicture}[baseline={(shifted)}]
				\coordinate (R) at (0,0);
				\fill (R) circle (0.6mm);
				\draw (R) circle (1.2mm);
				\coordinate (A) at (0,-0.5); 
				\draw (R) -- (A);
				\coordinate (B) at (0,-1);
				\fill (B) circle (0.6mm);
				\draw (A) -- (B);
				\filldraw[black,fill=white] (A) circle (0.6 mm);
				\coordinate (shifted) at ($(current bounding box.center)+(0,-2pt)$);
			\end{tikzpicture}=
			\begin{tikzpicture}[baseline={(shifted)}]
				\coordinate (R) at (0,0);
				\draw (R) circle (1.2mm);
				\coordinate (A) at (-0.3,-0.5);
				\coordinate (B) at (+0.3,-0.5);
				\draw (R) -- (A);
				\draw (R) -- (B);
				\filldraw[black,fill=white] (R) circle (0.6mm);
				\fill (A) circle (0.6 mm);
				\fill (B) circle (0.6 mm);
				\coordinate (shifted) at ($(current bounding box.center)+(0,-2pt)$);
			\end{tikzpicture}\qquad
			(iii)\quad\begin{tikzpicture}[baseline={(shifted)}]
				\coordinate (R) at (0,0);
				\fill (R) circle (0.6mm);
				\draw (R) circle (1.2mm);
				\coordinate (A) at (0,-0.5); 
				\draw (R) -- (A); 
				\coordinate (B) at (0,-1);
				\coordinate (C) at (0,-1.5);
				\fill (B) circle (0.6mm);
				\draw (A) -- (B);
				\draw (B) -- (C);
				\filldraw[black,fill=white] (A) circle (0.6 mm);
				\filldraw[black,fill=white] (C) circle (0.6 mm);
				\coordinate (shifted) at ($(current bounding box.center)+(0,-2pt)$);
			\end{tikzpicture}=
			\begin{tikzpicture}[baseline={(shifted)}]
				\coordinate (R) at (0,0);
				\draw (R) circle (1.2mm);
				\coordinate (A) at (-0.3,-0.5);
				\coordinate (B) at (+0.3,-0.5);
				\coordinate (C) at (-0.3,-1);
				\draw (R) -- (A);
				\draw (R) -- (B);
				\draw (A) -- (C);
				\filldraw[black,fill=white] (R) circle (0.6 mm);
				\fill (A) circle (0.6 mm);
				\fill (B) circle (0.6 mm);
				\filldraw[black,fill=white] (C) circle (0.6 mm);
				\coordinate (shifted) at ($(current bounding box.center)+(0,-2pt)$);
			\end{tikzpicture}\\
			&(i)\quad\begin{tikzpicture}[baseline={(mid.center)}]
				\coordinate (R) at (0,-0.15);
				\coordinate (A) at (0,-0.65);
				\fill (A) circle (0.6mm);
				\draw (R) -- (A);
				\coordinate (shifted) at ($(current bounding box.center)+(0,-2pt)$);
			\end{tikzpicture}\,\,\leq\,\,
			\begin{tikzpicture}[baseline={(shifted)}]
				\coordinate (R) at (0,0);
				\fill (R) circle (0.6mm);
				\draw (R) circle (1.2mm);
				\coordinate (shifted) at ($(current bounding box.center)+(0,-2pt)$);
			\end{tikzpicture}
			\qquad
			(ii)\quad\begin{tikzpicture}[baseline={(shifted)}]
				\coordinate (R) at (0,0);
				\fill (R) circle (0.6mm);
				\draw (R) circle (1.2mm);
				\coordinate (A) at (0,-0.5); 
				\draw (R) -- (A);
				\coordinate (B) at (0,-1);
				\coordinate (C) at (0,-1.5);
				\fill (B) circle (0.6mm);
				\draw (A) -- (B);
				\draw (B) -- (C);
				\filldraw[black,fill=white] (A) circle (0.6 mm);
				\filldraw[black,fill=white] (C) circle (0.6 mm);
				\coordinate (shifted) at ($(current bounding box.center)+(0,-2pt)$);
			\end{tikzpicture}
			\leq
			\begin{tikzpicture}[baseline={(shifted)}]
				\coordinate (R) at (0,0);
				\fill (R) circle (0.6mm);
				\draw (R) circle (1.2mm);
				\coordinate (A) at (0,-0.5); 
				\draw (R) -- (A);
				\coordinate (B) at (0,-1);
				\coordinate (C) at (0,-1.5);
				\fill (B) circle (0.6mm);
				\draw[dotted] (A) -- (B);
				\draw (B) -- (C);
				\filldraw[black,fill=white] (A) circle (0.6 mm);
				\filldraw[black,fill=white] (C) circle (0.6 mm);
				\coordinate (shifted) at ($(current bounding box.center)+(0,-2pt)$);
			\end{tikzpicture}\,\, =\,\, 
			\left(\,\,\begin{tikzpicture}[baseline={(shifted)}]
				\coordinate (R) at (0,0);
				\fill (R) circle (0.6mm);
				\draw (R) circle (1.2mm);
				\coordinate (A) at (0,-0.5);  
				\draw (R) -- (A);
				\filldraw[black,fill=white] (A) circle (0.6 mm);
				\coordinate (shifted) at ($(current bounding box.center)+(0,-2pt)$);
			\end{tikzpicture}\,\,\right)^2\\
			&(iii)\quad\begin{tikzpicture}[baseline={(shifted)}]
				\coordinate (P1) at (0,0);
				\coordinate (P2) at (-.3,-.5);
				\coordinate (P3) at (.3,-.5);
				\coordinate (P4) at (-.3,-1);
				\draw (P3) -- (P1) -- (P2) -- (P4);
				\filldraw[black,fill=white] circle (.6mm) (P4) circle (.6mm);
				\draw (P1) circle (1.2mm);
				\fill (P2) circle (.6mm) (P3) circle (.6mm);
				\coordinate (shifted) at ($(current bounding box.center)+(0,-2.5pt)$);
			\end{tikzpicture}\,\,\leq\,\,
			\begin{tikzpicture}[baseline={(shifted)}]
				\coordinate (P1) at (0,0);
				\coordinate (P2) at (-.3,-.5);
				\coordinate (P3) at (.3,-.5);
				\coordinate (P4) at (-.3,-1);
				\draw[dotted] (P1) -- (P2);
				\draw (P2) -- (P4);
				\draw (P1) -- (P3); 
				\filldraw[black,fill=white] (P1) circle (.6mm) (P4) circle (.6mm);
				\draw (P1) circle (1.2mm);
				\fill (P2) circle (.6mm) (P3) circle (.6mm);
				\coordinate (shifted) at ($(current bounding box.center)+(0,-2.5pt)$);
			\end{tikzpicture}=\left(\,\,\begin{tikzpicture}[baseline={(shifted)}]
				\coordinate (R) at (0,0);
				\fill (R) circle (0.6mm);
				\draw (R) circle (1.2mm);
				\coordinate (A) at (0,-0.5);
				\draw (R) -- (A);
				\filldraw[black,fill=white] (A) circle (0.6 mm);
				\coordinate (shifted) at ($(current bounding box.center)+(0,-2pt)$);
			\end{tikzpicture}\,\,\right)^2
		\end{aligned}$
	};
\end{tikzpicture}
	\label{fig:approximation} 
    \vspace{-10pt}
\caption{We show the diagrammatic \textit{swap} and \textit{cut} rules used in the planar approximation. (a) Swapping roots: the identity
\eqref{tree.rule.1} allows us to interchange the leading $J_A$ and $J_B$ sums, which in the tree representation corresponds to moving the marked root between a black and a white vertex without changing the value of the contribution. (b) Diagram cut inequalities: cutting an internal edge between neighbouring vertices (represented by dotted line) represents dropping a conditioning in the corresponding nested sums, leading to the upper bounds of the form \eqref{bound.a}. In the right column we have shown some simple examples of both cases. 
}
\end{figure*}

Since a summing index is only condition to one other summing index which is of the different type, e.g., $J_{B,r}\in\mathfrak{J}_B^{(J_{A,r'})}$ or $J_{A,r'}\in\mathfrak{J}_A^{(J_{B,r})}$, it happens that the graph is bipartite (the only neighbors of a type of vertex  are of the other type, only). Let us give the two simplest examples for $L=2,3$, where it is
\begin{widetext}
    \begin{align}\label{2nd.purity.a}
    \braket{\Tr\rho^2}&=\sum_{J_{A,1}}\sum_{J_{B,1}\in\mathfrak{J}_B^{(J_{A,1})}}\sum_{J_{B,2}\in\mathfrak{J}_B^{(J_{A,1})}}\Lambda_{{\bf J}_A,{\bf J}_B}n_{J_{A,1}}^An_{J_{B,1}}^Bn_{J_{B,2}}^B+\sum_{J_{A,1}}\sum_{J_{B,1}\in\mathfrak{J}_B^{(J_{A,1})}}\sum_{J_{A,2}\in\mathfrak{J}_A^{(J_{B,1})}}\Lambda_{{\bf J}_A,{\bf J}_B}n_{J_{A,1}}^An_{J_{B,1}}^Bn_{J_{A,2}}^A
\end{align}
and
\begin{align}\label{3rd.purity.a}
    \braket{\Tr\rho^3}&=\sum_{J_{A,1}}\sum_{J_{B,1}\in\mathfrak{J}_B^{(J_{A,1})}}\sum_{J_{B,2}\in\mathfrak{J}_B^{(J_{A,1})}}\sum_{J_{B,3}\in\mathfrak{J}_B^{(J_{A,1})}}\Lambda_{{\bf J}_A,{\bf J}_B}n_{J_{A,1}}^An_{J_{B,1}}^Bn_{J_{B,2}}^Bn_{J_{B,3}}^B\nonumber\\
    &\quad+\sum_{J_{A,1}}\sum_{J_{B,1}\in\mathfrak{J}_B^{(J_{A,1})}}\sum_{J_{B,2}\in\mathfrak{J}_B^{(J_{A,1})}}\sum_{J_{A,2}\in\mathfrak{J}_B^{(J_{B,2})}}\Lambda_{{\bf J}_A,{\bf J}_B}n_{J_{A,1}}^An_{J_{B,1}}^Bn_{J_{B,2}}^Bn_{J_{A,2}}^A\nonumber\\
    &\quad+\sum_{J_{A,1}}\sum_{J_{B,1}\in\mathfrak{J}_B^{(J_{A,1})}}\sum_{J_{A,2}\in\mathfrak{J}_B^{(J_{B,1})}}\sum_{J_{B,2}\in\mathfrak{J}_B^{(J_{A,1})}}\Lambda_{{\bf J}_A,{\bf J}_B}n_{J_{A,1}}^An_{J_{B,1}}^Bn_{J_{B,2}}^Bn_{J_{A,2}}^A\nonumber\\
    &\quad+\sum_{J_{A,1}}\sum_{J_{B,1}\in\mathfrak{J}_B^{(J_{A,1})}}\sum_{J_{A,2}\in\mathfrak{J}_A^{(J_{B,1})}}\sum_{J_{B,2}\in\mathfrak{J}_B^{(J_{A,2})}}\Lambda_{{\bf J}_A,{\bf J}_B}n_{J_{A,1}}^An_{J_{B,1}}^Bn_{J_{A,2}}^An_{J_{B,2}}^B\nonumber\\
    &\quad+\sum_{J_{A,1}}\sum_{J_{B,1}\in\mathfrak{J}_B^{(J_{A,1})}}\sum_{J_{A,2}\in\mathfrak{J}_A^{(J_{B,1})}}\sum_{J_{A,3}\in\mathfrak{J}_A^{(J_{B,1})}}\Lambda_{{\bf J}_A,{\bf J}_B}n_{J_{A,1}}^An_{J_{B,1}}^Bn_{J_{A,2}}^An_{J_{A,3}}^A+\text{NPT},
\end{align}
\end{widetext}
see Fig.~\ref{fig:tree-diagrams} for their diagrammatic representations. We have swapped the sum over $m$ with those against $J_{A,r}$ and $J_{B,r'}$ and applied the abbreviation $\Lambda_{{\bf J}_A,{\bf J}_B}$ in~\eqref{Lambda.def} where the set for $m$ surely depends on all previous summing indices $J_{A,r}$ and $J_{B,r'}$. We underline that the mean purity $\braket{\Tr\rho^2}$ has no non-planar diagrams and is, thus, exact.

The set of all these trees created by the recurrence relations~\eqref{recc.JA} and~\eqref{recc.JB} shall be denoted by $\mathcal{T}$ and the vertex set of a tree $t\in \mathcal{T}$ by $\mathcal{V}_t$. The ``color'' of a vertex $\widetilde J\in\mathcal{V}_t$ will be denoted by $c_{\widetilde J}\in\{A,B\}$. Then, we can write the higher purity in the compact form
\begin{equation}
     \braket{\Tr\rho^L}=\sum_{t\in\mathcal{T}}\sum_{\mathcal{V}_t}\Lambda_{{\bf J}_A,{\bf J}_B}\prod_{\widetilde J\in\mathcal{V}_t}n_{\widetilde J}^{c_{\widetilde J}}+\text{NPT}.
\end{equation}
The sum $\sum_{\mathcal{V}_t}$ shall represent the sum over all possible $J_{A,r}$ and $J_{A,r'}$ in this tree.
We underline that terms in $\text{NPT}$ have at least two dimensions $n_{\widetilde J}^{c_{\widetilde J}}$ less compared to the leading term and are, thus, exponentially smaller when $V\to\infty$.

There are several rules for the sums which carry over to their diagrammatic representation. For example, it holds
\begin{equation}\label{tree.rule.1}
    \sum_{J_A}\sum_{J_B\in\mathfrak{J}_B^{(J_A)}}=\sum_{J_B}\sum_{J_A\in\mathfrak{J}_A^{(J_B)}}
\end{equation}
for the leading sum (the root of the tree). Moreover, two sums not conditioned on each other commute, \ie
\begin{equation}\label{tree.rule.2}
\begin{split}
    \sum_{J_A\in\mathfrak{J}_B^{(J_B)}}\sum_{J'_B\in\mathfrak{J}_B^{(J'_A)}}&=\sum_{J'_B\in\mathfrak{J}_B^{(J'_A)}}\sum_{J_A\in\mathfrak{J}_B^{(J_B)}},\\
    \sum_{J_A\in\mathfrak{J}_A^{(J_B)}}\sum_{J'_A\in\mathfrak{J}_A^{(J'_B)}}&=\sum_{J'_A\in\mathfrak{J}_A^{(J'_B)}}\sum_{J_A\in\mathfrak{J}_A^{(J_B)}},\\
    \sum_{J_B\in\mathfrak{J}_B^{(J_A)}}\sum_{J'_B\in\mathfrak{J}_B^{(J'_A)}}&=\sum_{J'_B\in\mathfrak{J}_B^{(J'_A)}}\sum_{J_B\in\mathfrak{J}_A^{(J_B)}}.
\end{split}
\end{equation}

This construction allows us to identify two most important terms in the set of trees. There are always the two terms
\begin{equation}\label{leading.terms}
\begin{split}
    \mathfrak{R}_A^{(L)}&=\sum_{J_A}\sum_{J_{B,1},\ldots,J_{B,L}\in\mathfrak{J}_B^{(J_A)}}\Lambda_{{\bf J}_A,{\bf J}_B} n_{J_A}^{A}\prod_{l=1}^L n_{J_{B,l}}^{B},\\
    \mathfrak{R}_B^{(L)}&=\sum_{J_B}\sum_{J_{A,1},\ldots,J_{A,L}\in\mathfrak{J}_A^{(J_B)}}\Lambda_{{\bf J}_A,{\bf J}_B} n_{J_B}^{B}\prod_{l=1}^L n_{J_{A,l}}^{A}
\end{split}
\end{equation}
in the set of sums which we claim are exponentially dominating contributions. They correspond to trees with either only one black vertex ($J_A$ sum) and $L$ white ones ($J_{B,l}$ sum) or vice versa. Thence, they are the only ones that have this structure and any tree with only one white or black vertex can be transformed via the rules~\eqref{tree.rule.1} and~\eqref{tree.rule.2} into such a way that this single vertex in the tree becomes its root.

Using the fact that $\Lambda_{{\bf J}_A,{\bf J}_B}$ have only an algebraic behavior in $V$ instead of an exponential as it is the case for the dimensions, see subsection~\ref{sec:scaling}, we can approximate their inner sums at the lower terminals $|J-J_A|$ or $|J-J_B|$, respectively, like we have done it in the calculation for the total Hilbert space dimension $d$, see the discussion after~\eqref{eq:d-asymptotics}. This yields the approximations
\begin{eqnarray}
  \mathfrak{R}_A^{(L)}&=&\sum_{J_A}\sum_m\left(\widetilde\lambda_{m,f}^{(A;J_A)}\right)^L n_{J_{A}}^{A}  \left(n_{J_{B,\min}^{(J_A,m)}}^B\right)^L[1+O(V^{-1})],\nonumber\\
  \mathfrak{R}_B^{(L)}&=&\sum_{J_B}\sum_m\left(\widetilde\lambda_{m,f}^{(B;J_B)}\right)^L n_{J_{B}}^{B}  \left(n_{J_{A,\min}^{(J_B,m)}}^A\right)^L[1+O(V^{-1})].\nonumber\\
  &&\label{R.approx}
\end{eqnarray}
where we employed $J_{B,\min}^{(J_A,m)}=\min\{|m|,|J-J_A|\}$ and $J_{A,\min}^{(J_B,m)}=\min\{|m|,|J-J_B|\}$ as well as the functions
\begin{center}
\begin{figure}[t!]\label{fig:planar representations}
	\begin{tikzpicture}
		\draw (0,0) node[anchor=north west,text width=15cm]{
			$\begin{aligned}[t]
				\sum_{J_A} \mathcal{E}_{A,2}\,\, &=
				\begin{tikzpicture}[baseline={(shifted)}]
					\coordinate (P1) at (0,0);
					\coordinate (P2) at (-.3,-.5);
					\coordinate (P3) at (.3,-.5);
					\draw (P2) -- (P1) -- (P3);
					\filldraw[black,fill=white] (P1) circle (.6mm) ;
					\draw (P1) circle (1.2mm);
					\fill(P2) circle (.6mm) (P3) circle (.6mm);
					\coordinate (shifted) at ($(current bounding box.center)+(0,-2.5pt)$);
				\end{tikzpicture}
				+ 
				\begin{tikzpicture}[baseline={(shifted)}]
					\coordinate (P1) at (0,0);
					\coordinate (P2) at (-.3,-.5);
					\coordinate (P3) at (.3,-.5);
					\draw (P2) -- (P1) -- (P3);
					\fill (P1) circle (.6mm) ;
					\draw (P1) circle (1.2mm);
					\filldraw[black,fill=white] (P2) circle (.6mm) (P3) circle (.6mm);
					\coordinate (shifted) at ($(current bounding box.center)+(0,-2.5pt)$);
				\end{tikzpicture}\\
				\sum_{J_A} \mathcal{E}_{A,3}\,\, &=
				\underbrace{\begin{tikzpicture}[baseline={(shifted)}]
						\coordinate (P1) at (0,0);
						\coordinate (P2) at (-.3,-.5);
						\coordinate (P3) at (.3,-.5);
						\coordinate (P4) at (-.3,-1);
						\draw (P3) -- (P1) -- (P2) -- (P4);
						\filldraw[black,fill=white] (P1) circle (.6mm) (P4) circle (.6mm);
						\draw (P1) circle (1.2mm);
						\fill (P2) circle (.6mm) (P3) circle (.6mm);
						\coordinate (shifted) at ($(current bounding box.center)+(0,-2.5pt)$);
					\end{tikzpicture}
					+2\,\,
					\begin{tikzpicture}[baseline={(shifted)}]
						\coordinate (P1) at (0,0);
						\coordinate (P2) at (-.3,-.5);
						\coordinate (P3) at (.3,-.5);
						\coordinate (P4) at (-.3,-1);
						\draw (P3) -- (P1) -- (P2) -- (P4);
						\fill (P1) circle (.6mm) (P4) circle (.6mm);
						\draw (P1) circle (1.2mm);
						\filldraw[black,fill=white] (P2) circle (.6mm) (P3) circle (.6mm);
						\coordinate (shifted) at ($(current bounding box.center)+(0,-2.5pt)$);
				\end{tikzpicture}}_{\text{Sub-leading terms}}
				+
				\underbrace{\begin{tikzpicture}[baseline={(shifted)}]
						\coordinate (P1) at (0,0);
						\coordinate (P2) at (0,-.5);
						\coordinate (P3) at (-.3,-.5);
						\coordinate (P4) at (.3,-.5);
						\draw (P2) -- (P1) -- (P3) (P1) -- (P4);
						\filldraw[black,fill=white] (P1) circle (.6mm);
						\draw (P1) circle (1.2mm);
						\fill (P2) circle (.6mm)  (P3) circle (.6mm) (P4) circle (.6mm);
						\coordinate (shifted) at ($(current bounding box.center)+(0,-2.5pt)$);
					\end{tikzpicture}
					+
					\begin{tikzpicture}[baseline={(shifted)}]
						\coordinate (P1) at (0,0);
						\coordinate (P2) at (0,-.5);
						\coordinate (P3) at (-.3,-.5);
						\coordinate (P4) at (.3,-.5);
						\draw (P2) -- (P1) -- (P3) (P1) -- (P4);
						\fill (P1) circle (.6mm);
						\draw (P1) circle (1.2mm);
						\filldraw[black,fill=white] (P2) circle (.6mm)  (P3) circle (.6mm) (P4) circle (.6mm);
						\coordinate (shifted) at ($(current bounding box.center)+(0,-2.5pt)$);
				\end{tikzpicture}}_{\text{Leading terms}}\\
				&\leq
				3\left(\,\,\begin{tikzpicture}[baseline={(shifted)}]
					\coordinate (R) at (0,0);
					\fill (R) circle (0.6mm);
					\draw (R) circle (1.2mm);
					\coordinate (A) at (0,-0.6);  
					\draw (R) -- (A);
					\filldraw[black,fill=white] (A) circle (0.6 mm);
					\coordinate (shifted) at ($(current bounding box.center)+(0,-2pt)$);
				\end{tikzpicture}\,\,\right)^2+\begin{tikzpicture}[baseline={(shifted)}]
					\coordinate (P1) at (0,0);
					\coordinate (P2) at (0,-.5);
					\coordinate (P3) at (-.3,-.5);
					\coordinate (P4) at (.3,-.5);
					\draw (P2) -- (P1) -- (P3) (P1) -- (P4);
					\filldraw[black,fill=white] (P1) circle (.6mm);
					\draw (P1) circle (1.2mm);
					\fill (P2) circle (.6mm)  (P3) circle (.6mm) (P4) circle (.6mm);
					\coordinate (shifted) at ($(current bounding box.center)+(0,-2.5pt)$);
				\end{tikzpicture}
				+
				\begin{tikzpicture}[baseline={(shifted)}]
					\coordinate (P1) at (0,0);
					\coordinate (P2) at (0,-.5);
					\coordinate (P3) at (-.3,-.5);
					\coordinate (P4) at (.3,-.5);
					\draw (P2) -- (P1) -- (P3) (P1) -- (P4);
					\fill (P1) circle (.6mm);
					\draw (P1) circle (1.2mm);
					\filldraw[black,fill=white] (P2) circle (.6mm)  (P3) circle (.6mm) (P4) circle (.6mm);
					\coordinate (shifted) at ($(current bounding box.center)+(0,-2.5pt)$);
				\end{tikzpicture}
			\end{aligned}$};
	\end{tikzpicture}
\vspace{-10pt}
\caption{We illustrate the examples for $k=2$ and $k=3$ showing how the swap and cut rules are used to relate generic rooted trees to equivalent trees, where in addition to subleading terms we get the sum of two extremal trees $\mathfrak{R}_A^{(L)}$ and $\mathfrak{R}_B^{(L)}$ in~\eqref{leading.terms}, which dominate the large $\V$  behavior.}
\end{figure}
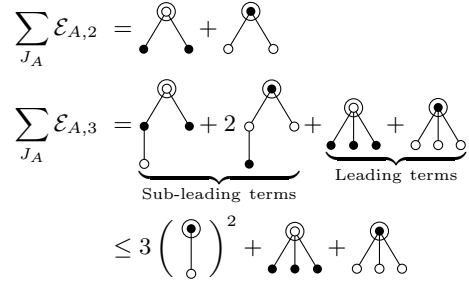
\end{center}

\begin{equation}\label{tildelambda}
\begin{split}
    \widetilde\lambda_{m,f}^{(A;J_A)}&=\sum_{k=0}^\infty \lambda_m^{(J_A,J_{B,\min}^{(J_A,m)}+k)}\left[\gamma\left(\frac{2J_{B,\min}^{(J_A,m)}}{V_B}\right)\right]^k,\\
    \widetilde\lambda_{m,f}^{(B;J_B)}&=\sum_{k=0}^\infty \lambda_m^{(J_{A,\min}^{(J_B,m)}+k,J_B)}\left[\gamma\left(\frac{2J_{A,\min}^{(J_B,m)}}{V_A}\right)\right]^k.
\end{split}
\end{equation}
The sums over $m$ go either between $\pm \min\{J_A,|J-J_A|\}$ or $\pm \min\{J_B,|J-J_B|\}$, respectively. The approximation~\eqref{tildelambda} will be derived and analyzed in subsection~\ref{sec:T2}.

What happens with the other sums? Let us pick a tree $t\in\mathcal{T}$ neither corresponding to $\mathfrak{R}_A^{(L)}$ nor to $\mathfrak{R}_B^{(L)}$. The aim is to bound this term from above by products of $\mathfrak{R}_A^{(l_r)}$ and $\mathfrak{R}_B^{(l_r)}$ with integer parameters $l_r\geq 1$ satisfying $\sum_r(1+l_r)=1+L$. Then, either kind of black and white vertices have at least two elements. Due to the bipartition every neighbor of a black vertex is white and vice versa. Hence, there is a pair of black and white neighboring vertices where each of them has at least two neighbors (including the one in the pair). This link between the two neighbors can be cut to obtain two subtrees which each comprise at least one black and one white vertex. This procedure can be continued until all subtrees have only one black and/or one white vertex but at least one of the other kind so that the corresponding sums represented by them are indeed of the form $\mathfrak{R}_A^{(l_r)}$ or $\mathfrak{R}_B^{(l_r)}$. See Fig.~\ref{fig:approximation} (first row) for the diagrammatic representation.

The cutting corresponds to the analytical upper bound by switching off a condition,  \ie
\begin{equation}\label{inequality}
\sum_{J_A\in\mathfrak{J}_A^{(J_B)}}\leq \sum_{J_A} \quad{\rm and}\quad \sum_{J_B\in\mathfrak{J}_B^{(J_A)}}\leq \sum_{J_B} .
\end{equation}
These inequalities indeed hold true as we only sum over non-negative summands so that the sum can only grow when forgetting a condition (refer to second row of~\ref{fig:approximation}). This is also true when splitting the CG terms into those only depending on the summing indices of the respective new sums, in particular
\begin{eqnarray}
    \Lambda_{({\bf J}_A,{\bf J}'_A),({\bf J}_B,{\bf J}'_B)}\leq\Lambda_{{\bf J}_A,{\bf J}_B}\Lambda_{{\bf J}'_A,{\bf J}'_B}.
\end{eqnarray}
For example, in the $L=3$ we have for the second sum in~\eqref{3rd.purity.a}
\begin{widetext}
\begin{equation}
\begin{split}
&\sum_{J_{A,1}}\sum_{J_{B,1}\in\mathfrak{J}_B^{(J_{A,1})}}\sum_{J_{B,2}\in\mathfrak{J}_B^{(J_{A,1})}}\sum_{J_{A,2}\in\mathfrak{J}_B^{(J_{B,2})}}\Lambda_{{\bf J}_A,{\bf J}_B}n_{J_{A,1}}^An_{J_{B,1}}^Bn_{J_{B,2}}^Bn_{J_{A,2}}^A\\
\leq\,& \sum_{J_{A,1}}\sum_{J_{B,1}\in\mathfrak{J}_B^{(J_{A,1})}}\sum_{J_{B,2}}\sum_{J_{A,2}\in\mathfrak{J}_B^{(J_{B,2})}}\Lambda_{J_{A,1},J_{B,1}}\Lambda_{J_{A,2},J_{B,2}}n_{J_{A,1}}^An_{J_{B,1}}^Bn_{J_{B,2}}^Bn_{J_{A,2}}^A=\mathfrak{R}_A^{(1)}\mathfrak{R}_B^{(1)}=(\mathfrak{R}_A^{(1)})^2.
\end{split}
\end{equation}
\end{widetext}
The last equality follows from the rule~\eqref{tree.rule.1}. Actually, it is even $\mathfrak{R}_A^{(1)}=\mathfrak{R}_B^{(1)}=d$ because of
\begin{equation}
    \Lambda_{J_{A},J_{B}}=\sum_m\lambda_m^{(J_A,J_B)}=1.
\end{equation} In general, we can bound for any tree $t\in\mathcal{T}$, not corresponding to $\mathfrak{R}_A^{(L)}$ nor to $\mathfrak{R}_B^{(L)}$,
\begin{equation}\label{bound.a}
     \sum_{\mathcal{V}_t}\Lambda_{{\bf J}_A,{\bf J}_B}\prod_{\widetilde J\in\mathcal{V}_t}n_{\widetilde J}^{c_{\widetilde J}}\leq \prod_{r=1}^{R_A} \mathfrak{R}_A^{(l_r^{A})}\prod_{r=1}^{R_B} \mathfrak{R}_B^{(l_r^{B})}
\end{equation}
with integers $l_r^{A},l_r^{B}\geq1$ satisfying
\begin{equation}
\sum_{r=1}^{R_A} (1+l_r^{A})+\sum_{r=1}^{R_B} (1+l_r^{B})=1+L.
\end{equation}

Now, we are ready to apply the exponential decay~\eqref{vanishing.ratios}. We divide the inequality by the sum $\mathfrak{R}_A^{(L)}+\mathfrak{R}_B^{(L)}$ we claim to be of leading order, especially we write
\begin{align}
    &\frac{\sum_{\mathcal{V}_t}\Lambda_{{\bf J}_A,{\bf J}_B}\prod_{\widetilde J\in\mathcal{V}_t}n_{\widetilde J}^{c_{\widetilde J}}}{\mathfrak{R}_A^{(L)}+\mathfrak{R}_B^{(L)}}\nonumber\\
    \leq\,& \prod_{r=1}^{R_A} \frac{\mathfrak{R}_A^{(l_r^{A})}}{(\mathfrak{R}_A^{(L)}+\mathfrak{R}_B^{(L)})^{(1+l_r)/(1+L)}}\nonumber\\
    &\times\prod_{r=1}^{R_B} \frac{\mathfrak{R}_B^{(l_r^{B})}}{(\mathfrak{R}_A^{(L)}+\mathfrak{R}_B^{(L)})^{(1+l_r)/(1+L)}}.
\end{align}
When combining this with the approximation~\eqref{R.approx} we get a sum of terms (note that this sum only runs at most over $V^2$ terms) of the form
\begin{eqnarray}
    &\left(\widetilde\lambda_{m,f}^{(A;J_A)}\right)^{l_r^A} \frac{n_{J_{A}}^{A}  \left(n_{|J-J_A|}^{B}\right)^{l_r^A}}{(\mathfrak{R}_A^{(L)}+\mathfrak{R}_B^{(L)})^{(1+l_r^A)/(1+L)}}\nonumber\\
    \leq\,&\left(\widetilde\lambda_{m,f}^{(A;J_A)}\right)^{\frac{Ll_r^A-1}{1+L}}\frac{n_{J_{A}}^{A}  \left(n_{|J-J_A|}^{B}\right)^{l_r^A}}{\left[n_{J_{A}}^{A}  \left(n_{|J-J_A|}^{B}\right)^{L}\right]^{\frac{1+l_r^A}{1+L}}}
\end{eqnarray}
and similarly when swapping the roles of $A$ and $B$.
With the help of~\eqref{vanishing.ratios} we see that those term vanish exponentially.

Furthermore, either $\mathfrak{R}_A^{(L)}$ exponentially dominates $\mathfrak{R}_B^{(L)}$ or vice versa, because of~\eqref{vanishing.ratios.b}. This surely depends on the subsystem ratio $f\neq1/2$. 
In summary, we have shown that for any fixed integer $L\geq1$, the moment is approximately
\begin{equation}\label{moment.asymptotic}
    \braket{\Tr\rho^L}=\begin{cases}
        \mathfrak{R}_A^{(L)}[1+O(e^{-\tilde{\varepsilon}V})], &f<1/2,\\
        \mathfrak{R}_B^{(L)}[1+O(e^{-\tilde{\varepsilon}V})], & f>1/2
    \end{cases}
\end{equation}
with some non-vanishing exponent $\tilde{\varepsilon}>0$.

For $f=1/2$  we obtain the sum $\mathfrak{R}_A^{(L)}+\mathfrak{R}_B^{(L)}$ as the leading order contribution for $L\geq 2$. It is important to note that $L\geq2$ excludes the case $L=1$ for $f=1/2$ since then the formula would break down because the sum would yield $2d$ for $\braket{\Tr\rho}$ instead of the correct value $d$. This shows that for $f=1/2$ the replica or moment method breaks down. We believe that the origin of this mismatch might be a non-analyticity in the exponent $L$ at $L=1$.

Assuming $f<1/2$, the result~\eqref{moment.asymptotic} implies that the (non-normalized) level density of $\rho$ can be approximated by a sum of Dirac delta functions,
\begin{equation}
    p(\lambda)=\sum_{J_A,m} n_{J_A}^A\delta\left(\lambda-\sum_{J_B\in\mathfrak{J}_{B,m}^{(J_A)}}\lambda_m^{(J_A,J_B)}n_{J_B}^B\right)+O(e^{-\tilde{\varepsilon}V}).
\end{equation}
The error is understood in the weak sense meaning after averaging over suitable test functions.
Hence, the entanglement entropy becomes
\begin{equation}\label{entanglement.after.planar}
\begin{split}
    \braket{S_A}&=-\sum_{J_A,m} n_{J_A}^A\left(\sum_{J_B\in\mathfrak{J}_{B,m}^{(J_A)}}\lambda_m^{(J_A,J_B)}\frac{n_{J_B}^B}{d}\right)\\
    &\quad\times\log\left(\sum_{J'_B\in\mathfrak{J}_{B,m}^{(J_A)}}\lambda_m^{(J_A,J'_B)}\frac{n_{J'_B}^B}{d}\right)+O(e^{-\tilde{\varepsilon}V}).
\end{split}
\end{equation}
Originally, the sum over $m$ is the last one (most outer) carried out with bounds listed in table~\ref{tab:bounds}. However, we can change the order of summation, so that we sum over $m$ before $J_A$. 

We can now ask for which values of $J_A$ this sum is dominated, considering the exponential growth of $d$ according to~\eqref{eq:d-asymptotics}, which appears in the denominator. Neither the logarithm term, nor $\lambda^{(J_A,J_B)}_{m}$ will be able to contribute such an exponential growth, so that the peak will be determined by $n^A_{J_A}$ and $n^B_{J_B}$ alone. Their product was already analyzed in the context of equation~\eqref{eq:ntilde}, where it was found that for a given $J_A$, the largest $n^B_{J_B}$ occurs for $J_B=|J-J_A|$ and then the dominating contribution in $J_A$ comes from $J_A=f J$ and $J_B=(1-f)J$. Moreover, we can use that the Clebsch-Gordon coefficients around this peak scale as
\begin{align}
    \lambda_m^{(fJ,(1-f)J)}=O\Bigl(\frac{1}{\sqrt{V}}\Bigl)\,,
\end{align}
so that we can factor out $\tilde{n}^B_{J_A}/[d\sqrt{\V}]$ inside the logarithm with $\tilde{n}^B_{J_A}$ given in~\eqref{eq:ntilde}

This discussion motivates us to split this expression~\eqref{entanglement.after.planar} into two terms, namely $\braket{S_A}=T_{\mathrm{S}}+T_{\mathrm{CG}}+O(e^{-\tilde{\varepsilon}V})$ given by
\begin{align}
    T_{\mathrm{S}}&=-\sum_{J_A,m}n^A_{J_A}\sum_{J_B\in\mathfrak{J}_{B,m}^{(J_A)}}\lambda_m^{(J_A,J_B)}\frac{n_{J_B}^B}{d}\log{\tfrac{\tilde{n}^B_{J_A}}{d\sqrt{\V}}}\,,\label{eq:TS}\\
    T_{\mathrm{CG}}&=-\sum_{J_A,m} n_{J_A}^A\sum_{J_B\in\mathfrak{J}_{B,m}^{(J_A)}}\lambda_m^{(J_A,J_B)}\frac{n_{J_B}^B}{d}\nonumber\\
    &\quad\times\log\left(\sum_{J'_B\in\mathfrak{J}_{B,m}^{(J_A)}}\sqrt{\V}\lambda_m^{(J_A,J'_B)}\frac{n_{J'_B}^B}{\tilde{n}^B_{J_A}}\right)\,.\label{eq:TCG}
\end{align}
We will see that $T_{\mathrm{S}}$ captures the relevant scaling in the volume, while only $T_{\mathrm{CG}}$ gives the leading contribution of the Clebsch-Gordon coefficients.

\subsection{Calculation of $T_{\text{S}}$:\\Volume law, logarithm and constant order}\label{sec:T1}

We consider the sum $T_{\mathrm{S}}$ and swap the sums in $m$ and $J_B$ to find
\begin{align}
    T_{\mathrm{S}}&=-\sum_{J_A}n^A_{J_A}\sum_{J_B\in\mathfrak{J}_{B}^{(J_A)}}\sum_{m}\lambda_m^{(J_A,J_B)}\frac{n_{J_B}^B}{d}\log{\frac{\tilde{n}^B_{J_A}}{d\sqrt{\V}}}\,,\nonumber\\
    &=-\sum_{J_A}\left(\frac{n^A_{J_A}\tilde{n}^B_{J_A}}{d}\right)\log\left(\frac{\tilde{n}^B_{J_A}}{d\sqrt{\V}}\right)\,,\label{eq:TS3}
\end{align}
where we used the normalization~\eqref{eq:CG-normalization} to perform the sum in $m$ for $-\min(J_A,J_B)\leq m\leq\min(J_A,J_B)$. This simplification enables us  to approximate the sum in $J_A$ by an integral of the form
\begin{align}
    T_{\mathrm{S}}=\int d(\delta J_A) \,\varrho(\delta J_A)\varphi(\delta J_A)\,,
\end{align}
where we introduced
\begin{align}
    \varrho(\delta J_A)&=\frac{2}{\sqrt{\V}}\,\frac{n^A_{J_A}\tilde{n}_{J_A}^B}{d}\,,\\
    \varphi(\delta J_A)&=\log{\frac{\tilde{n}^B_{J_A}}{d\sqrt{\V}}}\,,\\
    \delta J_A&=\tfrac{2(J_A-fJ)}{\sqrt{\V}}=\sqrt{\V}(j_A-fj)\,.\label{eq:deltaJA}
\end{align}
We already saw in~\eqref{gaussian-peak} that the probability weight $\varrho$ created by the dimensions approaches a Gaussian centered at $j_A=fj$ when written in terms of $j_A=2J_A/\V$. However, its variance scales $\sigma^2=O(\frac{1}{V})$ motivating the introduction of the variable $\delta J_A$ in~\eqref{eq:deltaJA}, as this will ensure that $\varrho$ approaches a Gaussian with fixed variance independent of $\V$ in the limit $\V\to\infty$. Specifically, we obtain the expansion in $\V$
\begin{align}\label{eq:varrho}
\varrho(\delta J_A)
	&=\sqrt{\tfrac{1}{2\pi\widetilde\sigma^2}}\e^{-\frac{1}{2}\frac{\delta J_A^2}{\widetilde\sigma^2}}(1+\tfrac{D_1\,\delta J_A+D_3\,\delta J_A^3}{\sqrt{V}}+O(\tfrac{1}{\V})).
\end{align}
The expansion coefficients are
\begin{align}
	\widetilde\sigma^2&=f(1-f)(1-j^2)\,,\\
	D_1&=\frac{1-j(1-j)-f(1-j+2j^2)}{fj(1-f)(1-j^2)}\,,\\
	D_3&=\frac{1}{3}\frac{(1-2f)j}{f^2(1-f)^2(1-j^2)^2}\,.
\end{align}

These were derived by writing $\rho(j_A)=X(j_A)\e^{\V Y(j_A)}$ with appropriate functions $X(j_A)$ and $Y(j_A)$, which can be found explicitly from the asymptotic formulas of $n^A_{J_A}$ in~\eqref{eq:n1-asymptotics}, $\tilde{n}^B_{J_A}$ in~\eqref{eq:ntilde} and $d$ in~\eqref{eq:d-asymptotics}. We then substituted $\delta J_A=\sqrt{\V}(j_A-fj)$ and expanded in powers of $\sqrt{V}$, yielding a Gaussian around the saddle point $\delta J_A=0$. The coefficients are then the derivatives $D_1=X'(jf)$ and $D_3=\frac{1}{2}Y'''(jf)$.

For the observable $\varphi$, we  start from~\eqref{eq:d-asymptotics} and~\eqref{eq:pre-final-n2} and similarly expand in $\V$ leading to
\begin{align}
	\varphi(\delta J_A)={}&f\V \beta(j)+\frac{1}{2}\log\V+\sqrt{\V}\beta'(j)\delta J_A\nonumber\\		&+\tfrac{\beta''(j)}{2(1-f)}\delta J_A^2+o(1). 
\end{align}
We would like to underline that there is no need to expand the weight $\rho$ to order $1/V$. Surely this term times the volume term in $\varphi$ is of order one. However, the leading term of $\varphi$ is constant in $\delta J_A$.  Combining this with the proper normalization of $\rho$ implies that the corresponding term must integrate to zero which it indeed does.

Finally, we evaluate the integral order by order to find
\begin{align}
\label{eq:T1-full-result}
	T_{\text{S}}={}&\V f\beta(j)+\tfrac{1}{2}\log(\V)+\tfrac{f+\log(1-f)}{2}+\log\left(\tfrac{2j}{1+j}\right)\nonumber\\
	&{}+\tfrac{(1-f)(1-j)}{2j}\log\left(\tfrac{1-j}{1+j}\right)+o(1)\,.
\end{align}

\subsection{Calculation of $T_{\text{CG}}$: Constant correction}\label{sec:T2}
We now evaluate the second piece given by~\eqref{eq:TCG}, which we rewrite as follows,
\begin{align}
    T_{\mathrm{CG}}&=-\sum_{J_A} \frac{n_{J_A}^A\tilde{n}^B_{J_A}}{\sqrt{\V}d}\sum_m g_m(J_A)\log{g_m}(J_A)\,,
\end{align}
where we introduced the function
\begin{align}
g_m(J_A)=\sum_{J'_B\in\mathfrak{J}_{B,m}^{(J_A)}}\sqrt{\V}\lambda_m^{(J_A,J'_B)}\frac{n_{J'_B}^B}{\tilde{n}^B_{J_A}}\label{eq:gm}
\end{align}
by factoring out $\tilde{n}^B_{J_A}$ and conveniently writing $\sqrt{\V} \lambda^{(J_A,J_B)}_m$. We will see that this has a well-defined non-vanishing limit for large $\V$, in particular the splitting $\braket{S_A}=T_{\mathrm{S}}+T_{\mathrm{CG}}+o(1)$ ensured $T_{\mathrm{CG}}=O(1)$.

We recall that only $n^A_{J_A}$, $n^B_{J_B}$ and $d$ scale exponentially in $\V$, which means that the additional contribution from $\lambda^{(J_A,J_B)}_m$ in the sum will not change the fact that the leading contribution stems from the terms around $J_A=fJ$, but to evaluate the sum correctly, we will still need to perform the sum over $J_B$ inside~\eqref{eq:gm}.

We can approximate the sum in $g_m(J_A)$ for large $\V$ in the same way as we have derived~\eqref{eq:pre-final-n2}. We introduce $k=J_B-|J-J_A|$ and find
\begin{align}
    g_m(J_A)=\sum^\infty_{k=0}e^{2\beta'(|j-j_A|/[1-f])k}\lambda_k(J_A,m)+O(\tfrac{1}{\V})\,,
\end{align}
where we used that the Clebsch-Gordon coefficients behave as
\begin{align}
\lambda_k(J_A,\mu\sqrt{\V})&=\lim_{\V\to\infty}\sqrt{\V}\lambda^{(J_A,|J-J_A|+k)}_{\mu \sqrt{V}}\nonumber\\
&=\left|\psi_k\left(\frac{2(|j-j_A|+|j_A|)}{j_A|j-j_A|},\mu\right)\right|^2,
\end{align}
see~\cite[Eq. (3)]{rowe2010shifted}.
Hence, their limits approach the wave functions of the harmonic oscillator
\begin{align}
    \psi_k(\omega,x)&=\frac{1}{\sqrt{2^kk!}}\left(\frac{\omega}{\pi}\right)^{1/4}\e^{-\frac{\omega}{2} x^2}H_k(\sqrt{\omega}x)\,.
\end{align}
The function $H_n(x)$ is the $n$-th Hermite polynomial satisfying the orthogonality condition with respect to the weight $e^{-x^2}$.

We recall the asymptotics of $n^B_{J_B}$ from~\eqref{eq:n2-asymptotics} and $\tilde{n}^B_{J_A}$ from~\eqref{eq:ntilde} to find
\begin{align}
	\frac{n^B_{J_B}}{\tilde{n}^B_{J_A}}=(1-\gamma)\gamma^k+O(\tfrac{1}{\V})\,,
\end{align}
where $\gamma=\gamma(\frac{j_B^{\min}}{1-f})$ and $\sum_{J_{B}\in\mathfrak{J}^{(J_A)}_B} n^B_{J_B}/\tilde{n}^B_{J_A}=1$.

Combining this with the limit of the CG coefficients we arrive at (for $m>0$)\footnote{Let us highlight an important subtlety: The sum over $J_B$ is over the set $\mathfrak{J}^{(J_A)}_{B,m}$ from~\eqref{JBm.def} rather than $\mathfrak{J}^{(J_A)}_{B}$. This has two consequences which both will not affect our calculation: (a) For $|m|>|J-J_A|$, the sum will not start at $|J-J_A|$, but we will see in a moment that the leading contributions comes from where $J-J_A=O(\V)$ and $m=O(\sqrt{\V})$, where the sum will start at $|J-J_A|$. (b) For $m=0$, the sum will skip every second $J_B$, as $J-J_A-J_B$ must be even. Therefore, Eq.~\eqref{eq:sum-prob} will be slightly modified, but as the case $m=0$ is a set of measure zero (in particular suppressed by $1/\sqrt{V}$) when turning the sum in $m$ to an integral over $\mu$, it will not affect $T_{\mathrm{CG}}$ at constant order in $\V$.
}
\begin{align}
	g_m(J_A)&=\sum_{J_B\in\mathfrak{J}^{(J_A)}_{B,m}}\frac{n^B_{J_B}}{\tilde{n}^B_{J_A}}\left|\psi_k\left(\tfrac{2}{j_Af(1-f)},\mu\right)\right|^2+O(\tfrac{1}{\V})\nonumber\\
    &=(1-\gamma)\sum^\infty_{k=0}\gamma^k\left|\psi_k\left(\tfrac{2}{j_Af(1-f)},\mu\right)\right|^2+O(\tfrac{1}{\V}).\label{eq:sum-prob}
\end{align}
\begin{figure*}[t]
  \centering
  \includegraphics[width=0.48\textwidth]{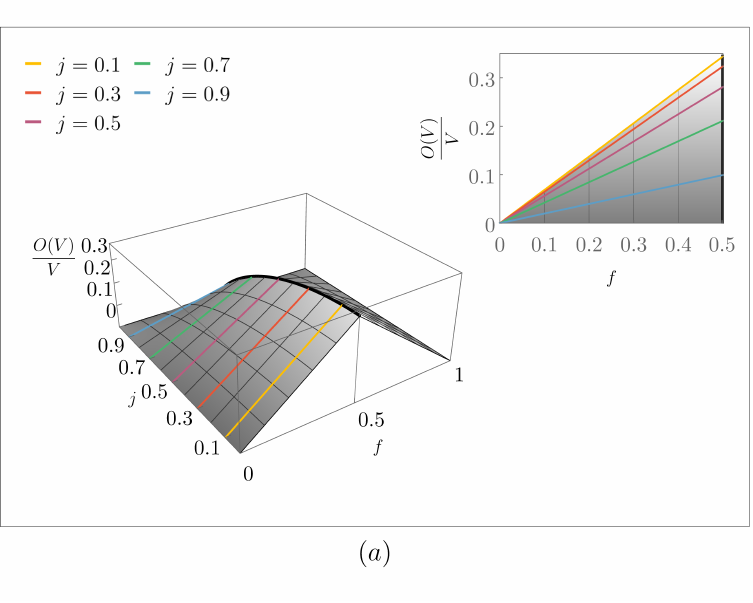}
  \hfill
  \includegraphics[width=0.48\textwidth]{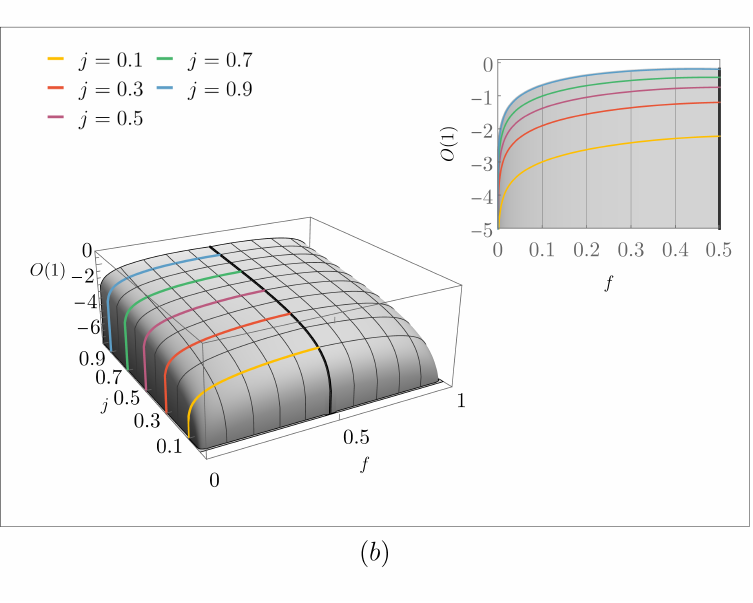}
  \caption{We plot the ($a$) leading and ($b$) subleading (constant) terms in average entanglement entropy as a function of subsystem fraction $f=\V_A/\V$ ($x$ axis) and spin density $j=2J/V$ ($y$ axis) from~\eqref{final result}. In ($a$) the leading order term $\V f\beta(j)$ is scaled by $1/\V$ so that we understand its behavior in the asymptotic limit $\V\to\infty$. One can see that both the plots are symmetric under the interchange $f\to 1-f$. In the respective insets, results for fixed $j$ values are plotted as function of $f$ which can be understood as overlapping sections of the main 3D plot along the fixed $j$ lines.}
  \label{fig:entanglemet entropy}
\end{figure*}

The sum over $k$ can be carried out via Mehler's formula~\cite{mehler1866ueber} for $|\gamma|<1$, \ie
\begin{align}
	\sum^\infty_{k=0}\frac{\gamma^k}{2^kk!}H_k(u)H_k(v)=\frac{1}{\sqrt{1-\gamma^2}}\exp\left(\tfrac{2\gamma uv-\gamma^2(u^2+v^2)}{1-\gamma^2}\right)\,,
\end{align}
which applies here as $\beta'(x)<0$, so $|\gamma(x)|=\e^{2\beta'(x)}<1$. Relevant for~\eqref{eq:sum-prob}, we find
\begin{equation}
	(1-\gamma)\sum^{\infty}_{k=0}\gamma^k|\psi_k(\omega,x)|^2=\left|\psi_0\left(\frac{1-\gamma}{1+\gamma}\omega,x\right)\right|^2\,,
\end{equation}
which means we get back the squared ground state wave function of the harmonic oscillator, only with a different frequency. This wave function is well-known to be a normal distribution so that
\begin{align}
	g(\mu)=g_m(J_A)=\frac{1}{\sqrt{2\pi \sigma^2_\mu}}\e^{-\mu^2/(2\sigma_\mu^2)}+O(\tfrac{1}{\V})\,.
\end{align}
When writing everything in terms of $\mu=m/\sqrt{\V}$, its variance is
\begin{equation}
	\sigma^2_\mu=\tfrac{1}{2}\left(\frac{1-\gamma}{1+\gamma}\omega\right)^{-1}=\frac{(1-f)j_A}{4j}\,.\label{eq:sigma-mu}
\end{equation}

We are now ready to compute $T_{\mathrm{CG}}$, where the sums in $m$ and $J_A$ approach integrals in $\mu$ and $\delta J_A$,
\begin{align}
    T_{\text{CG}}&=-\sum_{J_A}\frac{n^A_{J_A}\tilde{n}^B_{J_A}}{d}\sum_m\,g_m(J_A)\log g_m(J_A)\nonumber\\
	&=-\int d\delta J_A \varrho(\delta J_A)\int d\mu\,  \varrho_{\sigma_\mu}(\mu)\log\varrho_{\sigma_\mu}(\mu)+O(\tfrac{1}{\V})\,.
\end{align}

Here, we used the form of $\varrho(\delta J_A)$ from~\eqref{eq:varrho}. The integral over $\mu$ gives the entropy of a normal distribution that is well-known to be
\begin{align}
	-\int d\mu\,  g(\mu)\log g(\mu)=\frac{1}{2}\log(2\pi \e\sigma^2_\mu)+O(\tfrac{1}{\V})\,.
\end{align}

For the remaining integral over $\delta J_A$, we can use the explicit form from~\eqref{eq:varrho}, which yields
\begin{align}\label{eq:T2-full-result}
	T_{\text{CG}}=\frac{1}{2}\log(\tfrac{\pi \e f (1-f)}{2})+O(\tfrac{1}{\V})\,.
\end{align}
It is remarkable how the dependence on $j$ in $\sigma_{\mu}$ completely drops out for the leading order term.

\section{Final result and its relation to previous work}\label{sec:final result}
We add $T_{\text{S}}$ from~\eqref{eq:T1-full-result} and $T_{\text{CG}}$ from~\eqref{eq:T2-full-result} to find
\begin{align}
	\begin{split}
	\braket{S_A}_j={}&\V f \beta(j)+\tfrac{1}{2}\log(\V)+\tfrac{f+\log(1-f)}{2}\\
	&+\tfrac{(1-f)(1-j)}{2j}\log\left(\tfrac{1-j}{1+j}\right)+\log\left(\tfrac{2j}{1+j}\right)\\
	&+\tfrac{1}{2}\log(\tfrac{\pi \e f(1-f)}{2})+o(1)
	\end{split}\label{final result}
\end{align}
for the average entanglement entropy with $\beta(j)$ given in~\eqref{eq:beta}. We highlight again that this result is valid in the regime $0<f<\frac{1}{2}$ and $0<j<1$. For $f>1/2$, one needs to reflect $f\to1-f$ in the formula due to the symmetry in interchanging system $A$ and $B$ for the entanglement entropy.

The terms contributing to the entanglement entropy can be classified according to their distinct physical and mathematical origins as shown in table~\ref{tab:comparison}, where we also compare our finding~\eqref{final result} with two other results, namely the average entanglement entropy $\braket{S_A}_{\mathrm{SD}_2}$ in the so-called $\mathrm{SD}_2$ approximation from~\cite{PhysRevB.108.245101} and the average $G$-local entanglement entropy $\braket{S_A}_{G}$ studied in~\cite{Bianchi:2024aim}.

\textbf{Density of State (DS).} The leading volume-law contribution $\V f\,\beta(j)$ is related to the logarithm of the density of states for fixed $J=j\V/2$ and $J_z=0$, which itself is proportional to the Hilbert space dimension
\begin{align}
    \dim\mathcal{H}_{J,J_z=0}\sim e^{\beta(j)\V}\,,
\end{align}
which we saw in~\eqref{eq:d-asymptotics}. We have plotted the leading order term in figure~\ref{fig:entanglemet entropy}(left panel). Due to $f<\frac{1}{2}$,  $\V f=V_A$ is the size of the smaller subsystem, so that we can interpret $\beta(j)$ also as the average entanglement entropy density. Let us emphasize that this leading order behavior is well-understood and carefully studied for a wide range of different systems, starting from completely unconstrained Hilbert spaces as studied by Page~\cite{page}, systems with fixed particle number~\cite{PhysRevLett.119.220603,PRXQuantum.3.030201,Yauk:2023wbu}, Gaussian states~\cite{PhysRevB.103.L241118} and general fixed energy sectors~\cite{mishra2025quantum}.

\textbf{Abelian term (A).} The term $\frac{f+\log(1-f)}{2}$ was shown in~\cite{Yauk:2023wbu} to be a universal constant correction arising from fixing an Abelian $\mathrm{U}(1)$ symmetry sector. Specifically, this term arises when we have a global charge density $c$ that arises by coupling charge density $c_A$ in subsystem $A$ with charge density $c_B$ in subsystem $B$, such that $c=c_A+c_B$ and we have the dimensional scaling of the Hilbert space with charge $C$ given by
\begin{align}
    \dim \mathcal{H}^{C=cV}=\sqrt{\frac{|\beta''(c)|}{2\pi \V}}e^{\beta(c)\V}(1+O(\tfrac{1}{\V}))\,.\label{eq:Hc}
\end{align}
\emph{Note that~\cite{Yauk:2023wbu} focused on the most common case of an Abelian $\mathrm{U}(1)$ symmetry due to particle number conservation, but the finding is general.}

\textbf{Non-Abelian term (nA).} The additional non-Abelian term indicated in table~\ref{tab:comparison} arises from the fact that our system does \emph{not} have an Abelian symmetry, so that the dimensional scaling from~\eqref{eq:Hc} does not apply exactly. Indeed, if we look at the scaling of the dimension
\begin{align}
    d=\dim\mathcal{H}^{J=j\V/2,J_z=0}=\sqrt{\frac{2|\beta''(j)|}{\pi\V}}(1-\gamma(j))e^{\beta(j)\V}
\end{align}
from~\eqref{eq:d-asymptotics}, we notice an additional factor of $2(1-\gamma(j))$ (see~\eqref{eq:gamma} for the relation between $\gamma(j)$ and $\beta(j)$) due to the non-Abelian symmetry. This factor reflects that the states do not explore the whole Hilbert space but only subspaces of these tensor products restricted by the rules of the quantum numbers $J$, $J_A$, $J_B$ and $m$. Specifically, this factor arises arises as we do not just couple $j_A$ and $j_B=j-j_A$ to give a contribution towards spin density $j$ in the total system, but we also have larger $j_A+j_B>j$ contributing, based on the representation theory of $\mathrm{SU}(2)$. \emph{Going over to other non-Abelian groups, it is likely that this term still persists. Surely the function $\beta(j)$ may change but the mechanism should be very similar.}

\textbf{Clebsch-Gordon term (CG).} Finally, the Clebsch-Gordan (CG) terms originate from the specific normalization and coupling structure associated with angular-momentum addition for $\mathrm{SU}(2)$. We already saw that the term $T_{\mathrm{CG}}$ in~\eqref{eq:T2-full-result} arises as the entropy of a normal distribution over the magnetic quantum numbers $m$. Moreover, we had previously factored out the $\sqrt{\V}$ scaling when splitting the average entanglement entropy into $T_{\mathrm{S}}$ and $T_{\mathrm{CG}}$. The two (CG) terms in table~\ref{tab:comparison} add up to
\begin{align}
    S_{\mathrm{CG}}=\frac{1}{2}\log\frac{\pi e f(1-f)\V}{2}\,,
\end{align}
which is exactly the entropy of a normal distribution with variance $\frac{f(1-f)\V}{4}$ that scales linearly in the system size $\V$. It is an interesting observation that this term is independent of $j$, even though the original distribution $g(\mu)$ dependence on $j_A$ and $j$ through~\eqref{eq:sigma-mu}, but this dependence drops out when averaging over $j_A$. It remains to be seen if this is a generic feature for non-Abelian symmetries that the contribution from the specific coupling of different quantum numbers only depends on $f$. 

Behavior of all the constant contributions from abelian, non-abelian and CG terms with $f$ and $j$ is shown in figure~\ref{fig:approximation} (right panel). Additionally, we compare our findings with two related terms that were previously studied in the literature. 
\begin{center}
\begin{table*}[t!]
\centering
\begin{tabular}{p{1.7cm} p{1.8cm} p{1.7cm} p{12.2cm}}
\hline\hline
\textbf{Source} &
\textbf{$O(\V)$} &
\textbf{$O(\log \V)$} &
\textbf{$O(1)$} \\
\hline\hline

\symrow
$\braket{S_A}_{J,J_z=0}$ from~\eqref{final result}
&
\boxedwithlabelW{\OVboxW}{\text{DS}}{\V\, f\,\beta(j)}
&
\boxedwithlabelW{\OlogboxW}{\text{CG}}{\dfrac 12\log \V}
&
\OoneSumThree
  {\boxedwithlabelW{\OoneAboxW}{\text{A}}{\dfrac{f+\log(1-f)}{2}}}
  {\boxedwithlabelW{\OoneBboxW}{\text{nA}}{\dfrac{(1-f)(1-j)}{2j}\log\!\dfrac{1-j}{1+j}
  +\log\!\dfrac{2j}{1+j}}}
  {\boxedwithlabelW{\OoneCboxW}{\text{CG}}{\dfrac12 \log\!\dfrac{\pi\e f(1-f)}{2}}}
\\ \hline

\symrow
$\braket{S_A}_{\text{SD}_2}$ from~\cite{PhysRevB.108.245101}
&
\boxedwithlabelW{\OVboxW}{\text{DS}}{\V\, f\,\beta(j)}
&
\boxedwithlabelW{\OlogboxW}{\text{CG}}{\dfrac12 \log \V}
&
\OoneSumThree
  {\boxedwithlabelW{\OoneAboxW}{\text{A}}{\dfrac{f+\log(1-f)}{2}}}
  {\boxedwithlabelW{\OoneBboxW}{\text{\tiny{SD}}_2}{\dfrac{1-2f(1-j)}{2j}\log\!\dfrac{1-j}{1+j}
  +\log\!\dfrac{2j^{3/2}}{\sqrt{1-j^2}}~}}
  {\boxedwithlabelW{\OoneCboxW}{\text{CG}}{\dfrac12 \log\!\dfrac{\pi\e f(1-f)}{2}}}
\\ \hline

\symrow
$\braket{S_A}_G$ from~\cite{Bianchi:2024aim}
&
\boxedwithlabelW{\OVboxW}{\text{DS}}{\V\, f\,\beta(j)}
&
&
\OoneSumTwo
  {\boxedwithlabelW{\OoneAboxW}{\text{A}}{\dfrac{f+\log(1-f)}{2}}}
  {\boxedwithlabelW{\OoneBboxW}{\text{nA}}{\dfrac{(1-f)(1-j)}{2j}\log\!\dfrac{1-j}{1+j}
  +\log\!\dfrac{2j}{1+j}}}
\\ \hline\hline
\end{tabular}
\caption{Comparison of the contributions to the average entanglement entropy in~\eqref{final result} at orders $O(\V)$, $O(\log \V)$, and $O(1)$ (for $0<f<\tfrac12$) with two other quantities studied in the literature. The average entropy $\braket{S_A}_{\mathrm{SD}_2}$ in the so-called $\mathrm{SD}_2$ approximation was computed in~\cite{PhysRevB.108.245101}, where numerical tests indicated that this quantity captures correctly the orders $O(\V)$ and $O(\log{\V})$, but fails at order $O(1)$. The average $G$-local entanglement entropy $\braket{S_A}_G$ was introduced and computed in~\cite{Bianchi:2024aim}, where the entropy is computed exclusively with respect to gauge-invariant observables.}
\label{tab:comparison}
\end{table*}
\end{center}
\textbf{Average entanglement entropy $\braket{S_A}_{\mathrm{SD}_2}$ in the $\mathrm{SD}_2$ approximation from~\cite{PhysRevB.108.245101}.} In order to make analytical progress in computing the average entanglement entropy $\braket{S_A}_{J,J_z=0}$ also considered here,~\cite{PhysRevB.108.245101} introduced two approximations, referred to as `spin decomposition 1' ($\mathrm{SD}_1$) and `spin decomposition 2' ($\mathrm{SD}_2$). For ($\mathrm{SD}_1$), the reduced density matrix $\rho_A$ was approximated by retaining only its block-diagonal part ($J_A = J_A'$), while off-diagonal contributions coupling different representation sectors ($J_A \neq J_A'$) were neglected. This approximation was motivated by the expectation that such inter-sector couplings do not contribute to the leading entanglement entropy in the large-volume limit and it was shown numerically that for various cases with $f<\frac{1}{2}$ the error of this approximation is exponentially small in the volume. 

For the second approximation ($\mathrm{SD}_2$), it was then assumed that that only $J_B=J-J_A$ couples with $J_A$, effectively ignoring the sum over $J_B\in\mathfrak{J}^{J_A}_B$. Already the numerics in~\cite{PhysRevB.108.245101} indicated that this approximation will deviate at constant order, but the leading order volume term and also the logarithmic correction appeared to be correct. As can be seen in table~\ref{tab:comparison}, our random matrix model confirms this:
\begin{quote}
The ($\mathrm{SD}_2$) approximation from~\cite{PhysRevB.108.245101} for $f<\frac{1}{2}$ predicts the correct behavior at order $O(\V)$ and $O(\log{\V})$, but starts to deviate at constant order $O(1)$.
\end{quote}
The systematic difference at order $O(1)$ can be seen in table~\ref{tab:comparison}, when comparing the term (nA) with ($\mathrm{SD}_2$). This difference arises from the $\mathrm{SD}_2$ approximation where one imposes $J = J_A + J_B$ and effectively projects onto the dominant $J_B = J - J_A$ Hilbert space sector, thereby discarding other multiplicity contributions. While this is asymptotically justified for $J \simeq V/2$, it reshuffles the CG weight distribution and leads to a shifted $O(1)$ constant. Our approach keeps the full ${\rm SU}(2)$ decomposition without enforcing SD$_2$, and thus yields a constant term consistent with exact sector multiplicities.

In contrast to ($\mathrm{SD}_1$) and ($\mathrm{SD}_2$), the present work does not assume any block structure. Instead, starting from the full reduced density matrix, the asymptotic analysis of the moments shows that the contributions of non-planar diagrams is exponentially suppressed. Although these planar contributions still involve the full coupled structure of $\rho_A$, their net contribution to the entanglement entropy
in the thermodynamic limit coincides with that obtained within the $\mathrm{SD}_1$ approximation. Our analysis uncovers the mechanism for this. It is the very extreme rectangularities of the off-diagonal blocks that suppresses their impact. We called it dimensional selection which has been outlined in subsections~\ref{sec:dim.select} and~\ref{sec:selection}. In this sense, the $\mathrm{SD}_1$ block-diagonal treatment correctly captures the leading asymptotic entropic behavior, even though the underlying reduced density matrix is not block-diagonal, while the planar diagram framework provides a first principles justification of this picture and a systematic route to subleading corrections.

\textbf{Average $G$-local entanglement entropy $\braket{S_A}_G$ from~\cite{Bianchi:2024aim}.} This work introduces the framework to describe symmetry-resolved states with respect to a group action and illustrate their formalism for the group of $\mathrm{SU}(2)$. Therein, they derive a formula for the average $G$-local entanglement entropy, which quantifies the amount of entanglement in a symmetry-resolved state with respect to observables invariant under this group action. This quantity is \emph{distinct} from the average entanglement entropy $\braket{S_A}_{J,J_z=0}$ studied in the present manuscript, which the authors of~\cite{Bianchi:2024aim} refer to as $K$-local entanglement entropy. Most importantly, the average $\braket{S_A}_G$ is tractable for $\mathrm{SU}(2)$ and is computed up to constant order. At the time, the best approximation for $\braket{S_A}_{J,J_z=0}$ was $\braket{S_A}_{\mathrm{SD}_2}$ from~\cite{PhysRevB.108.245101}, which agrees at leading order $O(\V)$, but differs at $O(\log{\V})$ and $O(1)$.
\begin{quote}
    For $0<f<\frac{1}{2}$ and $0<j<1$, we find
    \begin{align}\label{eq:relationship}
    \braket{S_A}_{J,J_z=0}=\braket{S_A}_G+S_{\mathrm{CG}}
    \end{align}
    where we introduced $S_{\mathrm{CG}}=\frac{1}{2}\log\frac{\pi e f(1-f)\V}{2}$.
\end{quote}
This relationship is shown in table~\ref{tab:comparison} and consistent with the intuition that the $G$-local entanglement entropy is smaller, as it ignores the entropy associated to observables that are not gauge invariant. This finding is plausible, as this part is exactly the entropy of a Gaussian distribution over the magnetic quantum numbers $m$ with variance $\sigma^2=\frac{f(1-f)\V}{4}$, encoded in the Glebsch-Gordon coefficients. As can be seen from our derivation in section~\ref{sec:T2}, this distribution depends on $J$ and $J_A$, but this dependence drops out when averaging over $J_A$.

The relationship~\eqref{eq:relationship} also applies to the case $j=1$, as already reported in~\cite{Bianchi:2024aim}, but appears to fail at $j=0$. Specifically, when comparing the result~\eqref{case J=0} from~\cite{PhysRevB.108.245101} with the equivalent result\footnote{Note that the published version~\cite{Bianchi:2024aim} contains a typo for this equation. The correct asymptotics for $j=0$ is
\begin{align}
    \braket{S_A}^{j=0}_G={}&f\V \log{2}-\frac{1}{2}\log{\V}\nonumber\\
    &{}\frac{3(f-\log(1-f))}{2}-\frac{1}{2}\log(\frac{e^{2-\gamma_E}f(1-f)}{2})\,.
\end{align}
} from~\cite{Bianchi:2024aim}, one finds
\begin{align}
    \braket{S_A}_{J=0,J_z=0}=\braket{S_A}_G-S^{j=0}_{\mathrm{CG}}\,,
\end{align}
where we defined $S^{j=0}_{\mathrm{CG}}=\tfrac{1}{2}\log(\tfrac{e^{2-\gamma_E}f(1-f)\V}{2})$. This is also plausible, as there is a discontinuous behavior in the statistics when moving from $j=0$, where the peak scales as $J_A=O(\sqrt{\V})$, compared to $j>0$, where the peak scales as $J_A=f J=O(\V)$. For the remaining open case\footnote{Reference~\cite{Bianchi:2024aim} covers the full ranges $0\leq j\leq 1$ and $0\leq f \leq 1$, where we note that $\braket{S_A}_G$ is \emph{not} symmetric under $f\leftrightarrow (1-f)$, which is in contrast to $\braket{S_A}_{J,J_z=0}$.} of $f=\frac{1}{2}$, we conjecture that~\eqref{eq:relationship} with still applies, but we leave this analysis for future work.

\section{Discussion and Outlook}\label{sec:conclusion}
		
The present work develops a full analytical pathway to average entanglement in non-Abelian symmetry sectors by combining representation theory with random matrix theory methods.
We have analyzed the average entanglement entropy of Haar-random pure states constrained to irreducible ${\rm SU}(2)$ symmetry sectors of a spin-$1/2$ chain, with a fixed total spin $J$ and fixed magnetization $J_z=0$. A random pure state is characterized by random matrices chosen from a fixed trace ensemble so that it satisfies normalization condition though in the current setting the random matrix has much or structure than in the simple Page setting.

We began by decomposing the global Hilbert space into total-spin sectors together with their associated multiplicities, and then introduce a bipartition $A|B$ with subsystem fraction $f=\V_A/\V$.  
Under this bipartition, the reduced density matrix $\rho_A$ of subsystem $A$ acquires a highly nontrivial block matrix structure where each block corresponds to a pair of subsystem angular momentum quantum numbers $(J_A,J_B)$ and carries both Clebsch--Gordan weights and Hilbert space multiplicities that scale exponentially with the system size. In the next step, we lifted the fixed-trace constraint and mapped $\rho_A$ to a coupled and correlated Wishart random matrix $\ro$ reflecting the block structure. In this way the problem became more analytically tractable. The eigenvalue density of $\rho_A$ could be reconstructed from the asymptotic behavior of the moments $\braket{\Tr \ro^L}$ for integer $L$, in the spirit of Wigner's moment method~\cite{Wigner}.

Technically, we first used the Isserlis--Wick theorem to rewrite $\langle \Tr \rho^L \rangle$ as a sum over Gaussian pairings between rectangular blocks $W^{(J_A,J_B)}$ and their adjoints. This led to a diagrammatic expansion in terms of loops built from these blocks. By analogy with large-$N$ random matrix theory, we retained only \emph{planar contractions}, \ie pairings that can be drawn without crossings. Non-planar pairings necessarily introduce additional ``glueing'' operations between loops and reduce the overall contribution by factors of subsystem dimensions $n_{J_A}^A$, $n_{J_B}^B$, so they are exponentially suppressed and drop out in a planar diagram approximation. Within this planar sector, we obtained recursive relations for block-wise moments $\mathcal{E}_{A,k}^{(J_A)}$ and $\mathcal{E}_{B,k}^{(J_B)}$, which admitted a clean \emph{rooted tree} representation: vertices correspond to spin sectors $J_A$ or $J_B$, edges encode conditioned sums over admzissible partners determined by the sets $\mathfrak{J}_A^{(J_B)}$ and $\mathfrak{J}_B^{(J_A)}$, and each vertex is weighted by the corresponding multiplicity $n_{J}^A$ or $n_{J}^B$.

The key structural step is a \emph{dimensional selection principle} on this family of rooted trees. Using combinatorial \textit{swap} and \textit{cut} rules on the nested sums, every tree contribution can be bounded above by either of two extremal trees, $\mathfrak{R}_A^{(L)}$ and $\mathfrak{R}_B^{(L)}$, in which a single $J_A$  vertex is connected to $L$ vertices of $J_B$ and vice versa. Since the multiplicities scale exponentially in the system size $\V$, while the Clebsch--Gordan weights $\Lambda_{{\bf J}_A,{\bf J}_B}$ grow only algebraically, all non-extremal trees are exponentially suppressed compared to $\mathfrak{R}_A^{(L)}$ and $\mathfrak{R}_B^{(L)}$. For any fixed integer $L$, this yielded the asymptotic form $\langle \Tr \rho^L \rangle \approx \mathfrak{R}_A^{(L)}$ for subsystem fraction $f<\tfrac12$ (and $\approx \mathfrak{R}_B^{(L)}$ for $f>\tfrac12$), while at $f=\tfrac12$ both extremal trees contributed on equal footing and the moment method became non-analytic at $L=1$. From the resulting effective discrete spectrum, encoded in the extremal trees, we finally reconstructed the level density and entanglement entropy of $\rho_A$ in the thermodynamic limit.

In the final result, the entanglement entropy splits naturally as
$\langle S_A\rangle=T_{\mathrm S}+T_{\mathrm{CG}}+o(1)$, and this
separation has a clear physical meaning. The term $T_{\mathrm S}$
collects all contributions that arise from the \emph{overall scale} of
the dominant weights in the reduced density matrix. It contains the
leading volume law, which is fixed by the exponential growth of the
representation multiplicities $n^A_{J_A}$, $n^B_{J_B}$, and the total
dimension $d$, and which selects a single dominant sector
$J_A=fJ$, $J_B=(1-f)J$. Crucially, $T_{\mathrm S}$ also includes the
$\tfrac12\log V$ correction: this logarithmic term originates from the
universal $V^{-1/2}$ scaling of the Clebsch--Gordon coefficients near
the dominant sector,
$\lambda_m^{(fJ,(1-f)J)}\sim V^{-1/2}$. When this scale is factored out, it produces an explicit $\log\sqrt V$
contribution, which survives after summation and appears as $\tfrac12\log V$ in $T_{\mathrm S}$, which is why we annotate it as CG in table~\ref{tab:comparison}. Thus, although $T_{\mathrm S}$ is insensitive to the detailed \emph{shape} of the CG coefficients, it does retain their universal normalization scale and combines it with representation counting to generate all volume dependent terms and the remaining $j$--dependent constants. By contrast, $T_{\mathrm{CG}}$ captures only
the genuinely subleading effect of the CG structure. After the overall $V^{-1/2}$ scale has been removed, the rescaled coefficients define a normalized distribution over magnetic quantum numbers that converges to a Gaussian in the large volume limit. The contribution $T_{\mathrm{CG}}$ is precisely the entropy of this emergent Gaussian profile and therefore remains finite as $V\to\infty$. It depends only on the subsystem fraction $f$ and is independent of the global spin
density $j$. Together, this shows that the macroscopic scaling of entanglement is governed by symmetry and representation theory, with the CG coefficients entering only through their universal $V^{-1/2}$ normalization (captured in $T_{\mathrm S}$). The detailed structure of the CG coefficients affects the entropy only through a finite, size-independent universal correction $T_{\mathrm{CG}}$.
		
In summary, we established a general and tractable pathway to compute average entanglement entropy in the non-Abelian symmetry sector with arbitrary total spin.  The resulting closed form unifies the leading order and constant terms, clarifying how representation theory controls entanglement structure in spin systems. The presented computations also offer a broadly portable techniques for computing entropy in context of other entropic functionals or different dynamical setting. Several natural extensions follow from here. First, one can treat non-zero magnetization sectors $m\ne 0$ and re-derive the volume law within the same framework. Especially the case $m=O(V)$ would allow the study of entanglement entropy in the ferromagnetic setting. Secondly, the derivation can be generalized from global $\rm{SU}(2)$ to broader symmetry settings including $\rm{SU}(N)$, direct product groups and scenarios with multiple conserved charges to assess how the leading and sub-leading contributions to the volume law are modified under richer representation structures. 

\acknowledgements
LH thanks Marcos Rigol and Rohit Patil for helpful discussions. The research of AC and LH is supported by ‘The Quantum Information Structure of Spacetime’ Project (QISS), by grant $\#$62312 from the John Templeton Foundation. MK is funded by the Australian Research Council through the Discovery Project grant DP250102552.

\bibliography{references}

@ARTICLE{Segner,
       author = {Von Segner, Johann A.},
        title = "{Enumeratio modorum, quibus figurae planae rectilineae per diagonales dividuntur in triangula}",
      journal = {Novi commentarii academiae scientiarum Petropolitanae},
         year = 1758,
       volume = {7},
        pages = {203-209}
}

@ARTICLE{Wigner,
       author = {Wigner, Eugene P.},
        title = "{Characteristic Vectors of Bordered Matrices With Infinite Dimensions}",
      journal = {Annals of Mathematics},
         year = 1955,
       volume = {62},
        pages = {548-564},
          doi = {10.2307/1970079}
}

@ARTICLE{1999JMP....40.4782R,
       author = {{Reinsch}, Matthias W. and {Morehead}, James J.},
        title = "{Asymptotics of Clebsch-Gordan coefficients}",
      journal = {Journal of Mathematical Physics},
         year = 1999,
        month = oct,
       volume = {40},
       number = {10},
        pages = {4782-4806},
          doi = {10.1063/1.533000},
archivePrefix = {arXiv},
       eprint = {math-ph/9906007},
 primaryClass = {math-ph},
       adsurl = {https://ui.adsabs.harvard.edu/abs/1999JMP....40.4782R}
}

@article{page,
  title = {Average entropy of a subsystem},
  author = {Page, Don N.},
  journal = {Phys. Rev. Lett.},
  volume = {71},
  issue = {9},
  pages = {1291--1294},
  numpages = {0},
  year = {1993},
  month = {Aug},
  publisher = {American Physical Society},
  doi = {10.1103/PhysRevLett.71.1291},
  url = {https://link.aps.org/doi/10.1103/PhysRevLett.71.1291}
}

@article{PRXQuantum.3.030201,
  title = {Volume-Law Entanglement Entropy of Typical Pure Quantum States},
  author = {Bianchi, Eugenio and Hackl, Lucas and Kieburg, Mario and Rigol, Marcos and Vidmar, Lev},
  journal = {PRX Quantum},
  volume = {3},
  issue = {3},
  pages = {030201},
  numpages = {77},
  year = {2022},
  month = {Jul},
  publisher = {American Physical Society},
  doi = {10.1103/PRXQuantum.3.030201},
  url = {https://link.aps.org/doi/10.1103/PRXQuantum.3.030201}
}

@article{PhysRevB.108.245101,
  title = {Average pure-state entanglement entropy in spin systems with SU(2) symmetry},
  author = {Patil, Rohit and Hackl, Lucas and Fagan, George R. and Rigol, Marcos},
  journal = {Phys. Rev. B},
  volume = {108},
  issue = {24},
  pages = {245101},
  numpages = {15},
  year = {2023},
  month = {Dec},
  publisher = {American Physical Society},
  doi = {10.1103/PhysRevB.108.245101},
  url = {https://link.aps.org/doi/10.1103/PhysRevB.108.245101}
}

@article{PhysRevLett.44.301,
  title = {Is Black-Hole Evaporation Predictable?},
  author = {Page, Don N.},
  journal = {Phys. Rev. Lett.},
  volume = {44},
  issue = {5},
  pages = {301--304},
  numpages = {0},
  year = {1980},
  month = {Feb},
  publisher = {American Physical Society},
  doi = {10.1103/PhysRevLett.44.301},
  url = {https://link.aps.org/doi/10.1103/PhysRevLett.44.301}
}

@article{RevModPhys.82.277,
	title = {Colloquium: Area laws for the entanglement entropy},
	author = {Eisert, J. and Cramer, M. and Plenio, M. B.},
	journal = {Rev. Mod. Phys.},
	volume = {82},
	issue = {1},
	pages = {277--306},
	numpages = {0},
	year = {2010},
	month = {Feb},
	publisher = {American Physical Society},
	doi = {10.1103/RevModPhys.82.277},
	url = {https://link.aps.org/doi/10.1103/RevModPhys.82.277}
}

@article{PhysRevLett.119.220603,
	title = {Entanglement Entropy of Eigenstates of Quantum Chaotic Hamiltonians},
	author = {Vidmar, Lev and Rigol, Marcos},
	journal = {Phys. Rev. Lett.},
	volume = {119},
	issue = {22},
	pages = {220603},
	numpages = {6},
	year = {2017},
	month = {Nov},
	publisher = {American Physical Society},
	doi = {10.1103/PhysRevLett.119.220603},
	url = {https://link.aps.org/doi/10.1103/PhysRevLett.119.220603}
}

@article{PhysRevD.100.105010,
	title = {Typical entanglement entropy in the presence of a center: Page curve and its variance},
	author = {Bianchi, Eugenio and Don\`a, Pietro},
	journal = {Phys. Rev. D},
	volume = {100},
	issue = {10},
	pages = {105010},
	numpages = {13},
	year = {2019},
	month = {Nov},
	publisher = {American Physical Society},
	doi = {10.1103/PhysRevD.100.105010},
	url = {https://link.aps.org/doi/10.1103/PhysRevD.100.105010}
}

@article{PhysRevLett.130.140402,
	title = {Non-Abelian Eigenstate Thermalization Hypothesis},
	author = {Murthy, Chaitanya and Babakhani, Arman and Iniguez, Fernando and Srednicki, Mark and Yunger Halpern, Nicole},
	journal = {Phys. Rev. Lett.},
	volume = {130},
	issue = {14},
	pages = {140402},
	numpages = {8},
	year = {2023},
	month = {Apr},
	publisher = {American Physical Society},
	doi = {10.1103/PhysRevLett.130.140402},
	url = {https://link.aps.org/doi/10.1103/PhysRevLett.130.140402}
}

@article{PhysRevE.107.014130,
	title = {Eigenstate thermalization hypothesis in two-dimensional $XXZ$ model with or without SU(2) symmetry},
	author = {Noh, Jae Dong},
	journal = {Phys. Rev. E},
	volume = {107},
	issue = {1},
	pages = {014130},
	numpages = {9},
	year = {2023},
	month = {Jan},
	publisher = {American Physical Society},
	doi = {10.1103/PhysRevE.107.014130},
	url = {https://link.aps.org/doi/10.1103/PhysRevE.107.014130}
}

@article{Guryanova_2016,
	title={Thermodynamics of quantum systems with multiple conserved quantities},
	volume={7},
	ISSN={2041-1723},
	url={http://dx.doi.org/10.1038/ncomms12049},
	DOI={10.1038/ncomms12049},
	number={1},
	journal={Nature Communications},
	publisher={Springer Science and Business Media LLC},
	author={Guryanova, Yelena and Popescu, Sandu and Short, Anthony J. and Silva, Ralph and Skrzypczyk, Paul},
	year={2016},
	month=jul }

@article{Yunger_Halpern_2016,
	title={Microcanonical and resource-theoretic derivations of the thermal state of a quantum system with noncommuting charges},
	volume={7},
	ISSN={2041-1723},
	url={http://dx.doi.org/10.1038/ncomms12051},
	DOI={10.1038/ncomms12051},
	number={1},
	journal={Nature Communications},
	publisher={Springer Science and Business Media LLC},
	author={Yunger Halpern, Nicole and Faist, Philippe and Oppenheim, Jonathan and Winter, Andreas},
	year={2016},
	month=jul }

@article{PRXQuantum.3.010304,
	title = {Non-Abelian Quantum Transport and Thermosqueezing Effects},
	author = {Manzano, Gonzalo and Parrondo, Juan M.R. and Landi, Gabriel T.},
	journal = {PRX Quantum},
	volume = {3},
	issue = {1},
	pages = {010304},
	numpages = {14},
	year = {2022},
	month = {Jan},
	publisher = {American Physical Society},
	doi = {10.1103/PRXQuantum.3.010304},
	url = {https://link.aps.org/doi/10.1103/PRXQuantum.3.010304}
}

@article{PRXQuantum.4.020318,
	title = {Experimental Observation of Thermalization with Noncommuting Charges},
	author = {Kranzl, Florian and Lasek, Aleksander and Joshi, Manoj K. and Kalev, Amir and Blatt, Rainer and Roos, Christian F. and Yunger Halpern, Nicole},
	journal = {PRX Quantum},
	volume = {4},
	issue = {2},
	pages = {020318},
	numpages = {19},
	year = {2023},
	month = {Apr},
	publisher = {American Physical Society},
	doi = {10.1103/PRXQuantum.4.020318},
	url = {https://link.aps.org/doi/10.1103/PRXQuantum.4.020318}
}

@article{Ryu_2006,
	title={Aspects of holographic entanglement entropy},
	volume={2006},
	ISSN={1029-8479},
	url={http://dx.doi.org/10.1088/1126-6708/2006/08/045},
	DOI={10.1088/1126-6708/2006/08/045},
	number={08},
	journal={Journal of High Energy Physics},
	publisher={Springer Science and Business Media LLC},
	author={Ryu, Shinsei and Takayanagi, Tadashi},
	year={2006},
	month=aug, pages={045–045} }

@article{PhysRevE.100.062134,
	title = {Entanglement and matrix elements of observables in interacting integrable systems},
	author = {LeBlond, Tyler and Mallayya, Krishnanand and Vidmar, Lev and Rigol, Marcos},
	journal = {Phys. Rev. E},
	volume = {100},
	issue = {6},
	pages = {062134},
	numpages = {11},
	year = {2019},
	month = {Dec},
	publisher = {American Physical Society},
	doi = {10.1103/PhysRevE.100.062134},
	url = {https://link.aps.org/doi/10.1103/PhysRevE.100.062134}
}

@article{PhysRevE.105.014109,
	title = {Entanglement of midspectrum eigenstates of chaotic many-body systems: Reasons for deviation from random ensembles},
	author = {Haque, Masudul and McClarty, Paul A. and Khaymovich, Ivan M.},
	journal = {Phys. Rev. E},
	volume = {105},
	issue = {1},
	pages = {014109},
	numpages = {8},
	year = {2022},
	month = {Jan},
	publisher = {American Physical Society},
	doi = {10.1103/PhysRevE.105.014109},
	url = {https://link.aps.org/doi/10.1103/PhysRevE.105.014109}
}

@article{PhysRevE.107.064119,
	title = {Average entanglement entropy of midspectrum eigenstates of quantum-chaotic interacting Hamiltonians},
	author = {Kliczkowski, M. and Swietek, R. and Vidmar, L. and Rigol, M.},
	journal = {Phys. Rev. E},
	volume = {107},
	issue = {6},
	pages = {064119},
	numpages = {17},
	year = {2023},
	month = {Jun},
	publisher = {American Physical Society},
	doi = {10.1103/PhysRevE.107.064119},
	url = {https://link.aps.org/doi/10.1103/PhysRevE.107.064119}
}

@article{Yauk:2023wbu,
	author = {Yauk, Yale and Patil, Rohit and Zhang, Yicheng and Rigol, Marcos and Hackl, Lucas},
	title = {Typical entanglement entropy in systems with particle-number conservation},
	doi = {10.1103/PhysRevB.110.235154},
	journal = {Phys. Rev. B},
	volume = {110},
	number = {23},
	pages = {235154},
	year = {2024}
}

@article{mehler1866ueber,
  author       = {Mehler, F. Gustav},
  title        = {Ueber die Entwicklung einer Function von beliebig vielen Variablen nach Laplaceschen Functionen höherer Ordnung},
doi="https://eudml.org/doc/147987",
  journal = {Journal für die reine und angewandte Mathematik},
  year         = {1866},
  volume       = {66},
  pages        = {161--176},
}

@article{PhysRevB.107.045102,
	title = {Non-Abelian symmetry can increase entanglement entropy},
	author = {Majidy, Shayan and Lasek, Aleksander and Huse, David A. and Yunger Halpern, Nicole},
	journal = {Phys. Rev. B},
	volume = {107},
	issue = {4},
	pages = {045102},
	numpages = {13},
	year = {2023},
	month = {Jan},
	publisher = {American Physical Society},
	doi = {10.1103/PhysRevB.107.045102},
	url = {https://link.aps.org/doi/10.1103/PhysRevB.107.045102}
}

@article{Bianchi:2024aim,
	author = "Bianchi, Eugenio and Dona, Pietro and Kumar, Rishabh",
	title = "{Non-Abelian symmetry-resolved entanglement entropy}",
	eprint = "2405.00597",
	archivePrefix = "arXiv",
	primaryClass = "quant-ph",
	doi = "10.21468/SciPostPhys.17.5.127",
	journal = "SciPost Phys.",
	volume = "17",
	number = "5",
	pages = "127",
	year = "2024"
}

@article{Patil:2025ump,
	author = "Patil, Rohit and Rigol, Marcos",
	title = "{Eigenstate thermalization in spin-12 systems with SU(2) symmetry}",
	eprint = "2503.01846",
	archivePrefix = "arXiv",
	primaryClass = "quant-ph",
	doi = "10.1103/PhysRevB.111.205126",
	journal = "Phys. Rev. B",
	volume = "111",
	number = "20",
	pages = "205126",
	year = "2025"
}

@article{Swietek:2023fka,
	author = "Swiechtek, Rafal and Kliczkowski, Maksymilian and Vidmar, Lev and Rigol, Marcos",
	title = "{Eigenstate entanglement entropy in the integrable spin-12~XYZ model}",
	eprint = "2311.10819",
	archivePrefix = "arXiv",
	primaryClass = "cond-mat.stat-mech",
	doi = "10.1103/PhysRevE.109.024117",
	journal = "Phys. Rev. E",
	volume = "109",
	number = "2",
	pages = "024117",
	year = "2024"
}

@article{PhysRevE.82.031130,
	title = {Localization and the effects of symmetries in the thermalization properties of one-dimensional quantum systems},
	author = {Santos, Lea F. and Rigol, Marcos},
	journal = {Phys. Rev. E},
	volume = {82},
	issue = {3},
	pages = {031130},
	numpages = {12},
	year = {2010},
	month = {Sep},
	publisher = {American Physical Society},
	doi = {10.1103/PhysRevE.82.031130},
	url = {https://link.aps.org/doi/10.1103/PhysRevE.82.031130}
}

@article{PhysRevLett.124.050602,
  title = {Universal Entanglement of Typical States in Constrained Systems},
  author = {Morampudi, S. C. and Chandran, A. and Laumann, C. R.},
  journal = {Phys. Rev. Lett.},
  volume = {124},
  issue = {5},
  pages = {050602},
  numpages = {6},
  year = {2020},
  month = {Feb},
  publisher = {American Physical Society},
  doi = {10.1103/PhysRevLett.124.050602},
  url = {https://link.aps.org/doi/10.1103/PhysRevLett.124.050602}
}

@article{PhysRevX.8.041019,
	title = {Solution of a Minimal Model for Many-Body Quantum Chaos},
	author = {Chan, Amos and De Luca, Andrea and Chalker, J. T.},
	journal = {Phys. Rev. X},
	volume = {8},
	issue = {4},
	pages = {041019},
	numpages = {17},
	year = {2018},
	month = {Nov},
	publisher = {American Physical Society},
	doi = {10.1103/PhysRevX.8.041019},
	url = {https://link.aps.org/doi/10.1103/PhysRevX.8.041019}
}

@article{PhysRevLett.123.210603,
	title = {Spectral Statistics and Many-Body Quantum Chaos with Conserved Charge},
	author = {Friedman, Aaron J. and Chan, Amos and De Luca, Andrea and Chalker, J. T.},
	journal = {Phys. Rev. Lett.},
	volume = {123},
	issue = {21},
	pages = {210603},
	numpages = {6},
	year = {2019},
	month = {Nov},
	publisher = {American Physical Society},
	doi = {10.1103/PhysRevLett.123.210603},
	url = {https://link.aps.org/doi/10.1103/PhysRevLett.123.210603}
}

@article{PhysRevX.8.021062,
	title = {Many-Body Quantum Chaos: Analytic Connection to Random Matrix Theory},
	author = {Kos, Pavel and Ljubotina, Marko and Prosen, Toma\ifmmode \check{z}\else \v{z}\fi{}},
	journal = {Phys. Rev. X},
	volume = {8},
	issue = {2},
	pages = {021062},
	numpages = {11},
	year = {2018},
	month = {Jun},
	publisher = {American Physical Society},
	doi = {10.1103/PhysRevX.8.021062},
	url = {https://link.aps.org/doi/10.1103/PhysRevX.8.021062}
}

@article{PhysRevE.100.022131,
	title = {Structure of chaotic eigenstates and their entanglement entropy},
	author = {Murthy, Chaitanya and Srednicki, Mark},
	journal = {Phys. Rev. E},
	volume = {100},
	issue = {2},
	pages = {022131},
	numpages = {8},
	year = {2019},
	month = {Aug},
	publisher = {American Physical Society},
	doi = {10.1103/PhysRevE.100.022131},
	url = {https://link.aps.org/doi/10.1103/PhysRevE.100.022131}
}

@article{Huang:2017std,
	author = "Huang, Yichen",
	title = "{Universal eigenstate entanglement of chaotic local Hamiltonians}",
	eprint = "1708.08607",
	archivePrefix = "arXiv",
	primaryClass = "quant-ph",
	doi = "10.1016/j.nuclphysb.2018.09.013",
	journal = "Nucl. Phys. B",
	volume = "938",
	pages = "594--604",
	year = "2019"
}

@article{PhysRevD.110.L061901,
  title = {Entanglement asymmetry study of black hole radiation},
  author = {Ares, Filiberto and Murciano, Sara and Piroli, Lorenzo and Calabrese, Pasquale},
  journal = {Phys. Rev. D},
  volume = {110},
  issue = {6},
  pages = {L061901},
  numpages = {6},
  year = {2024},
  month = {Sep},
  publisher = {American Physical Society},
  doi = {10.1103/PhysRevD.110.L061901},
  url = {https://link.aps.org/doi/10.1103/PhysRevD.110.L061901}
}

@article{Russotto:2025cpn,
    author = "Russotto, Angelo and Ares, Filiberto and Calabrese, Pasquale",
    title = "{Symmetry breaking in chaotic many-body quantum systems at finite temperature}",
    eprint = "2504.06146",
    archivePrefix = "arXiv",
    primaryClass = "quant-ph",
    doi = "10.1103/kppn-3272",
    journal = "Phys. Rev. E",
    volume = "112",
    number = "3",
    pages = "L032101",
    year = "2025"
}

@article{PhysRevE.102.062113,
  title = {Eigenstate thermalization for observables that break Hamiltonian symmetries and its counterpart in interacting integrable systems},
  author = {LeBlond, Tyler and Rigol, Marcos},
  journal = {Phys. Rev. E},
  volume = {102},
  issue = {6},
  pages = {062113},
  numpages = {15},
  year = {2020},
  month = {Dec},
  publisher = {American Physical Society},
  doi = {10.1103/PhysRevE.102.062113},
  url = {https://link.aps.org/doi/10.1103/PhysRevE.102.062113}
}

@article{PhysRevX.14.031014,
  title = {Quantifying Quantum Chaos through Microcanonical Distributions of Entanglement},
  author = {Rodriguez-Nieva, Joaquin F. and Jonay, Cheryne and Khemani, Vedika},
  journal = {Phys. Rev. X},
  volume = {14},
  issue = {3},
  pages = {031014},
  numpages = {23},
  year = {2024},
  month = {Jul},
  publisher = {American Physical Society},
  doi = {10.1103/PhysRevX.14.031014},
  url = {https://link.aps.org/doi/10.1103/PhysRevX.14.031014}
}

@article{Buijsman:2025gle,
    author = "Buijsman, Wouter and Haque, Masudul and Khaymovich, Ivan M.",
    title = "{Power-law banded random matrix ensemble as a model for quantum many-body Hamiltonians}",
    journal = "arXiv: 2503.08825",
    archivePrefix = "arXiv",
    primaryClass = "cond-mat.dis-nn",
doi="https://doi.org/10.48550/arXiv.2503.08825",
    month = "3",
    year = "2025"
}

@article{Langlett:2025fam,
    author = "Langlett, Christopher M. and Jonay, Cheryne and Khemani, Vedika and Rodriguez-Nieva, Joaquin F.",
    title = "{Quantum chaos at finite temperature in local spin Hamiltonians}",
   journal = "arXiv:2501.13164",
    archivePrefix = "arXiv",
    primaryClass = "cond-mat.stat-mech",
doi="10.48550/arXiv.2501.13164",
    month = "1",
    year = "2025"
}

@article{Ruiz:2024mei,
    author = "Ruiz, Roberto and Sopena, Alejandro and Pozsgay, Bal{\'a}zs and L{\'o}pez, Esperanza",
    title = "{Efficient Eigenstate Preparation in an Integrable Model with Hilbert Space Fragmentation}",
    doi = "10.1103/g9f9-p8ks",
    journal = "PRX Quantum",
    volume = "6",
    number = "3",
    pages = "030316",
    year = "2025"
}

@article{Russotto:2024pqg,
    author = "Russotto, Angelo and Ares, Filiberto and Calabrese, Pasquale",
    title = "{Non-Abelian entanglement asymmetry in random states}",
    doi = "10.1007/JHEP06(2025)149",
    journal = "JHEP",
    volume = "06",
    pages = "149",
    year = "2025"
}

@article{Fitter:2022vpr,
    author = "Fitter, Khurshed P. and Lancien, Cecilia and Nechita, Ion",
    title = "{Estimating the entanglement of random multipartite quantum states}",
    eprint = "2209.11754",
    archivePrefix = "arXiv",
    primaryClass = "quant-ph",
    doi = "10.22331/q-2025-10-01-1870",
    journal = "Quantum",
    volume = "9",
    pages = "1870",
    year = "2025"
}

@article{Ma:2022vnd,
    author = "Ma, Ken K. W. and Volya, A. and Yang, Kun",
    title = "{Eigenstate thermalization and disappearance of quantum many-body scar states in weakly interacting fermion systems}",
    eprint = "2207.13688",
    archivePrefix = "arXiv",
    primaryClass = "cond-mat.stat-mech",
    doi = "10.1103/PhysRevB.106.214313",
    journal = "Phys. Rev. B",
    volume = "106",
    number = "21",
    pages = "214313",
    year = "2022"
}

@article{Kim:2023ykr,
    author = "Kim, MuSeong and Hwang, Mi-Ra and Jung, Eylee and Park, DaeKil",
    title = "{Average R{\'e}nyi entropy of a subsystem in random pure state}",
    eprint = "2301.09074",
    archivePrefix = "arXiv",
    primaryClass = "quant-ph",
    doi = "10.1007/s11128-023-04249-x",
    journal = "Quant. Inf. Proc.",
    volume = "23",
    number = "2",
    pages = "37",
    year = "2024"
}

@inbook{Bianchi:2023avf,
    author = "Bianchi, Eugenio and Livine, Etera R.",
    title = "{Loop Quantum Gravity and Quantum Information}",
    eprint = "2302.05922",
    archivePrefix = "arXiv",
    primaryClass = "gr-qc",
    doi = "10.1007/978-981-19-3079-9_108-1",
    year = "2023"
}

@article{Ashtekar:2021kfp,
    author = "Ashtekar, Abhay and Bianchi, Eugenio",
    title = "{A short review of loop quantum gravity}",
    eprint = "2104.04394",
    archivePrefix = "arXiv",
    primaryClass = "gr-qc",
    doi = "10.1088/1361-6633/abed91",
    journal = "Rept. Prog. Phys.",
    volume = "84",
    number = "4",
    pages = "042001",
    year = "2021"
}

@article{Ashtekar:2004eh,
    author = "Ashtekar, Abhay and Lewandowski, Jerzy",
    title = "{Background independent quantum gravity: A Status report}",
    eprint = "gr-qc/0404018",
    archivePrefix = "arXiv",
    doi = "10.1088/0264-9381/21/15/R01",
    journal = "Class. Quant. Grav.",
    volume = "21",
    pages = "R53",
    year = "2004"
}

@article{PhysRevB.103.L241118,
  title = {Page curve for fermionic Gaussian states},
  author = {Bianchi, Eugenio and Hackl, Lucas and Kieburg, Mario},
  journal = {Phys. Rev. B},
  volume = {103},
  issue = {24},
  pages = {L241118},
  numpages = {7},
  year = {2021},
  month = {Jun},
  publisher = {American Physical Society},
  doi = {10.1103/PhysRevB.103.L241118},
  url = {https://link.aps.org/doi/10.1103/PhysRevB.103.L241118}
}

@article{PhysRevLett.119.020601,
  title = {Entanglement Entropy of Eigenstates of Quadratic Fermionic Hamiltonians},
  author = {Vidmar, Lev and Hackl, Lucas and Bianchi, Eugenio and Rigol, Marcos},
  journal = {Phys. Rev. Lett.},
  volume = {119},
  issue = {2},
  pages = {020601},
  numpages = {6},
  year = {2017},
  month = {Jul},
  publisher = {American Physical Society},
  doi = {10.1103/PhysRevLett.119.020601},
  url = {https://link.aps.org/doi/10.1103/PhysRevLett.119.020601}
}

@article{PhysRevLett.78.2275,
  title = {Quantifying Entanglement},
  author = {Vedral, V. and Plenio, M. B. and Rippin, M. A. and Knight, P. L.},
  journal = {Phys. Rev. Lett.},
  volume = {78},
  issue = {12},
  pages = {2275--2279},
  numpages = {0},
  year = {1997},
  month = {Mar},
  publisher = {American Physical Society},
  doi = {10.1103/PhysRevLett.78.2275},
  url = {https://link.aps.org/doi/10.1103/PhysRevLett.78.2275}
}

@article{PhysRevA.54.3824,
  title = {Mixed-state entanglement and quantum error correction},
  author = {Bennett, Charles H. and DiVincenzo, David P. and Smolin, John A. and Wootters, William K.},
  journal = {Phys. Rev. A},
  volume = {54},
  issue = {5},
  pages = {3824--3851},
  numpages = {0},
  year = {1996},
  month = {Nov},
  publisher = {American Physical Society},
  doi = {10.1103/PhysRevA.54.3824},
  url = {https://link.aps.org/doi/10.1103/PhysRevA.54.3824}
}

@article{PhysRevLett.130.053602,
  title = {High-Dimensional Entanglement-Enabled Holography},
  author = {Kong, Ling-Jun and Sun, Yifan and Zhang, Furong and Zhang, Jingfeng and Zhang, Xiangdong},
  journal = {Phys. Rev. Lett.},
  volume = {130},
  issue = {5},
  pages = {053602},
  numpages = {7},
  year = {2023},
  month = {Feb},
  publisher = {American Physical Society},
  doi = {10.1103/PhysRevLett.130.053602},
  url = {https://link.aps.org/doi/10.1103/PhysRevLett.130.053602}
}

@article{rowe2010shifted,
  title={The shifted harmonic approximation and asymptotic SU (2) and SU (1, 1) Clebsch--Gordan coefficients},
  author={Rowe, David J and De Guise, Hubert},
  journal={Journal of Physics A: Mathematical and Theoretical},
  volume={43},
  number={50},
  pages={505307},
  year={2010},
 DOI={10.1088/1751-8113/43/50/505307},
  publisher={IOP Publishing}
}

@article{Pastur:1967zca,
    author = "Pastur, L. A. and Mar{\v{c}}enko, V. A.",
    title = "{Distribution of Eigenvalues for Some Sets of Random Matrices}",
    doi = "10.1070/SM1967v001n04ABEH001994",
    journal = "Math. USSR Sb.",
    volume = "1",
    number = "4",
    pages = "457",
    year = "1967"
}

@book{Eynard:2016yaa,
    author = "Eynard, B.",
    title = "{Counting Surfaces}",
    doi = "10.1007/978-3-7643-8797-6",
    isbn = "978-3-7643-8796-9, 978-3-7643-8797-6",
    publisher = "Springer",
    series = "Progress in Mathematical Physics",
    volume = "70",
    year = "2016"
}

@article{tHooft1974,
  author  = {’t Hooft, Gerard},
  title   = {A planar diagram theory for strong interactions},
  journal = {Nuclear Physics B},
  volume  = {72},
  pages   = {461--473},
doi = "10.1016/0550-3213(74)90154-0",
  year    = {1974}
}

@article{mishra2025quantum,
  title={Quantum thermalization and average entropy of a subsystem},
  author={Mishra, Smitarani and Shaon, Sahoo},
  journal={Physics Letters A},
  pages={131000},
doi = "10.1016/j.physleta.2025.131000",
  year={2025},
  publisher={Elsevier}
}

@article{lydzba2021single,
  title={Single-particle eigenstate thermalization in quantum-chaotic quadratic Hamiltonians},
  author={{\L}yd{\.Z}ba, Patrycja and Zhang, Yicheng and Rigol, Marcos and Vidmar, Lev},
  journal={Physical Review B},
  volume={104},
  number={21},
  pages={214203},
  year={2021},
  publisher={APS}
}

@article{lydzba2021entanglement,
  title={Entanglement in many-body eigenstates of quantum-chaotic quadratic Hamiltonians},
  author={{\L}yd{\.z}ba, Patrycja and Rigol, Marcos and Vidmar, Lev},
  journal={Physical Review B},
  volume={103},
  number={10},
  pages={104206},
  year={2021},
  publisher={APS}
}
\end{document}